\documentstyle[epsfig]{ar}
\newcommand{\klgg}    {\mbox{$K^\circ_L \! \rightarrow \!  \gamma\gamma$ }}

\newcommand{\klmm}    {\mbox{$K^\circ_L \! \rightarrow \! \mu^+ \mu^-$ }}
\newcommand{\klee}    {\mbox{$K^\circ_L \! \rightarrow \! e^+ e^-$ }}
\newcommand{\kpx}     {\mbox{$K^+ \! \rightarrow \! \pi^+ X^\circ$ }}
\newcommand{\klme}    {\mbox{$K^\circ_L \! \rightarrow \! \mu e$ }}
\newcommand{\klpp}    {\mbox{$K^\circ_L \! \rightarrow \! \pi^+ \pi^-$ }}

\newcommand{\klppg}   {\mbox{$K^\circ_L \! \rightarrow \! \pi^+ \pi^- \gamma$ }}
\newcommand{\klppee}  {\mbox{$K^\circ_L \! \rightarrow \! \pi^+ \pi^- e^+ e^- $ }}
\newcommand{\ksppee}  {\mbox{$K^\circ_S \! \rightarrow \! \pi^+ \pi^- e^+ e^- $ }}
\newcommand{\klpme}   {\mbox{$K^\circ_L \! \rightarrow \! \pi^\circ \mu e$ }}
\newcommand{\klpmng}  {\mbox{$K^\circ_L \! \rightarrow \! \pi^\pm \mu^\mp \nu_\mu\gamma$ }}
\newcommand{\klpnn}   {\mbox{$K^\circ_L \! \rightarrow \! \pi^\circ \nu \overline{\nu}$ }}
\newcommand{\klppnn}  {\mbox{$K^\circ_L \! \rightarrow \! \pi^+ \pi^- \nu \overline{\nu}$ }}
\newcommand{\klpoponn}{\mbox{$K^\circ_L \! \rightarrow \! \pi^\circ \pi^\circ \nu \overline{\nu}$ }}
\newcommand{\klpll}   {\mbox{$K^\circ_L \! \rightarrow \! \pi^\circ \ell^+ \ell^-$ }}
\newcommand{\kopll}   {\mbox{$K^\circ \! \rightarrow \! \pi^\circ \ell^+ \ell^-$ }}
\newcommand{\kpnn}    {\mbox{$K^+ \! \rightarrow \! \pi^+ \nu \overline{\nu}$ }}
\newcommand{\kppnn}   {\mbox{$K^+ \! \rightarrow \! \pi^+ \pi^\circ \nu \overline{\nu}$ }}
\newcommand{\kzpme}   {\mbox{$K \! \rightarrow \! \pi \mu e$}}
\newcommand{\kpme}    {\mbox{$K^+ \! \rightarrow \! \pi^+ \mu^+ e^-$}}
\newcommand{\kzppen}  {\mbox{$K \! \rightarrow \! \pi \pi e \nu_e$ }}
\newcommand{\kzppln}  {\mbox{$K \! \rightarrow \! \pi \pi \ell \nu_\ell$ }}
\newcommand{\klppen}  {\mbox{$K^\circ_L \! \rightarrow \! \pi^\pm \pi^\circ e^\mp \nu_e$ }}
\newcommand{\kppen}   {\mbox{$K^+ \! \rightarrow \! \pi^+ \pi^- e^+ \nu_e$ }}
\newcommand{\kppeng}  {\mbox{$K^+ \! \rightarrow \! \pi^+ \pi^- e^+ \nu_e\gamma$ }}
\newcommand{\kpopoen} {\mbox{$K^+ \! \rightarrow \! \pi^\circ \pi^\circ e^+ \nu_e$ }}
\newcommand{\kpopoeng}{\mbox{$K^+ \! \rightarrow \! \pi^\circ \pi^\circ e^+ \nu_e\gamma$ }}
\newcommand{\klppmn}  {\mbox{$K^\circ_L \! \rightarrow \! \pi^\circ \pi^\pm \mu^\mp \nu_\mu$ }}
\newcommand{\kppmn}   {\mbox{$K^+ \! \rightarrow \! \pi^+ \pi^- \mu^+ \nu_\mu$ }}

\newcommand{\kppp}    {\mbox{$K^+ \! \rightarrow \! \pi^+ \pi^+ \pi^-$ }}
\newcommand{\kpppg}   {\mbox{$K^+ \! \rightarrow \! \pi^+ \pi^+ \pi^- \gamma$ }}
\newcommand{\kppopog} {\mbox{$K^- \! \rightarrow \! \pi^- \pi^\circ \pi^\circ \gamma$ }}
\newcommand{\kmn}     {\mbox{$K^+ \! \rightarrow \! \mu^+ \nu_\mu$ }}
\newcommand{\kmng}    {\mbox{$K^+ \! \rightarrow \! \mu^+ \nu_\mu \gamma$ }}
\newcommand{\keng}    {\mbox{$K^+ \! \rightarrow \! e^+ \nu_\mu \gamma$ }}
\newcommand{\kmnee}   {\mbox{$K^+ \! \rightarrow \! \mu^+ \nu e^+ e^-$ }}
\newcommand{\kmnmm}   {\mbox{$K^+ \! \rightarrow \! \mu^+ \nu \mu^+ \mu^-$ }}
\newcommand{\kenee}   {\mbox{$K^+ \! \rightarrow \! e^+ \nu e^+ e^-$ }}
\newcommand{\kenmm}   {\mbox{$K^+ \! \rightarrow \! e^+ \nu \mu^+ \mu^-$ }}
\newcommand{\kln}     {\mbox{$K^+ \! \rightarrow \! \ell^+ \nu$ }}

\newcommand{\klnll}   {\mbox{$K^+ \! \rightarrow \! \ell^+ \nu 
\ell'^+ \ell'^-$ }}
\newcommand{\klng}    {\mbox{$K^+ \! \rightarrow \! \ell^+ \nu \gamma$ }}

\newcommand{\kleeg}   {\mbox{$K^\circ_L \! \rightarrow \! e^+ e^-\gamma$ }}
\newcommand{\kleegg}  {\mbox{$K^\circ_L \! \rightarrow \! e^+ e^-\gamma
\gamma$ }}
\newcommand{\klmmg}   {\mbox{$K^\circ_L \! \rightarrow \! \mu^+ \mu^- 
\gamma$ }}
\newcommand{\klmmgg}  {\mbox{$K^\circ_L \! \rightarrow \! \mu^+ \mu^- 
\gamma\gamma$ }}
\newcommand{\kleeee}  {\mbox{$K^\circ_L \! \rightarrow \! e^+ e^- e^+ e^-$ }}
\newcommand{\klbeemm} {\mbox{$K^\circ_L \! \rightarrow \! e^\pm e^\pm 
\mu^\mp \mu^\mp$ }}
\newcommand{\klmmee}  {\mbox{$K^\circ_L \! \rightarrow \! \mu^+ \mu^- 
e^+ e^-$ }}
\newcommand{\klmmmm}  {\mbox{$K^\circ_L \! \rightarrow \! \mu^+ \mu^- 
\mu^+ \mu^-$ }}
\newcommand{\klllg}   {\mbox{$K^\circ_L \! \rightarrow \! \ell^+ \ell^- 
\gamma$ }}
\newcommand{\klllee}  {\mbox{$K^\circ_L \! \rightarrow \! \ell^+ \ell^- 
e^+ e^-$ }}
\newcommand{\klllgg}  {\mbox{$K^\circ_L \! \rightarrow \! \ell^+ \ell^- \gamma\gamma$ }}

\newcommand{\bpsiks}  {\mbox{$B^\circ_d \! \rightarrow \! \psi K^\circ_S$ }}
\newcommand{\bsbd}    {\mbox{$\Delta M_{B_s}/\Delta M_{B_d}$ }}
\newcommand{\kspp}    {\mbox{$K^\circ_{\rm S} \! \rightarrow \! \pi^+ \pi^- $ }}
\newcommand{\ksppp}   {\mbox{$K^\circ_{\rm S} \! \rightarrow \! \pi^+ \pi^-\pi^\circ $ }}
\newcommand{\ksmm}    {\mbox{$K^\circ_{\rm S} \! \rightarrow \! \mu^+ \mu^- $ }}
\newcommand{\ksee}    {\mbox{$K^\circ_{\rm S} \! \rightarrow \! e^+ e^- $ }}
\newcommand{\ksgg}    {\mbox{$K^\circ_{\rm S} \! \rightarrow \! \gamma \gamma$ }}
\newcommand{\ksppg}   {\mbox{$K^\circ_{\rm S} \! \rightarrow \! \pi^+ \pi^- \gamma$ }}
\newcommand{\kspopo}  {\mbox{$K^\circ_{\rm S} \! \rightarrow \! \pi^\circ \pi^\circ $ }}
\newcommand{\kspopopo}{\mbox{$K^\circ_{\rm S} \! \rightarrow \! \pi^\circ \pi^\circ \pi^\circ$ }}
\newcommand{\kspgg}   {\mbox{$K^\circ_{\rm S} \! \rightarrow \! \pi^\circ \gamma \gamma$ }}
\newcommand{\kspee}   {\mbox{$K^\circ_{\rm S} \! \rightarrow \! \pi^\circ e^+ e^- $ }}
\newcommand{\klpopo}  {\mbox{$K^\circ_L \! \rightarrow \! \pi^\circ \pi^\circ$ }}
\newcommand{\klpopopo}{\mbox{$K^\circ_L \! \rightarrow \! \pi^\circ \pi^\circ \pi^\circ$ }}
\newcommand{\klpopog} {\mbox{$K^\circ_L \! \rightarrow \! \pi^\circ \pi^\circ \gamma$ }}
\newcommand{\kzpen}   {\mbox{$K\!\rightarrow\!\pi e \nu_e$ }}
\newcommand{\kpen}    {\mbox{$K^+\!\rightarrow\!\pi^\circ e^+\nu_e$ }}
\newcommand{\kpeng}   {\mbox{$K^+ \! \rightarrow \! \pi^\circ e^+ \nu_e \gamma$ }}

\newcommand{\kpmng}   {\mbox{$K^+ \! \rightarrow \! \pi^\circ \mu^+ \nu_\mu \gamma$ }}
\newcommand{\klpeng}  {\mbox{$K^\circ_L \! \rightarrow \! \pi^\pm e^\mp \nu_e\gamma$ }}
\newcommand{\kpp}     {\mbox{$K^+ \! \rightarrow \! \pi^+ \pi^\circ$ }}
\newcommand{\klpen}   {\mbox{$K^\circ_L \! \rightarrow \! \pi^\pm e^\mp \nu_e$ }}
\newcommand{\kpll}    {\mbox{$K^+ \! \rightarrow \! \pi^+ \ell^+ \ell^-$ }}
\newcommand{\kpmm}    {\mbox{$K^+ \! \rightarrow \! \pi^+ \mu^+ \mu^-$ }}
\newcommand{\kpee}    {\mbox{$K^+ \! \rightarrow \! \pi^+ e^+ e^-$ }}
\newcommand{\kppg}    {\mbox{$K^+ \! \rightarrow \! \pi^+ \pi^\circ \gamma$ }}
\newcommand{\kpgg}    {\mbox{$K^+ \! \rightarrow \! \pi^+ \gamma \gamma$ }}
\newcommand{\klpgg}   {\mbox{$K^\circ_L \! \rightarrow \! \pi^\circ \gamma\gamma$ }}
\newcommand{\kpeeg}   {\mbox{$K^+ \! \rightarrow \! \pi^+ e^+ e^-\gamma$ }}
\newcommand{\klpee}   {\mbox{$K^\circ_L \! \rightarrow \! \pi^\circ e^+ e^-$ }}
\newcommand{\klpeeg}  {\mbox{$K^\circ_L \! \rightarrow \! \pi^\circ e^+ e^- \gamma $ }}
\newcommand{\klpmm}   {\mbox{$K^\circ_L \! \rightarrow \! \pi^\circ \mu^+ \mu^-$ }}
\newcommand{\kzpnn}    {\mbox{$K \! \rightarrow \! \pi \nu \overline{\nu}$ }}

\newcommand{\kzppg}    {\mbox{$K \! \rightarrow \! \pi \pi \gamma$ }}
\newcommand{\kzpgg}    {\mbox{$K \! \rightarrow \! \pi \gamma \gamma$ }}

\newcommand{\vtd}     {\mbox{$V_{td}$ }}
\newcommand{\vts}     {\mbox{$V_{ts}$ }}

\newcommand{\vus}     {\mbox{$V_{us}$ }}

\newcommand{\vcb}     {\mbox{$V_{cb}$ }}

\begin{document}

\title{ 
{
\vspace{-1.25cm}
\begin{flushright}
\small\rm 
BNL-67590\\
August 1, 2000\\
Revised November 8, 2000
\end{flushright}
\vspace{.5cm}
}
Developments in Rare Kaon Decay Physics\footnote
{With permission from the Annual Review of Nuclear and Particle Science.
Final version of this material is scheduled to appear in the Annual Review
of Nuclear and Particle Science, Vol. 50, to be published in December 2000
by Annual Reviews ( http://www.AnnualReviews.org )}
 }
\markboth{Barker \& Kettell}{Rare Kaon Decay Physics}

\author{A.R.~Barker\affiliation{Department of Physics, University of Colorado, 
Boulder, Colorado 80309;\\ e-mail: tonyb@cuhep.colorado.edu}
S.H.~Kettell\affiliation{Physics Department, Brookhaven National Laboratory, 
Upton, New York 11973;\\ e-mail: kettell@bnl.gov}}


\begin{abstract}
We review the current status of the field of rare kaon decays. The study of
rare kaon decays has played a key role in the development of the 
standard model, and the field continues to have significant impact. The two 
areas of greatest import are the search for physics beyond the standard 
model and the determination of fundamental standard-model parameters.
Due to the exquisite sensitivity of
rare kaon decay experiments, searches for new physics can 
probe very high mass scales. 
Studies of the \kzpnn modes in particular, where the first event has
recently been seen, will permit tests 
of the standard-model picture 
of quark mixing and {\it CP} violation.

\end{abstract}

\maketitle

\section{INTRODUCTION}

This article reviews the status of
rare kaon decays, with emphasis on the progress made since the 
1993 review in this series~\cite{litt}. 
Several other excellent review articles are 
available focusing on rare kaon decays~\cite{hagelin},
theoretical studies of rare kaon 
decays~\cite{buchalla,bf,buras,dambrosio}, and non-rare 
kaon decays~\cite{winstein}.
Due to limited space, we cannot cover many interesting topics, such as
{\it CP} violation in \klpp decays---$\epsilon$ and $\epsilon'$, or
{\it T/CPT} violation in \klpp decays, or searches for {\it T}~violation
in the transverse polarization of the $\mu^+$ in \kpmng and \kmng.

Kaons have a relatively long lifetime because they decay only through
the weak interaction.  As a result, studies of their decays provide
key insights into the behavior of the weak interaction under the
three fundamental symmetry operators {\it C}, {\it P}, and {\it T}.
The first of these, {\it C} or charge conjugation, is a unitary operator that
replaces particles by anti-particles.  Thus, in one possible sign convention, 
$C\vert K^\circ\rangle = -\vert\overline K^\circ\rangle$ and
$C\vert K^+\rangle = -\vert\overline K^-\rangle.$
The parity operator, {\it P}, inverts spatial directions, replacing
left by right and vice-versa.  The kaons are pseudoscalar particles
which are odd under the action of {\it P}.  Under the combined operator
{\it CP}, then
\begin{eqnarray}
 CP\vert K^\circ\rangle &= \vert \overline K^\circ\rangle \\ \nonumber
 CP\vert \overline K^\circ\rangle &= \vert K^\circ\rangle  \label{eqn:k0k0bar}
\end{eqnarray}
Even and odd eigenstates of {\it CP} called $K_1$ and $K_2$ can then be formed 
from the symmetric
and antisymmetric combinations of the $K^\circ$ and $\overline K^\circ.$
If {\it CP} were an exact symmetry of the weak interaction, these
combinations would be identified with
the observed eigenstates of mass and lifetime, called $K^\circ_S$ and $K^\circ_L.$
The famous discovery in 1964~\cite{fitch} of the decay 
$K^\circ_L\rightarrow\pi\pi$ implied that {\it CP} symmetry is violated in weak decays,
since the $K_L,$ which decays mostly to {\it CP}-odd final states, can also
decay to $\pi\pi,$ which is {\it CP}-even.  We have since learned that
nearly all of the $K_L\rightarrow\pi\pi$ decay can be explained by
so-called {\sl indirect} {\it CP} violation, in which the mass and lifetime
eigenstates $K_L$ and $K_S$ are mixtures of the {\it CP} eigenstates given by
\begin{eqnarray}
 \vert K^\circ_S\rangle &= \left(\vert K_1\rangle 
      + \epsilon\,\vert K_2\rangle\right)/\sqrt{1 + \vert\epsilon\vert^2}
      \\ \nonumber
 \vert \overline K^\circ_L \rangle &= \left(\vert K_2\rangle 
      + \epsilon\,\vert K_1\rangle\right)/\sqrt{1 + \vert\epsilon\vert^2},
       \label{eqn:klksdef}
\end{eqnarray}
and the decay proceeds via $\epsilon\,\vert K_1\rangle$.
A question that has been open until recently is whether there is also
{\sl direct} {\it CP} violation, in which the {\it CP}-odd eigenstate $K_2$ decays to
{\it CP}-even final states such as $\pi\pi.$  The traditional method of
searching for this phenomenon, which is expected in the Standard Model,
is to look for a small deviation from unity of the so-called double ratio
\begin{equation}
 R = {\Gamma(K_L\rightarrow\pi^\circ\pi^\circ)/\Gamma(K_S\rightarrow\pi^\circ\pi^\circ)
       \over
       \Gamma(K_L\rightarrow\pi^+\pi^-)/\Gamma(K_S\rightarrow\pi^+\pi^-)}. 
\end{equation}
The value of this ratio has been reported~\cite{ktev_e,na48_e} 
to be significantly
different from unity, establishing the existence of direct
{\it CP} violation in weak interactions.
The Standard Model predicts a variety of other direct-{\it CP}-violating effects
in rare kaon decays; measurements of these processes, which are addressed
in this article, can provide additional tests of the Standard Model
picture of {\it CP} violation.

The anti-unitary operator {\it T} reverses the arrow of time.  In
Lorentz-invariant local field theories, like the Standard Model, 
the combined operator {\it CPT} is an exact symmetry of the Lagrangian.
Thus the observed {\it CP} violation in kaon decays would imply the
existence of {\it T} violation.  However, it is also interesting to
search for more direct evidence of {\it T} violation, and a number of
kaon-decay experiments have also played a central role in this effort.

The field of rare kaon decays has a long and rich history: the
discovery of the kaon in 1949~\cite{brown},  the postulation of
``strangeness''~\cite{pais}, the $\tau$--$\theta$ puzzle~\cite{dalitz} and 
the understanding of parity violation~\cite{lee}, 
the understanding of quark mixing~\cite{cabibbo,kobayashi}, the discovery of
{\it CP} violation~\cite{fitch}, 
the small rate for \klmm and 
flavor-changing neutral currents (FCNCs) in
general, and the development
of the Glashow, Iliopoulos, Maiani (GIM) mechanism~\cite{glashow}
and the prediction of the charm quark mass~\cite{GL}.  
As the field has evolved, so has the definition of ``rare'' decays, from 
branching ratios of $\sim$10$^{-3}$ to the current levels of $\sim$10$^{-12}$.

This article will, in general,  cover modes with branching ratios below 
$\sim10^{-5}$,
with one measured to be less than $10^{-11}$.
The two areas of
greatest interest have been the very sensitive searches for physics
beyond the standard model through lepton flavor-violating (LFV)
decays and the studies of the standard-model picture of the Cabibbo, Kobayashi,
Maskawa (CKM) mixing~\cite{kobayashi} and {\it CP}
violation that have recently begun to bear fruit.

A large number of results 
from  experiments at Brookhaven National Laboratory (BNL) 
(E787, E865, E871), Fermi National Accelerator Laboratory (FNAL) (E799-II: KTeV),
and the European Laboratory for Particle Physics (CERN) (NA48)
have been reported at recent 
conferences 
\cite{ichep98,dpf99,moriond99,panic99,epshep99,kaon99,hf99,lp99,daphne99,bconf99,lathuile00,moriond00,ichep00,dpf00}.
Many of these results have not yet been published.

\subsection{Standard Model ``Golden Modes'' and CKM Matrix}

The weak decay of quarks is described through the unitary
CKM matrix. This matrix and the Wolfenstein
parameterization~\cite{wolfenstein,buras2} are shown below:
\begin{eqnarray}
 V_{\rm CKM} & = & \left( \begin{array}{ccc} V_{ud} & V_{us} & V_{ub} \\
                          V_{cd} & V_{cs} & V_{cb} \\
                          V_{td} & V_{ts} & V_{tb} \end{array} \right) \\
 \nonumber
 & \simeq & \left( \begin{array}{ccc} 1-\lambda^2/2 
& \lambda  & A\lambda^3(\rho-i\eta) \\
                          -\lambda      &  1-\lambda^2/2 & A\lambda^2 \\
                    A\lambda^3(1-\rho-i\eta) & -A\lambda^2 
& 1 \end{array} \right) + {\cal O}(\lambda^4) \\ \nonumber
 \vspace{2mm} & \simeq & 
\left( \begin{array}{ccc} 1-\frac{\lambda^2}{2}-\frac{\lambda^4}{8} 
& \lambda  
& A\lambda^3(\rho-i\eta) \\ 
\vspace{2mm}

\!\!\!-\lambda+\frac{A^2\lambda^5(1-2\rho-2i\eta)}{2} 
&  1-\frac{\lambda^2}{2}-\frac{\lambda^4}{8}(1+4A^2) 
& A\lambda^2 \\ 
A\lambda^3(1-\overline{\rho}-i\overline{\eta}) 
& -A\lambda^2[1 - \lambda^2\frac{1-2\rho-i2\eta}{2}] 
& 1 - \frac{A^2\lambda^4}{2} \end{array} \right),
\label{eqn:ckm}
\end{eqnarray}
where $\lambda$ is the sin of the Cabibbo angle, $\lambda\equiv\sin \theta_C \simeq 0.22$,
and $\overline{\rho}$ and $\overline{\eta}$ are related to the Wolfenstein parameters
$\rho$ and $\eta$ by $\overline{\rho} \equiv \rho(1-\lambda^2/2)$ and
$\overline{\eta} \equiv \eta(1-\lambda^2/2)$.

\begin{figure}[htbp]
\epsfig{file=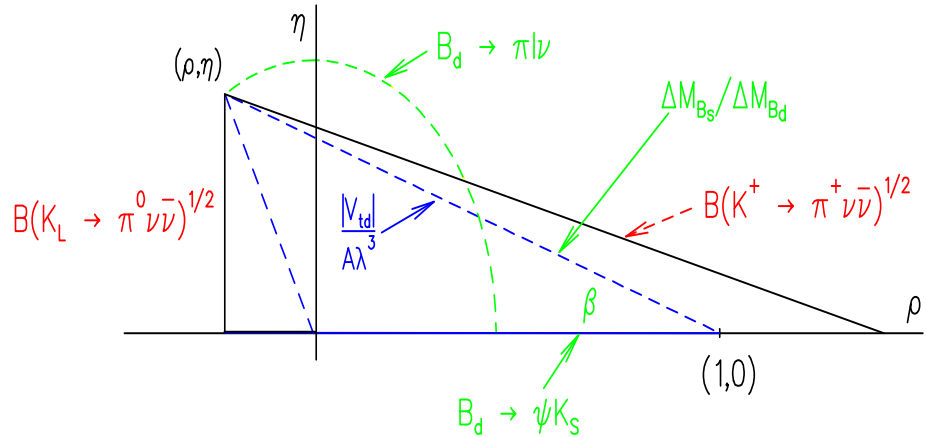,height=4.75in,width=2.25in,angle=90}
\caption{Traditional representation of the unitarity triangle. Measurements of
B meson decays introduce constraints shown in green,  contributions from the two 
golden kaon decay modes are marked in red.
\label{fig:tri_norm}}
\end{figure}

The unitarity of this matrix can be expressed in terms of six unitarity 
conditions which can be represented graphically in the form of triangles,
all of which have the same area. The area of these triangles is
equal to one half of the Jarlskog invariant, $J_{CP}$~\cite{jarlskog}.
This is the fundamental measure of {\it CP} violation in the standard model.
One of the possible unitarity relations that is frequently cited in the literature
is
\begin{equation}
V_{ub}^*V_{ud} + V_{cb}^*V_{cd} + V_{tb}^*V_{td} = 0.
\label{eqn:tri_b}
\end{equation}
This equation can be represented graphically, as in
Figure~\ref{fig:tri_norm}, where we have divided all sides by
$V_{cb}^*V_{cd}$, which is a real quantity to ${\cal O}(\lambda^6)$.
This particular representation provides a convenient display, with the apex 
of the triangle given
by the two least well-known of the Wolfenstein parameters, 
$\overline{\rho}$ and $\overline{\eta}$.
The best information currently comes from several measurements of
B meson decays, as well as the measured value of $\epsilon$ from 
kaon decays.
All of the unitarity triangles should be tested; it is desirable to
overconstrain each of the unitarity relations and to measure $J_{CP}$
in each of the triangles.

The most powerful tests of our understanding of {\it CP}-violation and
quark mixing will come from comparison of the results from B meson and kaon 
decays with little theoretical ambiguity. The two  premier tests
are expected to be:
\begin{itemize}
  \item Comparison of the angle $\beta$ from the ratio B(\klpnn)/B(\kpnn)
and the {\it CP}-violating asymmetry in the decay \bpsiks~\cite{sinb,bergmann}.
  \item  Comparison of the magnitude $|\vtd|$ from \kpnn and the ratio 
of the mixing frequencies of $B_s$ to $B_d$ mesons, expressed in terms of
the mass difference ratio \bsbd~\cite{bb3,bergmann}.
\end{itemize}

The current value of the fundamental level of {\it CP} violation in the SM,
$J_{CP}$ = $(2.7\pm1.1)\times10^{-5}$, is known, primarily from measurements
of B meson decays, with about 40\% 
uncertainty~\cite{marciano2}.
Measurement of $J_{CP}$
 in the kaon system is very clean theoretically (uncertainty of $\sim$2\%)
and can be expected to be measured to $\sim$8\% within a decade. 
While measurement of $J_{CP}$ in the B system is  difficult and is plagued
by theoretical uncertainties, it is likely that a 15\% measurement is
possible and if this could be pushed to the level expected from
the kaon system, the comparison of these values will also be an 
important test of the SM.

\subsection{Form Factor Measurements}

        Interest in rare kaon decays extends well beyond their
        potential to determine standard-model parameters.  
        Dozens of different medium-rare (branching ratios in
        the range $10^{-5}$ to $10^{-8}$) kaon decays have been
        measured.  With the ever-increasing sensitivity of
        experiments designed to search for the very-rare
        modes that probe standard-model parameters or search for new physics,
        the statistics available for these medium-rare
        decays have increased to the point where both
        precision branching ratio measurements and form factor
        studies are possible.

        Both the branching ratios and the form factors provide
        excellent tests of chiral perturbation theory 
	(ChPT)~\cite{chpt,chpt_gl},
        which should work well at the relatively low momentum
        scales characteristic of kaon decays.  The wide variety of
        different modes and form factors can be used to test
        ChPT.

        Measurements of a number of modes, such as \kleegg
        and \klpgg, are directly relevant to the determination
        of standard-model parameters because these modes can be backgrounds
        to more interesting decays, such as \klpee or \klpnn. They
        can also provide information necessary to disentangle different
        amplitudes contributing to the signal mode, such as
        the $\pi^\circ\gamma^*\gamma^*$ intermediate state for
        \klpee or the $\gamma^*\gamma^*$ intermediate state for
        \klmm.

        The study of these ``non-marquee'' decay modes is thus more
        than a beneficial by-product of experiments designed to search 
        for the more significant decays. Their properties are often
        of vital importance to the determination of backgrounds, 
        the extraction of standard-model parameters, or  tests of the 
        reliability of theoretical tools like ChPT.

\subsection{Searches for New Physics}

        A major thread in the history of the study of rare
        kaon decays is the search for exotic phenomena, often
        referred to as ``beyond the standard model'' 
        (BSM).  The quintessential example is the long search
        for the decay \klme.  This decay is absolutely forbidden
        in the standard model with massless neutrinos; 
        specifically, it
        is forbidden by the symmetry of conserved lepton flavor
        number, for which no fundamental reason is known.
        Grand unified theories or other extensions to the
        standard model often contain heavy vector bosons that connect the
        standard-model lepton families---for example, coupling muons to
        electrons (horizontal gauge bosons) or quarks to
        leptons (leptoquarks).  Both types of particles could
        mediate the otherwise forbidden decays, such as \klme or \kzpme.
        
        Because these decays simply do not happen in the standard model
        and are relatively simple to detect, they provide
        exceptional sensitivity to BSM physics.
        With experiments now probing branching ratios at the level of
        $10^{-12}$, even a very 
        heavy exotic boson (on the order of 100~TeV, for the 
        usual electroweak coupling) would lead to
        a detectable signal.

\section{{\it CP} VIOLATION AND THE CKM MATRIX}

\begin{figure}[htbp]
\epsfig{file=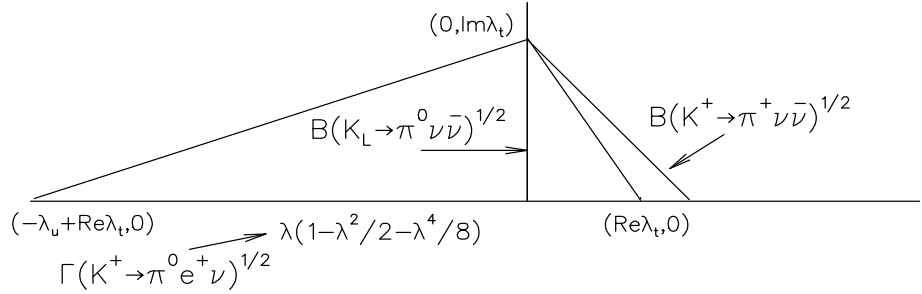,height=4.75in,width=1.5in,angle=90}
\caption{Unitarity triangle for the K system (not to scale).
\label{fig:tri_k}}
\end{figure}

The unitarity triangle is most readily expressed for the kaon system as follows:
\begin{eqnarray}
V_{us}^*V_{ud} + V_{cs}^*V_{cd} + V_{ts}^*V_{td} & = & 0 \\ \nonumber
{\rm or } & & \\ \nonumber
\lambda_u + \lambda_c + \lambda_t & = & 0, 
\label{eqn:tri_k}
\end{eqnarray}
with the three vectors $\lambda_i \equiv V^*_{is}V_{id}$ converging to
form a very elongated triangle in the complex plane. 
This is illustrated graphically in Figure~\ref{fig:tri_k}.
The first vector, $\lambda_u=V^*_{us}V_{ud}$, is well known.
The height will be measured by \klpnn and the third vector,
$\lambda_t=V^*_{ts}V_{td}$, will be measured by the decay \kpnn.  The theoretical
ambiguities in interpreting all of these measurements are very small.  It
may be possible to extract additional constraints on the height of the triangle
from \klpll decays and on $Re(\lambda_t)$ from \klmm decays.

The base of this triangle has the length $b \equiv \lambda_u = V^*_{us}V_{ud}$, 
determined from
the decay rate of \kzpen and nuclear beta decay. If we assume unitarity then
$b$ is determined completely from \kzpen 
and $b= |V_{us}|$  to very good approximation.  
To even better accuracy it is expressed as
\begin{equation} 
b = \lambda -\frac{\lambda^3}{2}-\frac{\lambda^5}{8}.
\end{equation}
The value of $\lambda$,
the best-known of the Wolfenstein parameters,
is extracted~\cite{leutwyler} from the
measurement of the \kzpen rate~\cite{pdg}. 
The height of the triangle, $h \equiv Im(\lambda_t)$
can be derived from a measurement of the \klpnn branching ratio.
The area of the triangle, $a$, is then given by
two kaon decay measurements as
\begin{equation}
J_{CP} = 2a = b\times h = \lambda_u\times Im(\lambda_t) = 0.976\times\lambda\times Im(\lambda_t);
\end{equation}
the ultimate
uncertainty on $Im(\lambda_t)$ and $a$ will be limited, not by theoretical
ambiguities, but by experimental uncertainties on B(\klpnn), to ${\cal O}(5-10\%)$
from the next round of \klpnn experiments.
This compares favorably to the B system, where three (four without the unitarity
assumption) measurements are needed.

\renewcommand{\thefootnote}{\alph{footnote}}
Table~\ref{tab:ckm} lists current values~\cite{pdg,rosner,ckmfit} for 
the magnitudes of the CKM matrix elements and Wolfenstein parameters.
\begin{table}[ht]
\centering
\begin{minipage}[h]{\linewidth}
\caption{Magnitudes of CKM matrix parameters. The current
values for the matrix elements
$V_{ji}$ are listed, where $i$ loops over the $d$-type and 
$j$ represents the $u$-type
quarks, as are the
$\lambda_j\equiv V^*_{js}V_{jd}$ values as defined earlier in the text and the 
Wolfenstein parameters ($\lambda$, A, $\rho$ and $\eta$).}
\vskip 0.1 in
\begin{tabular}{|l||l|l|l||l|} \hline
V$_{ji}$  &  \multicolumn{1}{|c|}{i=d} & \multicolumn{1}{|c|}{i=s} & 
\multicolumn{1}{|c||}{i=b} & \multicolumn{1}{|c|}{$\lambda_j\equiv V^*_{js}V_{jd}$} \\ \hline\hline
V$_{ui}$  & $0.9740\pm0.0010$ & $0.2196\pm0.0023$ & $0.0032\pm0.0008$ 
& $0.2139\pm0.0026$\\ \hline
V$_{ci}$  & $0.224\pm0.016$  & $1.04\pm0.16$  & $0.0395\pm0.0017$ 
& $0.233\pm0.040$\\ \hline
 V$_{ti}$  & $0.0084\pm0.0018$\footnote[1]{The entries for \vtd and \vts
assume a 3 generation unitary matrix.}  & \multicolumn{1}{|c|}
{$\sim$\vcb\footnotemark[1]}
& $0.99\pm0.29$ 
& $.00033\pm.00009$\\ \hline\hline
$\lambda$ & \multicolumn{4}{|c|}{$0.2196\pm0.0023$~\cite{pdg}} \\ \hline
A         & \multicolumn{4}{|c|}{$0.819\pm0.039$~\cite{pdg}} \\ \hline
$\rho$    & \multicolumn{4}{|c|}{$0.14\pm0.15$~\cite{rosner} \hspace{0.5cm}
($0.18\pm0.04$~\cite{ckmfit})\footnote[2]{The uncertainties on $\rho$ and $\eta$
are conservative, generally accepted values. Values from more aggressive treatment of errors are given in parentheses.}} \\ \hline
$\eta$    & \multicolumn{4}{|c|}{$0.38\pm0.13$~\cite{rosner} \hspace{0.5cm} 
($0.36\pm0.03$~\cite{ckmfit})\footnotemark[2]} \\ \hline
\end{tabular}
\label{tab:ckm}
\end{minipage}
\end{table}
\renewcommand{\thefootnote}{\arabic{footnote}}

\subsection{\klmm}

\label{sec:klmm}
\begin{figure}[htb]
\epsfig{file=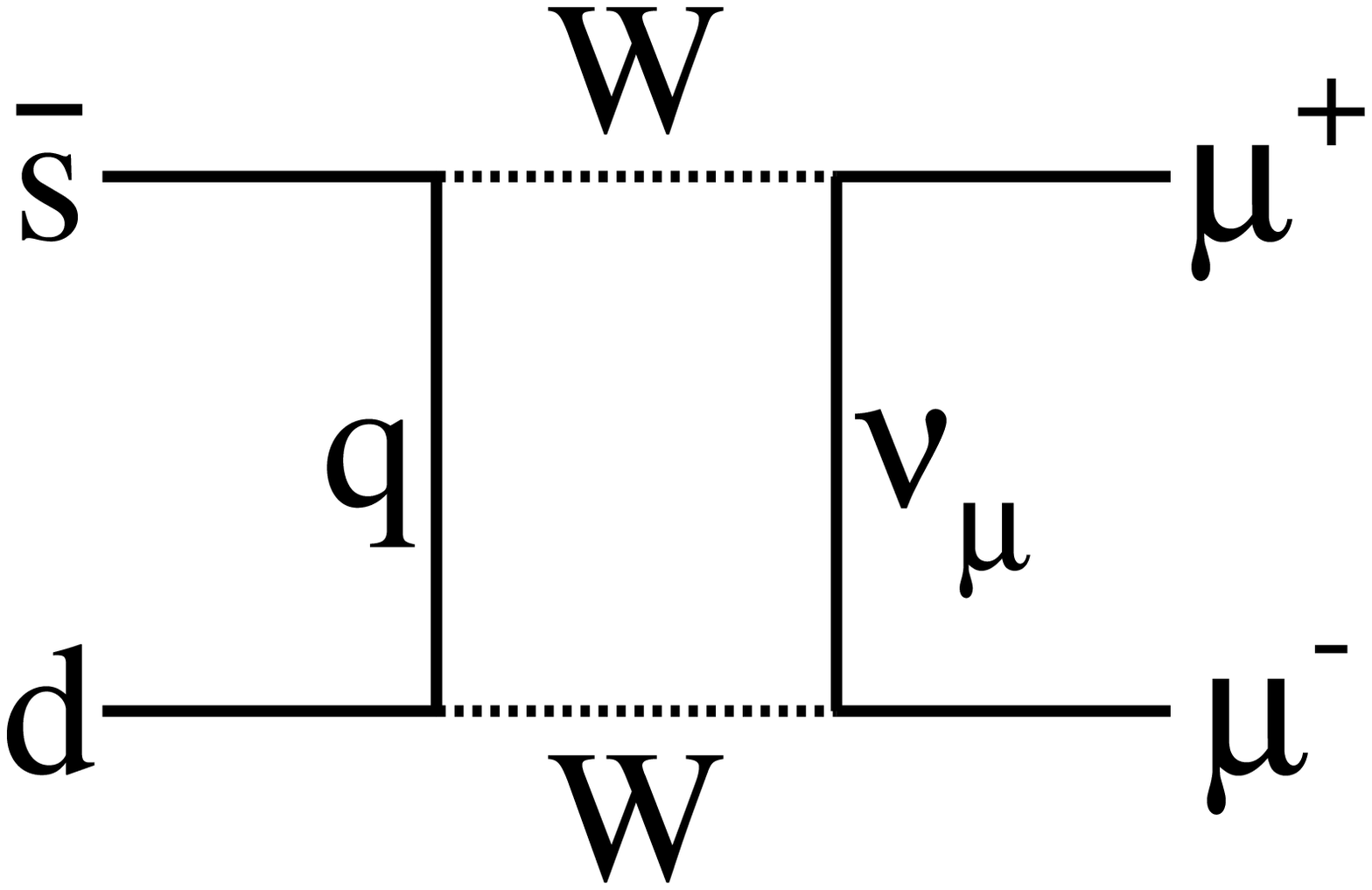,height=1.15in,width=1.55in,angle=0} \hspace{0.25cm}
\epsfig{file=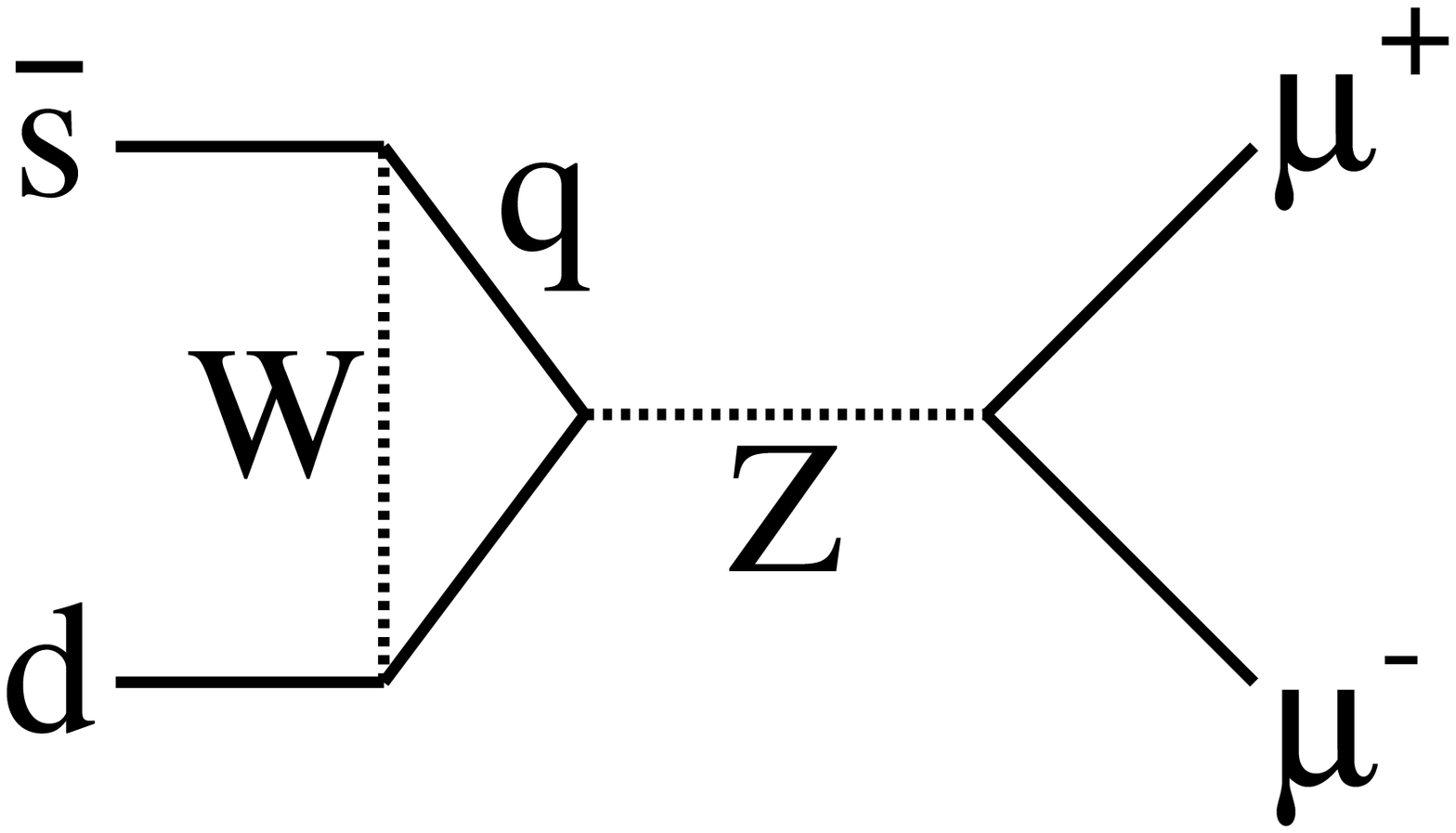,height=1.15in,width=1.55in,angle=0} \hspace{0.25cm}
\epsfig{file=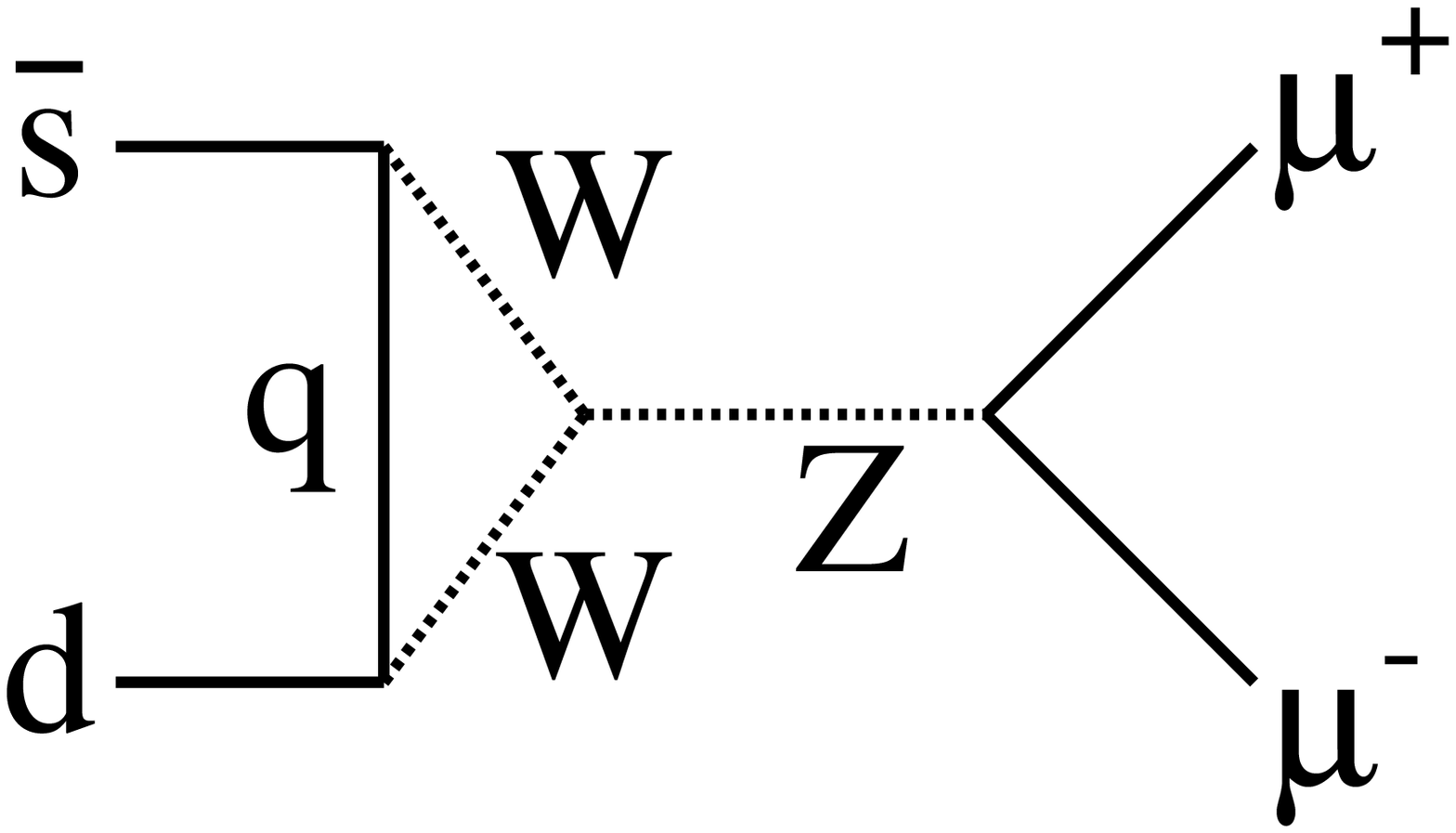,height=1.15in,width=1.55in,angle=0}
\caption{Feynman diagrams for the short-distance component of the decay \klmm.
\label{fig:feyn_klmm}}
\end{figure}

\begin{figure}[htb]
\epsfig{file=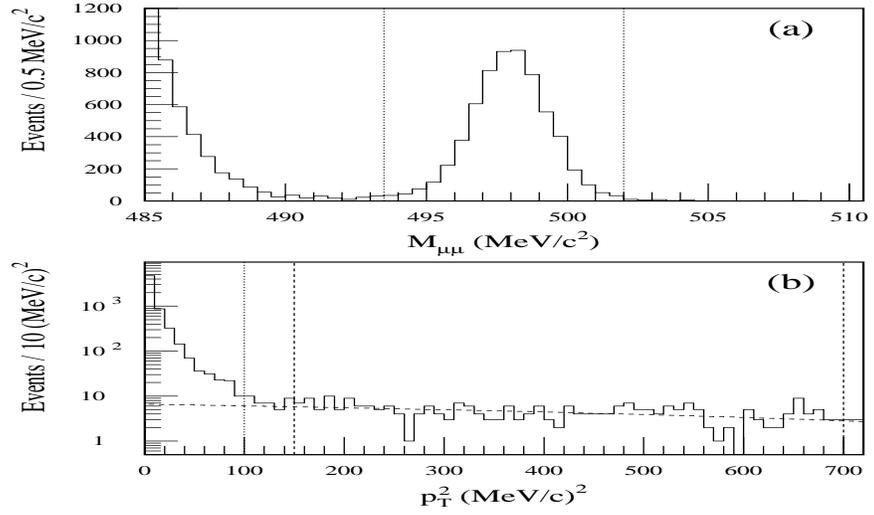,height=2.75in,width=5.25in,angle=0}
\caption{Final sample of \klmm decays from experiment E871 at BNL.
A total of $\sim$6200 \klmm events are observed in the peak. a) 
Reconstructed mass of the $\mu^+\mu^-$ pair, $M_{\mu\mu}$ and  b) the momentum of the
reconstructed $\mu^+\mu^-$ pair relative to that of the parent kaon ($p_T$), where
the direction of the parent kaon is derived from the locations of the target and the
decay vertex.
\label{fig:e871_mm}}
\end{figure}

The decay \klmm is dominated by the process of \klgg with the two real photons converting
to a $\mu^+\mu^-$ pair. This contribution can be precisely calculated
in QED~\cite{sehgal} 
based on a measurement of the \klgg branching ratio. However, there is
also a long-distance dispersive contribution, through off-shell
photons. This contribution needs additional input from 
ChPT~\cite{dambrosio1,dumm},
which may be aided by new, improved measurements of the decays 
\kleeg, \klmmg, \kleeee
and \klmmee (see Section~\ref{sec:klgg}), although there is some dispute as to
the reliability of such calculations~\cite{valencia,derafael}.  Most interesting
is the short-distance contribution which proceeds through 
internal quark loops,
dominated by the top quark (see Figure~\ref{fig:feyn_klmm}).
This contribution is sensitive to the real part of the
poorly known CKM matrix element \vtd or equivalently to $\rho$~\cite{buras,geng}.
If this
were the only contribution to the decay, the branching ratio $B_{SD}(\klmm)$
could be written as
\begin{eqnarray}
B_{SD}(\klmm) & = & \frac{\tau_L}{\tau_{K^+}}\frac{\alpha^2 B(K_{\mu2})}
		{\pi^2 \sin^4 \theta_W |V_{us}|^2}
[Y_c Re(\lambda_c) + Y_t Re(\lambda_t)]^2 \\ \nonumber
              & = & 1.51 \times 10^{-9} A^4 (\rho_0-\overline{\rho})^2,
\end{eqnarray}
with $\rho_0 = 1.2$ and the Inami-Lim functions~\cite{buras,inami}, $Y_q$,
are functions of $x_q \equiv$ $M_q^2$/$M_W^2$ where $M_W$ is the mass
of the $W$ boson and $M_q$ is the mass of the quark $q$.
This mode has now been measured
with impressively high statistics~\cite{e871_mm} 
(see Figure~\ref{fig:e871_mm}) by the E871
collaboration (see Section~\ref{sec:e871}).
The branching ratio, B(\klmm) $= (7.18 \pm 0.17) \times 10^{-9}$, is a
factor of three more precise than previous measurements, and the error on
the rate relative to \klpp,
\begin{equation}
\frac{\Gamma(\klmm)}{\Gamma(\klpp)} = (3.474\pm0.054)\times10^{-6},
\end{equation}
no longer dominates the 
error on the ratio
\begin{equation}
\frac{\Gamma(\klmm)}{\Gamma(\klgg)} = (1.213\pm0.030)\times10^{-5},
\end{equation}
contributing only $\sim$1.5\%
of the 2.5\% error. The remaining significant sources of uncertainty,
\begin{equation}
\frac{\Gamma(\klgg)}{\Gamma(\klpopo)} = 0.632\pm0.009 \; , \qquad
\frac{\Gamma(\kspp)}{\Gamma(\kspopo)} = 2.186\pm0.028,
\end{equation}
will probably be improved in the near future by the KLOE experiment
at Frascati.

This measured ratio is only slightly above the unitarity bound from the
on-shell two-photon contribution
\begin{equation}
\frac{\Gamma(\klmm)}{\Gamma(\klgg)} = 1.195 \times 10^{-5}
\end{equation}
and limits possible short-distance contributions.
With a recent  estimate of the long-distance dispersive
contribution~\cite{dambrosio1}, a limit on $\rho$ was extracted:
 $\rho >-0.33$ 
at 90\% confidence level (CL)~\cite{e871_mm}.

Unlike \klmm, which is predominantly
mediated by two real photons,
the decay \klee proceeds primarily via two off-shell
photons. The relative contribution from short-distance top loops
is significantly smaller than in \klmm. However, the recent 
observation  by
E871~\cite{e871_ee} of four events, with a branching ratio of
 B(\klee) $= (8.7^{+5.7}_{-4.1})
\times 10^{-12}$, is consistent with ChPT
predictions~\cite{dumm,valencia} and is the smallest branching ratio
ever measured for any elementary particle decay.

\subsection{\klpll}

\label{sec:klpee}
     The decays \klpee and \klpmm can proceed via the
direct-{\it CP}-violating components of the diagrams in
Figure~\ref{fig:feyn_klmm} and the  $s\rightarrow d\gamma^*$
amplitude, where $\gamma^*$ represents a  virtual photon. 
These processes are calculable with high precision 
within the Standard Model since they are dominated by top-quark exchange.
If these were the only contributions to this decay, the branching
ratio for \klpee would be related to the 
CKM matrix elements by~\cite{DonoghueGabbiani}
\begin{eqnarray}
B_{SD}(\klpee) & = &  \frac{\tau_L\alpha^2 B(K_{e3})}{\tau_{K^+}4\pi^2 |V_{us}|^2} 
(y_{7A}^2 + y_{7V}^2)| Im(\lambda_t) |^2 \\ \nonumber
   & = & 6.91\times10^{-11} A^4 \eta^2, 
\end{eqnarray}
With the current best-fit value of $1.38\times 10^{-4}$ for
$Im(\lambda_t)$~\cite{bosch}, 
this implies a branching ratio of about
$5\times 10^{-12}.$  The related decay $\klpmm$ is expected to be suppressed relative
to the electron mode by about a factor of five owing to the reduced
phase space.

     Unfortunately, the decay \klpee
can occur in two other ways.  First, there is an indirect-{\it CP}-violating
contribution from the {\it CP}-even, $K^\circ_1$ component of 
$K^\circ_L.$  This contribution could be determined from a measurement
of \kspee, but the current upper limit~\cite{kspee} of
B(\kspee) $<1.6\times 10^{-7}$ is far from the
expected level of less than $10^{-8}$.  A $K_S$
branching ratio of $10^{-9}$  would imply an indirect-{\it CP}-violating
contribution to \klpee of about $3\times 10^{-12}$,
comparable to the expected
direct-{\it CP}-violating contribution.  In addition, a significant contribution is
expected from the interference between the direct and 
indirect-{\it CP}-violating amplitudes.

     Further complicating the picture is the presence of a {\it CP}-conserving
amplitude involving a $\pi^\circ\gamma^*\gamma^*$ intermediate state, from
which the virtual photons materialize into an $e^+e^-$ pair.  A model for
the $K^\circ_L\pi^\circ\gamma^*\gamma^*$ vertex is needed to determine the
size of this contribution.  This vertex can be studied by measuring the
\klpgg decay and the related decay \klpeeg. These decays are discussed in
Section~\ref{sec:kzpgg}.  Based on recent
measurements of these modes, by the KTeV experiment at Fermilab (see Section~\ref{sec:e799}),
the {\it CP}-conserving contribution to \klpee
has been estimated at 1--2$\times 10^{-12}$,
comparable to the expected direct-{\it CP}-violating part.

     Given these three contributions, it will be difficult to
extract CKM matrix parameters from even a precision measurement of
\klpee. But there is a still more
formidable roadblock to progress on these modes, first pointed out
by Greenlee~\cite{Greenlee}.  As is discussed in Section~\ref{sec:klgg},
the radiative Dalitz decay \kleegg has a rather large branching ratio 
($\sim6\times 10^{-7}$). The two
photons may have an invariant mass near that of the
$\pi^\circ$, so that the final state is indistinguishable from
the $\pi^\circ e^+e^-$ mode.  Two strategies can be used to deal
with this background.  First, a high-precision calorimeter can be
used to minimize the size of the region in $M_{\gamma\gamma}$ where 
confusion can occur;  second, the difference in the kinematic distributions
expected in the radiative Dalitz decay can be used
to remove most of the background events, at a cost of some
acceptance for the signal mode \klpee. These techniques reduce, but
cannot eliminate, this background, so that the present searches for
\klpee are background-limited at the level of $10^{-10}$.

\begin{figure}[htbp]
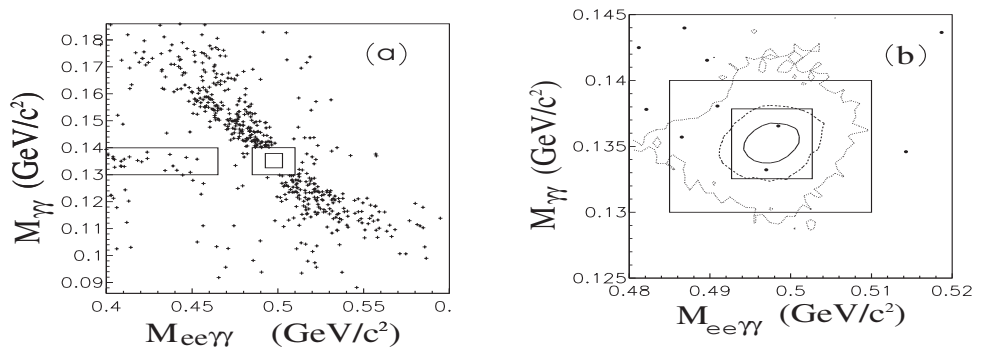

   \epsfig{file=barkf05a.eps,height=2.5in,width=2.5in,angle=0}
   \epsfig{file=barkf05b.eps,height=2.3in,width=2.5in,angle=0}
 \caption{KTeV:
    Reconstructed $\pi^\circ e^+e^-$ mass plotted vs reconstructed 
     $\gamma\gamma$ mass from the \klpee analysis of 1996-1997 data. 
     a) shows the situation before
    the final set of kinematic cuts, and the
     diagonal band passing through the signal region is due to
     the background mode \kleegg.  Events in the exclusion box
     surrounding the signal box are not shown.
     b) shows the two events remaining in the signal
     region (the smallest box) after all cuts are applied.  The
     contours contain 68\% and 
     95\% of any real \klpee signal.
\label{fig:e799_pee} }
\end{figure}

     The most recent limit on \klpee
comes from the KTeV experiment~\cite{klpee}.
The analysis selects on the direction of the photons with
respect to the electrons to minimize the background from 
radiative Dalitz decays while preserving as much sensitivity as possible.
Figure~\ref{fig:e799_pee}$a$ shows the $ee\gamma\gamma$ mass vs
the $\gamma\gamma$ mass for the KTeV data with the signal box excluded, and
Figure~\ref{fig:e799_pee}$b$ shows an expanded view around the signal box.
KTeV found two events that passed all cuts, compared with
an expected background level of 1.1$\pm$0.4 events.  This finding leads to an upper 
limit $B(\klpee) < 5.1\times 10^{-10} \; \; (90\%\,{\rm CL})$.
Although this limit represents a significant improvement over previous
results~\cite{E731piee,BNL845piee,E799oldpiee}, it is still two
orders of magnitude above the standard-model prediction for the
direct-{\it CP}-violating component of this decay.
 
\begin{figure}[htbp]
\epsfig{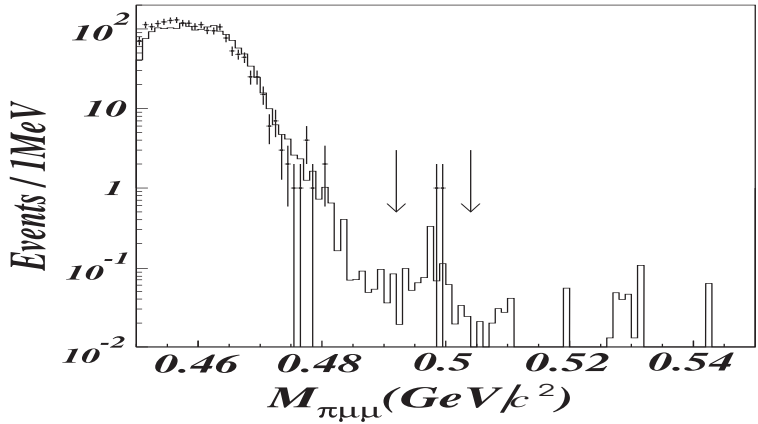}
 \caption{KTeV: Invariant mass distribution, $M_{\pi^\circ\mu\mu}$, for
 events passing all other cuts from the 1996-1997 data set. \label{fig:e799_pmm} }
\end{figure}

     A similar analysis of the related muon mode by KTeV resulted
in a slightly smaller upper limit~\cite{klpmm},
$B(\klpmm) < 3.8\times 10^{-10} \; \; (90\%\,{\rm CL}) $.
Figure~\ref{fig:e799_pmm} shows a plot of the $\pi^\circ\mu\mu$ mass,
with two events in the signal region and an expected background of $0.87\pm0.15$.
The greater sensitivity for this mode results from making looser
kinematic cuts.  Due to the reduced phase space for this decay, the
branching ratio of the muon mode is expected to be a factor of five
smaller than that 
for the electron mode, so that this limit is farther from the
expected level than the limit for $\pi^\circ e^+e^-$.
An improvement of roughly a factor of two
may be expected in both these limits when the analysis of the 1999
KTeV data set is complete.

Table~\ref{tab1_klpll} summarizes the experimental measurements of \kopll.
\begin{table}[ht]
\centering
\caption{Summary of \kopll results}
\vskip 0.1 in
\begin{tabular}{|l||l|l|} \hline
Decay Mode  &  \multicolumn{1}{|c|}{Branching Ratio} & Experiment 
\\ \hline\hline
\klpee  & $< 5.1\times10^{-10}$ & KTeV (2000)~\cite{klpee}\\ \hline
\klpmm  & $< 3.8\times10^{-10}$ & KTeV (2000)~\cite{klpmm}\\ \hline
\kspee & $< 1.6\times10^{-7}$ & NA48 (2000)~\cite{kspee}\\ \hline
\end{tabular}
\label{tab1_klpll}
\end{table}
     The obstacles to determining CKM matrix elements from measurements of
the \klpll\
 modes are formidable.  Although
work on these modes is likely to continue, future efforts will focus
on the related decay \klpnn,
which is free of the problems affecting the $\pi^\circ e^+e^-$ and 
$\pi^\circ\mu^+\mu^-$ modes.

\subsection{\kzpnn}

\label{sec:kzpnn}

The decay modes \kpnn and \klpnn are the golden modes for
determining the CKM parameters $\rho$ and $\eta$. Together with the
other golden mode \bpsiks, and perhaps the ratio
\bsbd, they provide the best opportunity
to test the Standard Model explanation of {\it CP} violation
and to search for new physics.
The \kzpnn decays are sensitive to the magnitude and imaginary part of \vtd. 
From these two modes, the unitarity triangle can be completely
determined.

\begin{figure}[htbp]
  \epsfig{file=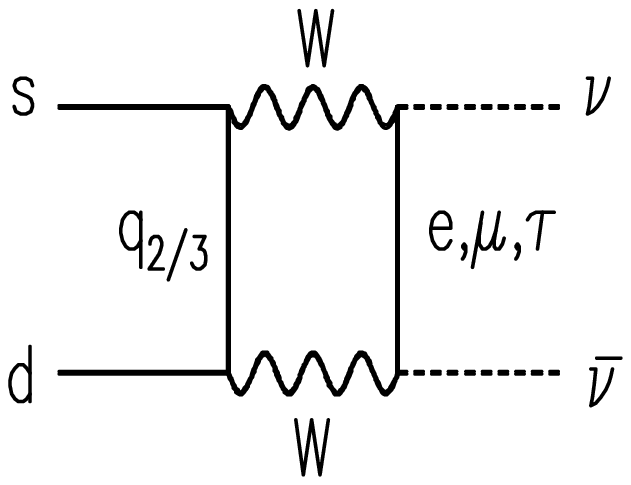,height=1.0in,width=1.5in,angle=0} \hspace{0.25cm}
  \epsfig{file=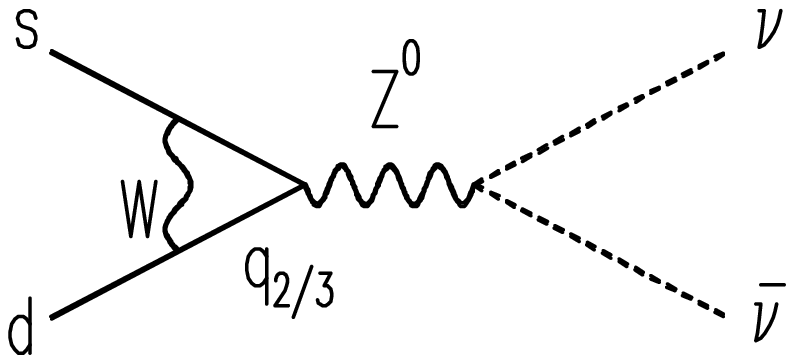,height=0.8in,width=1.5in,angle=0} \hspace{0.25cm}
  \epsfig{file=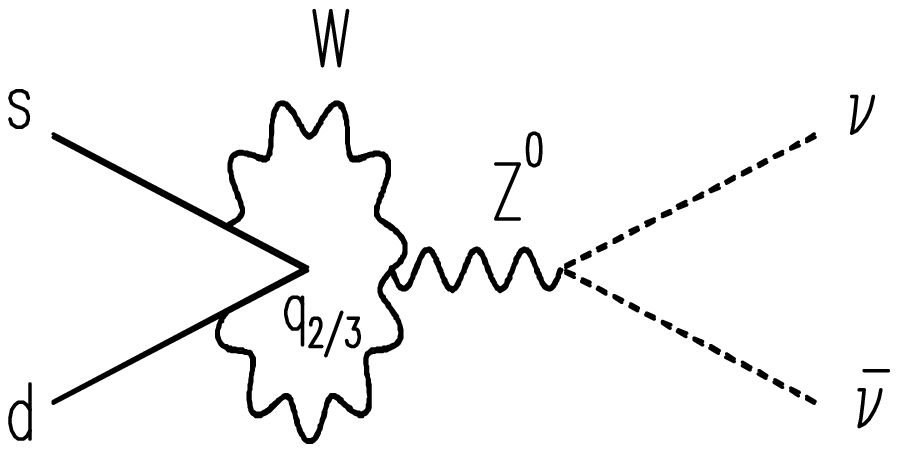,height=0.9in,width=1.5in,angle=0} 
\caption{Feynman diagrams for the decays \kzpnn.
\label{fig:feyn_kzpnn}}
\end{figure}

These modes proceed through loops dominated by the top quark, as shown in
Figure~\ref{fig:feyn_kzpnn}.
The hadronic matrix element for these
decays can be extracted from the well-measured \kpen (K$_{e3}$) decay.
The branching ratios have been calculated in the next-to-leading-log
approximation~\cite{bb1}, complete with corrections for isospin 
violation~\cite{marciano} and two-loop-electroweak
effects~\cite{bb2}. They can be expressed as follows~\cite{bb3}:
\begin{eqnarray}
B(\klpnn) & = & \frac{\tau_L}{\tau_{K^+}}
        \frac{\kappa_L\alpha^2 B(K_{e3})}{2\pi^2 \sin^4 \theta_W |V_{us}|^2}
         \sum_l |Im(\lambda_t)X_t|^2 \\ \nonumber
B(\kpnn) & = &\frac{\kappa_+\alpha^2 B(K_{e3})}
        {2\pi^2 \sin^4 \theta_W |V_{us}|^2}
      \sum_l  | X_t\lambda_t + X_c\lambda_c |^2.
\end{eqnarray}
The factors $\kappa_L$ and $\kappa_+$ refer to  the isospin corrections relating
\kzpnn to \\
\kpen. The Inami-Lim functions~\cite{buras,inami}, $X_q$, 
are also functions of $x_q;$ 
these contain QCD corrections. 
The sum is over the three neutrino
generations.  (See Reference~\ref{ref:bb3} for more information).  The
intrinsic theoretical uncertainty in the branching ratio B(\kpnn) is 7\%, 
predominantly from the
next-to-leading-log calculation of $X_c$. The intrinsic theoretical
uncertainty for \klpnn is even smaller, $\sim$2\%, coming from the 
uncertainties in $\kappa_L$ and $X$.  
These equations can be rewritten in terms of the Wolfenstein parameters, and
based on our current understanding of standard-model parameters, the
branching ratios are predicted to be
\begin{eqnarray}
B(\klpnn) & = & 4.08\times10^{-10}A^4\eta^2 \\ \nonumber
         & = & (3.1\pm1.3)\times10^{-11} \\
B(\kpnn) & = & 8.88 \times 10^{-11} A^4 
[(\overline{\rho}_0-\overline{\rho})^2 + (\sigma\overline{\eta})^2] 
\\ \nonumber
         & = & (8.2\pm3.2)\times10^{-11} ,
\end{eqnarray}
where $\sigma = (1-\frac{\lambda^2}{2})^{-2}$ and 
$\overline{\rho}_0 = 1.4$~\cite{bb3}.
In addition, it is possible to place a theoretically 
unambiguous upper limit on \kpnn
from the limit on \bsbd \\ derived
from $\Delta M_{B_s} < 14.3~ps^{-1}$~\cite{bsbd}.
This limit is~\cite{bb3}
\begin{equation}
B(\kpnn) < 1.67\times10^{-10}.
\end{equation}
The decay amplitude \klpnn is direct-{\it CP}-violating, and offers the best opportunity
for measuring the Jarlskog invariant $J_{CP}$.
Although the B meson system will provide clean measures of some
angles of the triangle, the determination of its area will be much less precise.

Two sets of recent experimental results have stimulated a lot of
theoretical activity on various BSM contributions~\cite{bosch,buras3}
to \kzpnn. The value of $Re(\epsilon'/\epsilon) = 
(19.3\pm2.4)\times10^{-4}$, including the most recent experimental
results~\cite{ktev_e,na48_e}, is larger than most previous theoretical
calculations had predicted~\cite{buchalla,buras,bosch,theory_e}.  In
addition, the first observation of \kpnn from E787 based data
collected in 1995~\cite{e787_pnn1}, with B(\kpnn) =
$4.2^{+9.7}_{-3.5}\times10^{-10}$, was a factor of five larger than
the standard-model prediction.  The theoretical calculation of
$\epsilon'/\epsilon$ has large uncertainty, so although the measured
value was higher than most calculations, this did not necessarily
imply the need for new physics.  The same is true of the \kpnn
measurement; although the standard-model prediction is unambiguous and
the result was high, the branching ratio based on one event was
entirely consistent with the standard model.
Reference~\ref{ref:buras3} points out that in a generic supersymmetric
extension to the standard model, an enhanced $Zds$ or $sdg$ vertex
that contributes to $\epsilon'/\epsilon$ will also enhance either
\klpnn or \klpee. The enhancement of \kpnn is smaller, as it is
limited to some extent by the measured value of \klmm. An enhancement
in these rare modes would be much easier to interpret than in
$\epsilon'/\epsilon$.

\subsubsection{\kpnn}

\label{sec:kpnn}

\begin{figure}[htbp]
\epsfig{file=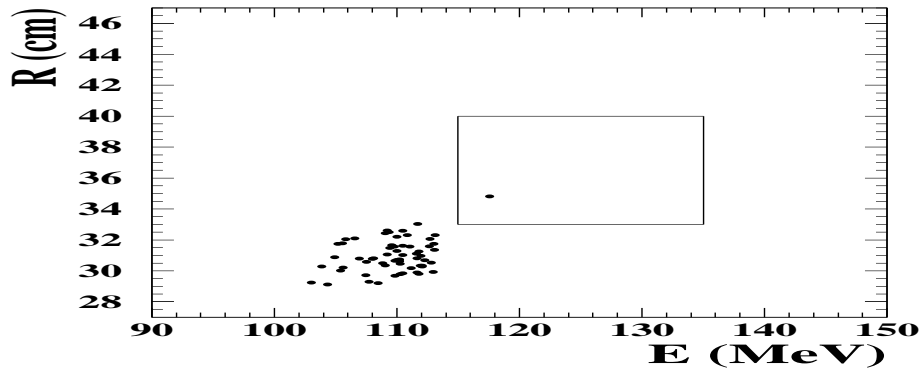,height=2.in,width=4.75in,angle=0}
\caption{E787: Final data sample collected in 1995--1997 after all cuts. One
clean \kpnn event is seen in the signal box. The remaining events are \kpp
background.
\label{fig:e787_pnn}}
\end{figure}

Although the decay \kpnn is attractive theoretically,
it is quite challenging experimentally. Not only is the branching ratio
expected to be less than $10^{-10}$, it is a three-body decay with two
undetectable neutrinos. The key to a convincing measurement of this decay
is a thorough understanding of the background at a level of $10^{-11}$.
The E787 experiment (see Section~\ref{sec:e787}) previously reported 
results of the analysis of the 1995 data sample~\cite{e787_pnn1}.
This experiment  employs two guiding
principles for determining the background:
\begin{itemize}
  \item The background is measured from the same data as the \kpnn signal. 
In this manner, any hardware problems, changes in rates, or changes in detector performance
are automatically considered.
  \item For all background from kaon decays, two independent sets of selection criteria
are devised, with large rejection (e.g.\ typical rejections are $R>100$)
for that background type. This allows a measurement of background levels
at a sensitivity $R$ times greater than the signal
by reversing one set of 
selection criteria.
Because one set of criteria is always reversed, the criteria are
devised without any bias from examining events in the signal region.
\end{itemize}
The three major sources of background, \kmn, \kpp, and 
pions from the beam, are all measured, with a total background
of $0.08\pm0.02$ events from the analysis of the data collected 
during 1995--1997. One clean \kpnn event
was found (see Figure~\ref{fig:e787_pnn}), and
based on this
one event~\cite{e787_pnn2}, which was also seen in the earlier data,
 the branching ratio is B(\kpnn) =
$1.5^{+3.4}_{-1.2} \times 10^{-10}$. This measurement
places a limit on $| \vtd |$---and without any
reference to measurements from B meson decays,
limits on $\lambda_t$ can be derived: 
\begin{eqnarray}
0.002 < & |\vtd| & < 0.04 \\ \nonumber
 & | Im(\lambda_t) | & < 1.22\times10^{-3} \\ \nonumber
-1.10\times10^{-3} < & Re(\lambda_t) &  < 1.39\times10^{-3} \\ \nonumber
1.07\times10^{-4} < & | \lambda_t | & < 1.39\times10^{-3}.
\end{eqnarray}
The E787 experiment, with all data recorded
should reach a factor of two higher sensitivity---to the
level of the standard-model expectation for \kpnn.

A new experiment, E949 (see Section~\ref{sec:e787}), 
is under construction at BNL and will run from
2001 through 2003. Taking advantage of the large AGS 
proton flux and the
experience gained with the E787 detector, E949 
should observe 10 standard-model events in a two-year run. The background is
well understood and based on E787 measurements is expected to be 10\% of the 
standard-model signal.
A proposal for an experiment which promises a  further factor of 10 
improvement has been prepared
at FNAL. The CKM experiment (E905) is designed to collect 100 
standard-model events, with an estimated  background of approximately  10\% of the signal, in a
 two-year run starting in about 2005.  This
experiment will use a new technique, with $K^+$ decay-in-flight
and momentum/velocity spectrometers.

\subsubsection{\klpnn}

\label{sec:klpnn}

The decay \klpnn is even cleaner theoretically and is
purely direct-{\it CP}-violating. Unfortunately, it is even more difficult 
experimentally, because all particles involved in the initial and final 
states are neutral.

Presently, the best limit on \klpnn is derived in a model-independent
way~\cite{grossman} from the E787 measurement of \kpnn:
\begin{eqnarray}
B(\klpnn) & < & 4.4\times B(\kpnn) \\ \nonumber
         & < & 2.6\times10^{-9} \; \; (90\%\,{\rm CL}). 
\label{eq:pnn}
\end{eqnarray}

\begin{figure}[htbp]
\epsfig{file=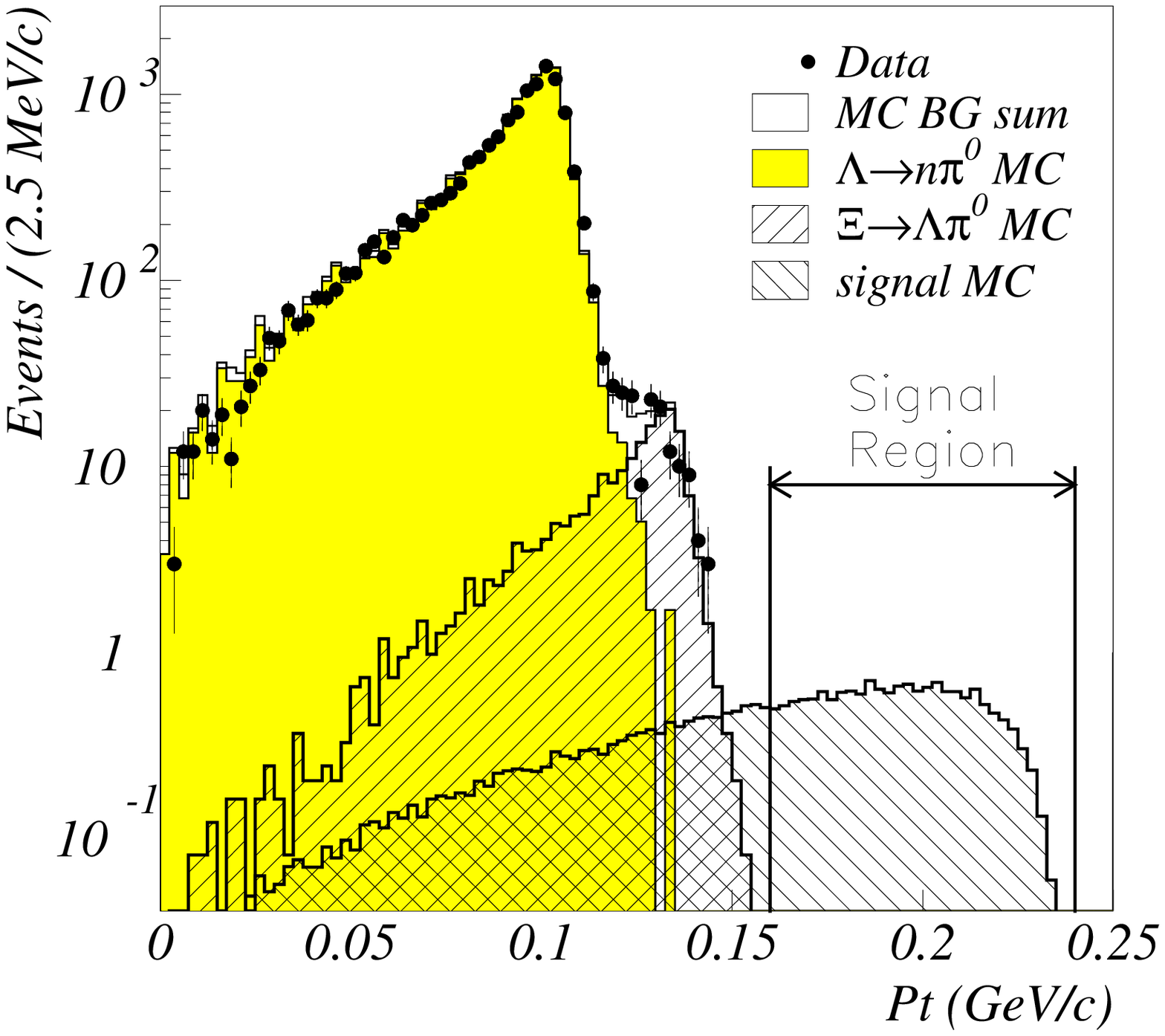,height=2.25in,width=5.25in,angle=0}
\caption{KTeV: Final \klpnn data sample collected during 1996--1997 after all cuts. No
\klpnn events are seen above $P_T = 160$ MeV/$c$. 
\label{fig:e799_pnn}}
\end{figure}

The goal is to observe this mode directly in order
to extract a second of the CKM matrix parameters.  Here we concentrate
on the high-transverse-momentum technique for making this measurement, as used in
the existing KTeV results; future efforts toward measuring \klpnn
may also include center-of-mass experiments.
The \klpnn decay is identified by two photons from the common decay
$\pi^\circ\rightarrow\gamma\gamma.$   
$K_L$ decays such as \klpopopo and \klpopo
can easily produce background if all but
two of the final-state photons are unobserved.  Background can also
arise from $\pi^\circ$'s
produced by $\Lambda$ or $\Xi^\circ$ hyperon decays to final states
such as $n\pi^\circ$ with the neutron undetected, if the beam contains
large numbers of hyperons.
	An excellent system of photon veto detectors can 
        substantially reduce these backgrounds, but additional
        kinematic cuts will also be necessary.  In the center of
        mass experiments, a simple invariant mass cut can be
        made to reject \klpopo and \klpopopo
        backgrounds.  In experiments like KTeV where the kaon 
        momentum is unknown, one can exploit the fact that the 
        neutrinos recoiling against the $\pi^\circ$ in \klpnn
        are massless, so that the transverse momentum of the
        $\pi^\circ$ extends to larger values than are possible in 
        the background modes, as shown in Figure~\ref{fig:e799_pnn}.

KTeV (see Section~\ref{sec:e799}) does not measure the kaon momentum; 
in order to determine the transverse momentum of the $\pi^\circ,$
the decay vertex must be known.  The longitudinal position of the vertex
can be determined from the invariant mass constraint, but the transverse
position can only be known within the size of the kaon beam.  Thus a
narrow ``pencil'' beam is needed, which limits the available intensity.
KTeV tried this approach in a one-day test run and observed one
background event, probably from a neutron interaction.  
From this special run, a 90\%-CL limit~\cite{e799_pnn_gg} of
B(\klpnn) $< 1.6\times 10^{-6}$ was determined.
An alternative is to use the rarer $\pi^\circ\rightarrow e^+e^-\gamma$
decay.  This is a factor of 80 less sensitive but has several 
advantages.  First, the location of the decay vertex can be determined from
the charged tracks, so that a high-intensity, wide neutral beam can be used.
This allowed the KTeV data for this mode to be taken in the standard
configuration with standard triggers.  Second, 
this approach allows
determination of the transverse momentum with better precision, reducing
the background level.  
The $P_T$ distribution of $\pi^\circ$ events
passing all other cuts can be seen in Figure~\ref{fig:e799_pnn}.
The backgrounds nearest the search region come from
the decays $\Lambda\rightarrow n\pi^\circ$ and 
$\Xi^\circ\rightarrow\Lambda\pi^\circ.$  In this search using the full 1997
KTeV data set, with an expected background of $0.12^{+0.05}_{-0.04},$ 
no events were seen, and at the 90\% confidence level,
$B(\klpnn) < 5.9\times 10^{-7}~\cite{e799_pnn},$
still more than four orders of magnitude from the standard-model prediction.

The next generation of \klpnn experiments will start with E391a 
(see Section~\ref{sec:e391a}) at 
the High Energy Accelerator Research Organization (KEK)
in Tsukuba (Japan),
which hopes to reach a sensitivity of $\sim10^{-10}$. This 
experiment will use a technique similar to KTeV, with a pencil beam,
high quality calorimetry and very efficient photon vetos. This 
experiment would eventually move to the Japanese Hadron Facility (JHF), a new
50 GeV proton accelerator that is expected to be built around 2006,
and attempt to push to a
sensitivity of ${\cal O} (10^{-14})$.  
Two other future experiments propose to reach sensitivities of ${\cal O}
(10^{-13})$: E926 (KOPIO/RSVP) at BNL (see Section~\ref{sec:e926})
and E804 (KAMI) at FNAL (see Section~\ref{sec:kami}).
The KOPIO technique is significantly different from the others; all possible
initial and final state quantities will be measured, including the $K_L$ 
momentum and the photon times, energies and directions.
A pre-radiator is used to reconstruct the directions of the two photons
and a low momentum bunched beam is used to derive the kaon momentum 
from time of flight. This technique, which also relies on a very efficient
photon veto system, has additional tools to reject backgrounds:
the quality of the $\pi^\circ$
vertex and the $\pi^\circ$ momentum in  the kaon center of mass.

The decays \klppnn, \klpoponn and \kppnn are also very clean theoretically 
and measurements
of these branching ratios could be used to determine $\eta$ and 
$\rho$~\cite{ppnn}. Unfortunately,
the SM expectations for these branching ratios are ${\cal O}(10^{-13})$,
${\cal O}(10^{-13})$ and
${\cal O}(10^{-14})$ respectively and are
not accessible for precision measurements in current or next generation 
experiments. A limit on \kppnn
has recently been derived by E787: B(\kppnn) $< 4.3\times10^{-5}$~\cite{e787_ppnn}.
Table~\ref{tab_kpnn} summarizes the current experimental status of 
\kzpnn.
\begin{table}[ht]
\centering
\caption{Summary of \kzpnn results}
\vskip 0.1 in
\begin{tabular}{|l||l|r|l|} \hline
Decay Mode  &  \multicolumn{1}{|c|}{Branching Ratio} & events& Experiment  
\\ \hline\hline
\kpnn&$(1.5^{+3.4}_{-1.2})\times10^{-10}$&1&E787 (2000)~\cite{e787_pnn2}\\ \hline
\klpnn&$< 1.6\times 10^{-6}$&0&KTeV (2000)~\cite{e799_pnn_gg}\\ \hline
\klpnn(ee$\gamma$) &$<5.9\times 10^{-7}$& 0 & KTeV (2000)~\cite{e799_pnn}\\ \hline
\klpnn($\pi^+\nu\overline{\nu}$) &$<2.6\times 10^{-9}$& ---  & E787 (2000)~\cite{e787_pnn2}\\ \hline
\kppnn &$<4.3\times 10^{-5}$& 0  & E787 (2000)~\cite{e787_ppnn}\\ \hline
\end{tabular}
\label{tab_kpnn}
\end{table}

\section{FORM FACTORS}

        In addition to the rare kaon decays that directly
        probe standard-model parameters (as discussed in the preceding
        section), and those that are sensitive to BSM physics
        (the subject of the following section), there is an
        impressively broad array of other decay modes on which
        substantial experimental progress has been made in recent
        years.  Although these results receive less attention, 
	they provide critical information in a variety of areas.

        For example, there has historically been strong interest
        in the \klmm decay as a probe of weak interaction dynamics,
        specifically $Re(V_{td}),$ through its short-distance amplitude.
        But the short-distance amplitude is known to be quite
        small compared with the long-distance part, involving the
        $\gamma\gamma$ and $\gamma^*\gamma^*$ intermediate states.
        An accurate determination of the $K_L\gamma^*\gamma^*$
        form factor is  needed in order to 
        evaluate the long-distance contribution, 
	which is needed in turn  to extract $Re(V_{td})$ from B(\klmm).
        There are various theoretical models for the 
        form factor, each involving parameters that must be
        determined from experiments discussed in Section~\ref{sec:klgg}.

        The measurement of such modes as \klpgg and \kleegg is
        important for a different reason.  As discussed
        in Section~\ref{sec:klpee}, \klllgg is an important background
        in the search for \klpll. Both the
        absolute number of events from this process and the
        kinematic distributions of those events are thus
        important to the effort to learn about standard-model parameters from
        \klpee and \klpmm.  In particular, large samples of these
        events must be studied to determine the effectiveness
        of kinematic cuts necessary to observe this extremely rare decay.
        The decay \klpgg, though not a background to \klpee,
        can be used to determine the {\it CP}-conserving part of the  amplitude.
        Additional contributions from the $\pi^\circ\gamma^*\gamma^*$
        intermediate state with off-shell photons are also
        important.  These can be determined from ChPT models, but again
        there are undetermined parameters that must be extracted
        by studying kinematic distributions in \klpgg and the
        related mode \klpeeg.
        
        Because of their low energy release and
	wide variety of final states, kaon decays
        provide an excellent testing ground for the predictions of
        ChPT.  For example, the \kzpgg modes and
        the direct-emission component of radiative semileptonic decay
        modes have proven to be good testing grounds for comparing
        ${\cal O}(p^4)$ to ${\cal O}(p^6)$ calculations. 
        ChPT calculations of $\pi\pi$ scattering can likewise be tested
        by measuring the form factors of \kzppln (K$_{\ell 4}$) decays.

        Sometimes the study of these less well-known modes can turn
        up new phenomena of considerable interest.  For example,
	an interesting observation of a 
        {\it CP}-violating and {\it T}-odd angular asymmetry in the \klppee
        decay has been made by the NA48 and KTeV experiments.
        This is the largest {\it CP}-violating effect yet seen, and the
        first {\it CP}-violating effect ever observed in an angular
        distribution.

Recent reviews of the current state of radiative and semileptonic
kaon decay measurements are available from the DA$\Phi$NE workshop
\cite{kettell,lowe}.

\subsection{\klgg and Related Decays}

\label{sec:klgg}

     Like the $\pi^\circ,$ the neutral kaons couple to two photons.
The effective interaction term for the {\it CP}-conserving interaction 
between a pseudoscalar meson field $P$ of mass $M_P$ and the electromagnetic 
field $F_{\mu\nu}$ is given by
\begin{equation}
 {\cal L} = {if_{P\gamma\gamma}\over 4M_P} \,\epsilon_{\mu\nu\lambda\sigma}
F^{\mu\nu}F^{\lambda\sigma}P. 
\end{equation}
In terms of the polarizations $\epsilon_i$ and momenta $k_i$ of the two
photons, this vertex becomes
\begin{equation}
{\cal L} = -{2f_{P\gamma\gamma}\over M_P} \,\epsilon_{\mu\nu\lambda\sigma}
    k_1^\mu\epsilon_1^\nu k_2^\lambda\epsilon_2^\sigma, 
\end{equation}
which leads to a $\gamma\gamma$ partial width of
\begin{equation}
\Gamma_{\gamma\gamma}(P) = {f^2_{P\gamma\gamma}M_P\over 16\pi}. 
\end{equation}
The coupling $f_{K_S\gamma\gamma}$ is determined in ${\cal O}(p^4)$
ChPT, without any free parameters, leading to the prediction~\cite{chpt_klpgg}
B(\ksgg) = $2.0\times10^{-6}.$  This is in good agreement with
the experimental value~\cite{kspee} (see Table~\ref{tab_kgg}), although
the experimental errors are still rather large.
\begin{table}[ht]
\centering
\caption{Summary of results of kaon decays to two photons and related 
modes}
\vskip 0.1 in
\begin{tabular}{|l||l|r|l|} \hline
Decay Mode  &  \multicolumn{1}{|c|}{Branching Ratio} & events& Experiment
\\ \hline\hline
\ksgg & $(2.6\pm0.4\pm0.2)\times10^{-6}$ & 148 & NA48-00~\cite{kspee}\\ \hline
\klgg & $(5.92\pm0.15)\times10^{-4}$ & 110000 & NA31-87~\cite{klgg}\\ \hline
\klmm & $(7.24\pm0.17)\times10^{-9}$ & 6200 & E871-00~\cite{e871_mm}\\ \hline
\klee & $(8.7^{+5.7}_{-4.1})\times10^{-12}$& 4 & E871-98~\cite{e871_ee}\\ \hline
\kleeg &$(1.06\pm0.02\pm0.02\pm0.04)\times10^{-5}$& 6854 & NA48-99~\cite{kleeg}
\\ \hline
\klmmg & $(3.66\pm0.04\pm0.07)\times10^{-7}$ & 9105& KTeV-00~\cite{klmmg} 
\\ \hline
\kleeee & $(3.77\pm0.18\pm0.13\pm 0.21)\times10^{-8}$ & 436 & KTeV-00~\cite{klmmg}
\\ \hline
\klmmee & $(2.50\pm0.41\pm0.15)\times10^{-9}$ & 38 & KTeV-00~\cite{klmmg}\\ \hline
\klmmmm & no limit &  & \\ \hline
\ksmm & $< 3.2\times10^{-7}$ & 0 & CERN-73~\cite{ksmm}\\ \hline
\ksee & $< 1.4\times10^{-7}$ & 0 & CPLEAR~\cite{ksee}\\ \hline
\kleegg & $(5.84\pm0.15\pm0.32)\times10^{-7}$ & 1543 & KTeV-00~\cite{kleegg}
\\ \hline
\klmmgg & $(1.42^{+1.0}_{-0.8}\pm0.10)\times10^{-9}$ & 4 & 
KTeV-00~\cite{klmmgg}\\ \hline
\end{tabular}
\label{tab_kgg}
\end{table}

The interaction term for two real photons can be extended to off-shell 
photons with nonzero $k^2.$  In this case, the coupling
$f_{P\gamma^*\gamma^*}$ can depend on the two $k^2$ values.  Typically, a
form factor $F_{P\gamma\gamma}$ is introduced, so that
\begin{equation}
f_{P\gamma^*\gamma^*}(k_1^2,k_2^2)
   = f_{P\gamma\gamma}\,F_{P\gamma\gamma}(k_1^2, k_2^2), 
\end{equation}
and the form factor is consequently normalized to the point 
$F_{P\gamma\gamma}(0,0) = 1.$ 
The form factor $F_{K_L\gamma\gamma}$ is needed to accurately
calculate the long-distance contribution to \klmm and \klee,
which both have important contributions from the $\gamma^*\gamma^*$ 
intermediate state.
As is discussed in Section~\ref{sec:klmm}, this long-distance 
contribution
must be subtracted from the precise experimental measurement of
\klmm in order to determine the interesting
short-distance part of the amplitude for that decay, which can be 
related to the real part of the CKM matrix element \vtd or, equivalently,
to the standard-model parameter $\rho.$

A number of models are available for the form factor. 
The simplest approach is to determine
the coefficient of the first few
terms in a Taylor series expansion in the parameters $x_i = k_i^2/M_K^2.$
An alternative parameterization~\cite{dambrosio1} assumes that the form factor
can be written in terms of vector-meson poles with arbitrary residues:
\begin{equation}
F(k_1^2, k_2^2) = 1 + \alpha\left({k_1^2\over k_1^2 - M_V^2}
   + {k_2^2\over k_2^2 - M_V^2}\right) + \beta\,{k_1^2 k_2^2\over
     (k_1^2 - M_V^2)(k_2^2 - M_V^2)}. 
\end{equation}
Rare kaon decays can be used to study the $K\gamma\gamma$ form factors
in several regions.  For example, the electron and muon Dalitz decays
\kleeg and \klmmg\
are sensitive to the form factor with $k_2^2 = 0$ and 
$4m^2 < k_1^2 < M_K^2,$ where $m^2$ is the lepton mass. 
From lepton universality~\cite{numao} the form factors obtained
in the electron and muon modes should be the same.
A new, high-precision result 
from KTeV (see Section~\ref{sec:e799}) for \klmmg is available; 
the mass distribution of the final event sample 
collected during 1996--1997 is shown in Figure~\ref{fig:e799_mmg}.
\begin{figure}[htb]
\epsfig{file=barkf10.eps,height=2.25in,width=5.25in,angle=0}
\caption{KTeV: Final \klmmg data sample collected in 1996--1997 after all cuts.
A total of 9105 \klmmg events are observed in the peak.
\label{fig:e799_mmg}}
\end{figure}
As Table~\ref{tab_kgg} shows, substantial statistics are now available in 
both of these $\ell\ell\gamma$ modes.
Experiments analyzing $l^+l^-\gamma$ data have usually fit for the
$\alpha_{K^*}$ parameter in the Bergstr{\"o}m, Mass{\'o} \& Singer
 form factor model~\cite{BMS},
\begin{equation}
 F(k^2, 0) = {k^2\over k^2 - M_\rho^2}
   + {2.5\alpha_{K^*}k^2\over k^2 - M_{K^*}^2}
    \left({4\over 3} - {k^2\over k^2 - M_\rho^2}
          - {k^2/9\over k^2 - M_\omega^2}
          - {2k^2/9\over k^2 - M_\phi^2}\right).  
\end{equation}
This model is based on a vector-dominance picture of
pseudoscalar-pseudoscalar transitions (the first term) and
vector-vector transitions involving $K^*V$ vertices (the second term).
This form provides an acceptable fit to the data.  Two recent
fits based on high-statistics analyses yield rather different
values, though.  A recent fit~\cite{kleeg} to the NA48 
(see Section~\ref{sec:na48}) \kleeg data yields
$\alpha_{K^*} = -0.36\pm 0.06,$ somewhat more negative than the best
fit values from earlier experiments studying the same mode.
A new KTeV result~\cite{klmmg} based on the \klmmg mode finds
$\alpha_{K^*} = -0.157\pm 0.027,$ a value 
about three sigma different from the new NA48 result.  The 
$k^2$ region sampled by the \klmmg data is much more heavily weighted
to large $k^2$ values, such that a different form factor model might
reduce the discrepancy.

The rarer double-internal-conversion modes, where the final state
consists of two lepton pairs, are sensitive to the form factor in the
region
 $4m_1^2 < k_1^2 < (M_K - 2m_2)^2$ and $4m_2^2 < k_2^2 < (M_K -
2m_1)^2,$ where $m_1$ and $m_2$ are the two lepton masses.  Three such
modes are expected: \kleeee, \klmmee and \klmmmm. The production of
muon pairs requires virtual photons at much higher $k^2$ and is strongly
suppressed.  The largest sample of $e^+e^-e^+e^-$ decays reported to
date is from KTeV, with 436 events in the 1997 data
sample.  KTeV has also reported seeing 
38 $e^+e^-\mu^+\mu^-$ events. 
The predicted branching ratio
for the \klmmmm mode is below $10^{-12},$ two orders of magnitude
beyond the sensitivity of KTeV.

Two decay modes related to \klgg, though not especially
interesting in themselves, have significant implications for the attempt
to observe direct {\it CP} violation in \klpee and \klpmm.
These are the radiative Dalitz
decays \kleegg and \klmmgg.
With a typical infrared cutoff of
5~MeV for the photon energies in the kaon center of mass, the
electron mode, \kleegg, has a branching 
ratio of about $6\times 10^{-7},$ five orders of magnitude higher
than the expected rate for \klpee. Moreover,
the peak of the $\gamma\gamma$ invariant mass spectrum in observed events
is near the $\pi^\circ$ mass.  With a good calorimeter, experiments can
limit the $\pi^\circ$ mass range to a few MeV, but the number of
$e^+e^-\gamma\gamma$ events in this range still swamps the expected
$\pi^\circ e^+e^-$ signal.  Further reduction in this background can be achieved by
cutting on kinematic variables to remove, for example, events in which the
momentum of one
of the photons is nearly parallel to that of
one of the leptons;  this topology is
typical of radiative Dalitz decays but uncommon for $\pi^\circ e^+e^-$ events.
But even an optimal set of cuts cannot reduce this background to the level
of the expected signal, a major impediment to the measurement of 
$B(\klpee).$  KTeV has identified a sample of over
1500 $e^+e^-\gamma\gamma$ events and has verified that their kinematic
distributions are generally in agreement with those predicted.

The decay \klmmgg is likewise a serious
background to the measurement of \klpmm.
The absolute rate for this decay is much less than 
the rate for the corresponding
electron mode (see Table~\ref{tab_kgg}).  
Unfortunately, the part of the phase space where this
decay can be a background to \klpmm, after
the $M_{\gamma\gamma}$ and other kinematic cuts, is not particularly
suppressed.  Thus, the \klpmm mode does not
allow experiments to eliminate the radiative Dalitz background.

Table~\ref{tab_kgg} summarizes results of kaon decays to two real or off-shell photons. 
With completion of the KTeV analysis, the 
\klllg modes  should be improved by more than a factor of two.  
The \klllee  data should triple. 
The $K_S$ modes may be improved by NA48 in a dedicated experiment after the
$\epsilon'/\epsilon$ running.  The \klgg and \ksgg as well as several other
modes will be improved by KLOE.  No improvements are expected
for \klmm or \klee in the foreseeable future.

\subsection{\kzpgg}

\label{sec:kzpgg}
The decay rate and spectral shape of \klpgg are  
calculated at ${\cal O}(p^4)$ of ChPT, without any free
parameters~\cite{chpt_klpgg}. The prediction of the spectral shape is a
striking success of ChPT. However, the decay rate is a factor of three too
small. To match the experimental value, a model-dependent contribution
from ${\cal O}(p^6)$ is needed, which is usually parameterized with a
constant $a_V$~\cite{chpt_av}, that measures the vector meson
exchange contribution to the amplitude.  This parameter is of particular
importance because the {\it CP}-conserving contribution to
\klpee depends on the value
of $a_V$.  Based on half of the total data sample, KTeV 
(see Section~\ref{sec:e799}) has recently
measured $a_V = -0.72\pm0.05\pm0.06$~\cite{klpgg}, implying a
contribution of 1--2$\times10^{-12}$ to \klpee.
NA48 has also reported a preliminary result~\cite{na48_klpgg}, based on almost 1400 events from
part of the 1998 and 1999 runs, of B(\klpgg) = $(1.51\pm0.05\pm0.20)\times10^{-6}$,
with $a_V = -0.45$.

Calculation for the charged mode \kpgg is more
complicated, requiring an unknown parameter, 
$\hat{c}$~\cite{chpt_ecker,chpt_ecker2}, even at ${\cal
O}(p^4)$; however, it also provides a good test of ChPT~\cite{chpt_kpgg}. 
Both the decay rate and spectral shape are predicted with
this single parameter. As with \klpgg, the two-photon invariant mass
($M_{\gamma\gamma}$) peaks above the two-pion mass, $M_{\pi^+\pi^-}$,
implying an intermediate \kppp decay.
This observation has led to improved ChPT predictions, normalizing
to the \kppp measurement---the so-called unitarity 
corrections~\cite{chpt_unitarity}.
E787 (see Section~\ref{sec:e787}) has measured a branching ratio~\cite{kpgg} of
$B(\kpgg) = (6.0\pm1.5\pm0.7)\times10^{-7} (100<P_{\pi^+}<180$MeV/$c$) 
and $\hat{c} = 1.8\pm0.6$. The data favor 
unitarity corrections.

Table~\ref{tab_kzpgg} summarizes the experimental measurements of \kzpgg.
\begin{table}[ht]
\centering
\caption{Summary of $\kzpgg$ results}
\vskip 0.1 in
\begin{tabular}{|l||l|r|l|} \hline
Decay Mode&\multicolumn{1}{|c|}{Branching Ratio}& events& Experiment\\ \hline\hline
\klpgg & $(1.68\pm0.07\pm0.08)\times10^{-6}$ & 884 & KTeV (1999)~\cite{klpgg}\\ \hline
\kpgg &  $(6.0\pm1.5\pm0.7)\times10^{-7}$& 26 & E787 (1997)~\cite{kpgg}\\ \hline
\klpeeg &$(2.20\pm0.48\pm0.11)\times10^{-8}$ & 18 & KTeV (1999)~\cite{klpeeg} \\ \hline
\end{tabular}
\label{tab_kzpgg}
\end{table}
The KTeV measurement of \klpeeg
should improve by a factor of three; the measurement of \klpgg
should improve by a factor of two.  Additional improvements will await
the next round of experiments, including a possible
search for \kspgg by NA48 after
the $\epsilon'/\epsilon$ running is completed.

\subsection{\kpll}

\label{sec:kpee}
The \kpll decays are suppressed in the standard model, since they proceed via 
a flavor-changing neutral current:  an effective $Zds$ coupling
that is forbidden in the SM at tree-level, but permitted at the one-loop level.
However, they are dominated by long-distance effects and proceed 
electromagnetically
through single-photon exchange, so it is not possible to extract short-distance
physics from these modes (unless one measures the lepton polarization).
These decays have been extensively studied within 
ChPT~\cite{DonoghueGabbiani,chpt_ecker,chpt_ecker2,chpt_pee}.
To ${\cal O}(p^4)$ in ChPT, the rate and di-lepton invariant mass 
spectra ($M_{\ell^+\ell^-}$) for both \kpee and \kpmm are described by
one free parameter, $w_+$. This parameter has been calculated in various 
models~\cite{chpt_ecker}.
At  ${\cal O}(p^6)$ in ChPT, additional parameters are needed;
$w_+$ is replaced by $a_+$ and $b_+$~\cite{chpt_damb}.

The \kpee decay was first observed in the 1970s~\cite{cern_pee}, and since 
that time
a series of experiments at BNL has increased the event sample
substantially~\cite{e777_pee,e851_pee,e865_pee}. The most recent experiment, 
E865~\cite{e865_pee} (see Section~\ref{sec:e865}), reported a branching ratio of
$B(\kpee) = (2.94\pm0.05\pm0.13\pm0.05)\times10^{-7}$, based on 10,175 events.
(The errors are statistical, systematic, 
and from the theoretical uncertainty of the spectral shape.)
A fit to the $M_{e^+e^-}$ spectra gives values for the form factor 
parameters of
\begin{equation}
a_+ = -0.587\pm0.010 , \qquad b_+ = -0.655\pm0.044.
\end{equation}

The first observation of the decay \kpmm was reported by E787 
(see Section~\ref{sec:e787}) in
1997~\cite{e787_pmm}.  A total of 200 events were recorded during the
1989--1991 running period.  The branching ratio was measured to be
$B(\kpmm) = (5.0\pm0.4\pm0.7\pm0.6)\times10^{-8}$. 
This mode has subsequently been observed by
E865~\cite{e865_pmm}, with 430 events and a measured branching ratio of
$B(\kpmm) = (9.22\pm0.60\pm0.49)\times10^{-8}$. Because all events are
fully reconstructed, a measurement of the $\mu\mu$ invariant mass
($M_{\mu\mu}$) is possible. This most recent measurement disagrees
with the previous one by more than 3$\sigma$, for reasons that are not yet
understood.

Table~\ref{tab_kpll} summarizes the experimental measurements of \kpll.
\begin{table}[ht]
\centering
\caption{Summary of \kpll results}
\vskip 0.1 in
\begin{tabular}{|l||l|r|l|} \hline
Decay Mode  &  \multicolumn{1}{|c|}{Branching Ratio} & events& Experiment  
\\ \hline\hline
\kpee&$(2.94\pm0.05\pm0.13)\times10^{-7}$&10300&E865 (1999)~\cite{e865_pee}\\ \hline
\kpmm&$(9.22\pm0.60\pm0.49)\times10^{-8}$&430&E865 (2000)~\cite{e865_pmm}\\ \hline
\kpeeg & \multicolumn{1}{|c|}{---} & $\sim$30 & E865 (1999)~\cite{kpeeg}\\ \hline
\end{tabular}
\label{tab_kpll}
\end{table}

\subsection{\kzppg}

\label{sec:kppg}

The radiative K$_{\pi2}$ decays---\kppg, \klppg, and \ksppg---have two
contributions.  In the inner bremsstrahlung (IB) process, a photon is
radiated from one of the charged particles. In the direct emission
(DE) process, the photon is radiated from an intermediate state.  The
branching ratio of the IB contribution scales with the underlying
K$_{\pi2}$ decay rate. The DE decay
probes the kaon structure and has been studied extensively in 
ChPT~\cite{dambrosio2}. In $K_L$, IB is highly suppressed because
the underlying decay is {\it CP} violating; in $K^+$, it is somewhat
suppressed due to the $\Delta I = 1/2$ rule; in $K_S$, it is not
suppressed.

The decay \klppg(DE)~\cite{carroll,ksppg_de,ksppg} has a long history.
A new result~\cite{klppg} from KTeV
(see Section~\ref{sec:e799}) has recently been reported.
  The branching ratio for the DE
component is B(\klppg;DE, E$_{\gamma}^{*}>$20 MeV) =
$(2.92\pm0.07)\times 10^{-5}$.  The fraction of DE
is DE/(DE+IB) = $0.683\pm 0.011$.  This result is based
on $\sim$5\% of the total KTeV data for this mode.

\begin{figure}[htbp]
   \epsfig{file=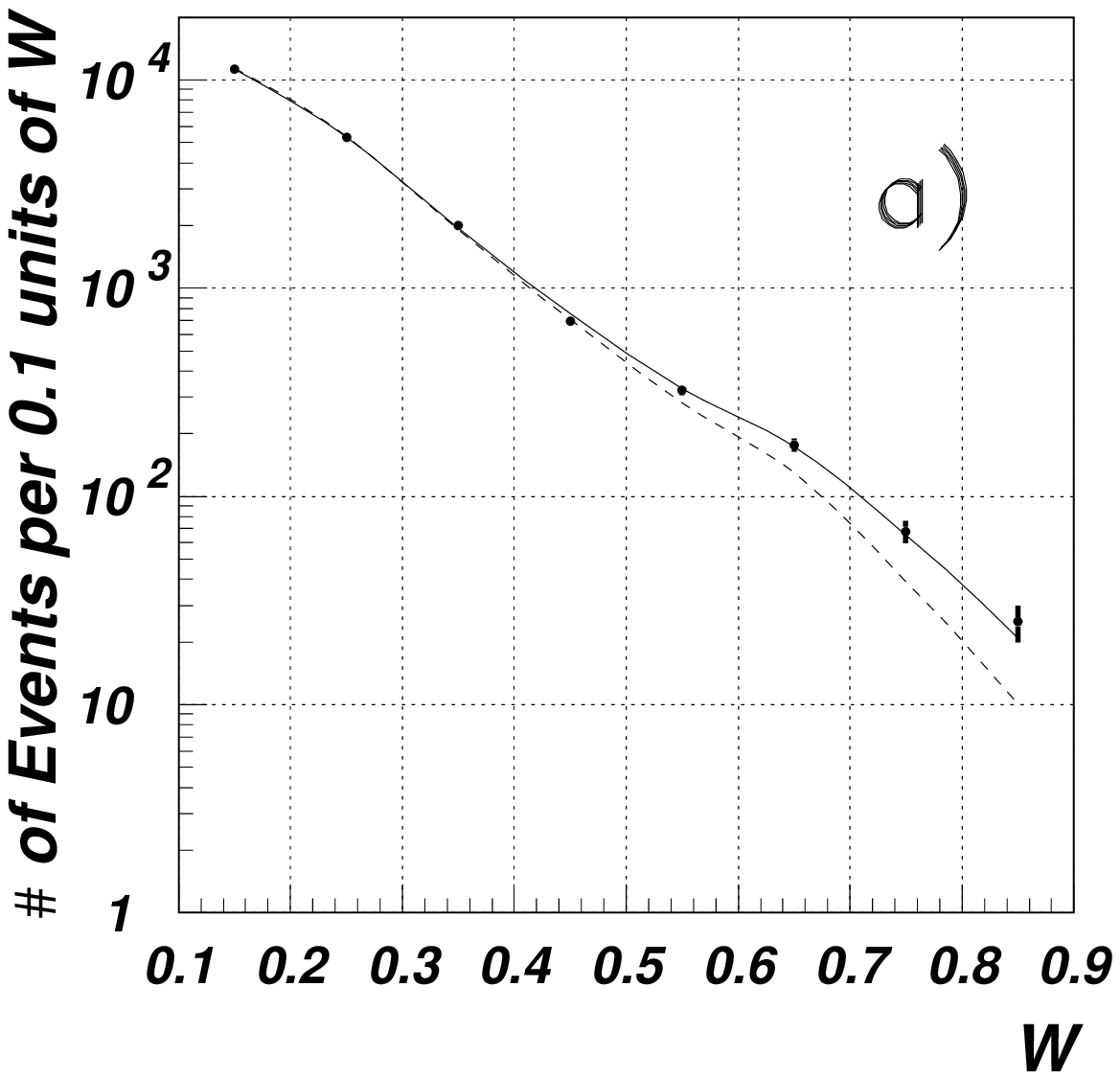,height=2.25in,width=2.6in,angle=0} 
   \epsfig{file=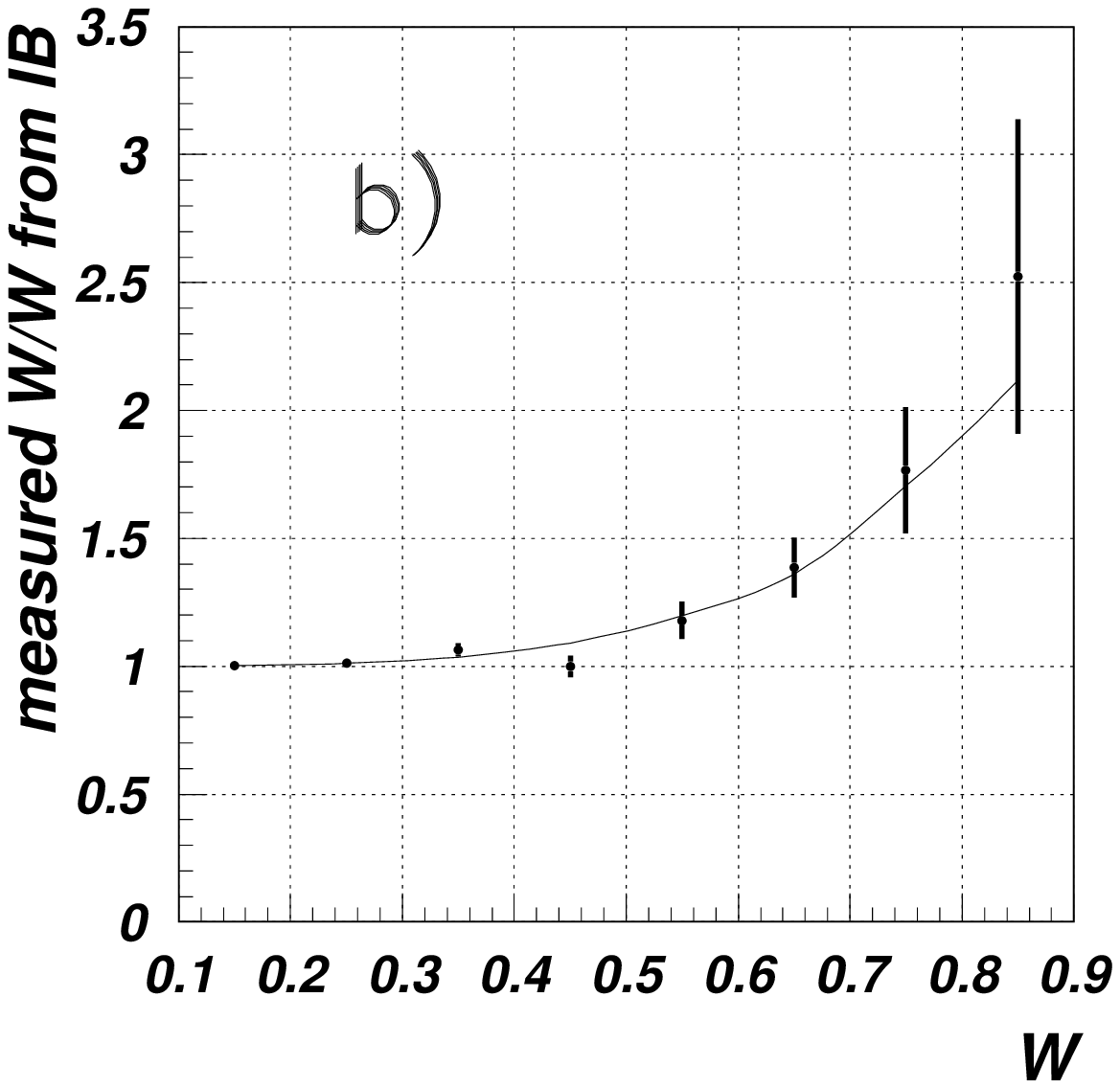,height=2.25in,width=2.6in,angle=0}
 \caption{ E787: a) The measured $W$ spectrum for signal events compared to best fits to
 IB$+$DE (solid curve) and IB alone (dashed curve); b) The ratio of the measured
$W$ spectrum to the predicted IB spectrum.  \label{fig_kppg} }
\end{figure}

The charged mode, \kppg, also has a long history~\cite{abrams,smith,bolotov}.
New results from E787~\cite{kppg} (see Section~\ref{sec:e787})
are striking in that the DE branching ratio is a factor of four lower than
the previous value.  The data are traditionally expressed in terms of
the variable $W$ that behaves similarly to the photon energy, and which is defined as
\begin{eqnarray}
 W^{2} & \equiv & (p\cdot q)/{m_{K^+}^{2}} \times
(p_{+}\cdot q)/{m_{\pi^+}^{2}} \\ \nonumber
           &   =  & E_{\gamma}^2\times (E_{\pi^+} - P_{\pi^+}\times
\cos{\bf\theta_{\pi^+\;\gamma}})/({m_{K^+}}\times{m_{\pi^+}^{2}}),
\label{eqn_w}
\end{eqnarray}
where $p$ is the four momentum of the kaon, $q$ is the four momentum of
the photon and $p_{+}$ is the four momentum of the $\pi^+$.
The new result from E787, based on half of the total data set,
shown in Figure~\ref{fig_kppg},
has about eight times higher statistics than previous results.  
The branching
ratio for the DE component, from a fit to IB and DE, is
B(\kppg;DE, $55<\!T_{\pi^+}\!<90$ MeV) = $(4.72\pm0.77\pm0.28)\times 10^{-6}$.
The interference term is small, $(-0.4\pm1.6)$\%, and the DE
is $(1.85\pm0.30)$\% compared with the IB term. The decay rate, 
corrected to full phase
space,\footnote{This correction assumes that the form factor has no
energy dependence.} is now measured to be similar to that for $K_L$:
$\Gamma(\kppg;DE)=808\pm132
s^{-1}$ vs. $\Gamma(\klppg;DE)=617\pm18 s^{-1}$.

In the neutral kaon decay, \klppg, the DE part of the decay can be either
{\it CP}-violating or {\it CP}-conserving, but experiments show that the DE decay
is consistent with a {\it CP}-conserving M1 radiative transition.
There is also a {\it CP}-odd interference term. These {\it CP}-odd
and {\it CP}-even terms manifest themselves 
in a {\it CP}-violating asymmetry in the polarization of
the photon, which is not observable in these experiments.
However, a {\it CP}- and {\it T}-odd angular asymmetry is expected
in the  related decay \klppee,
in which the photon internally converts to an $e^+e^-$ pair, since the
angular distribution of the leptons preserves information about the photon
polarization.
This effect, predicted in 1992 by Sehgal \& Wanninger~\cite{ppee_asym1},
is an asymmetry in the distribution of the angle $\phi$ between the
two planes formed by the lepton momenta and the pion momenta in the
$K_L$ rest frame.  The predicted asymmetry~\cite{ppee_asym1,ppee_asym2} 
is quite large 
because the two interfering amplitudes are of comparable size.

\begin{figure}[htbp]
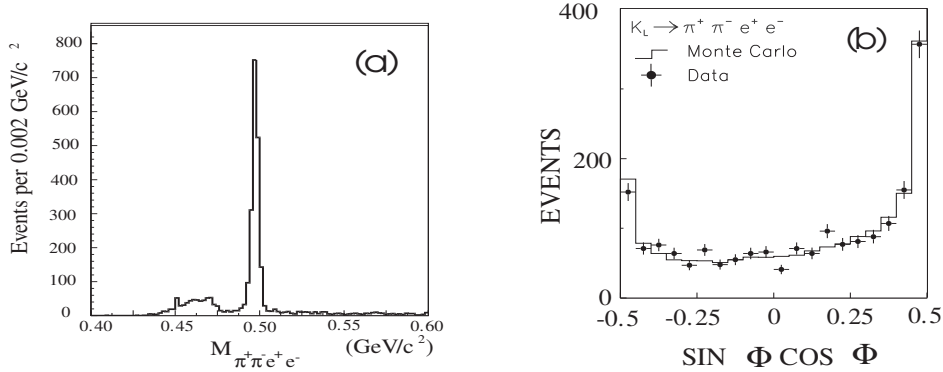

\epsfig{file=barkf13a.eps,height=2.5in,width=2.6in,angle=0}
\epsfig{file=barkf13b.eps,height=2.5in,width=2.6in,angle=0}
\caption{KTeV: a) Distribution of invariant mass for $\pi^+\pi^- e^+e^-$ events. 
b) Distribution of the angle $\phi$ between the $e^+e^-$ and $\pi^+\pi^-$
planes in the $K_L$ rest frame.  The asymmetry
observed between negative and positive values of $\sin\phi\cos\phi$ is
{\it CP}-violating and {\it T}-odd. \label{fig:e799_ppee}}
\end{figure}

NA48 reports~\cite{lubranohf8} (see Section~\ref{sec:na48}) 
a signal of 458 events, over 37 background events, 
giving a preliminary branching ratio of B(\klppee) = $(2.90\pm 0.15)\times 10^{-7}$.
In 1998, KTeV published a branching ratio~\cite{kpipieebr} based on a
small subset of the 1997 data.  A new result from the
full 1997 data set,
with over 1500 events, has now been reported~\cite{barkerhf8},
B(\klppee) = $(3.63\pm 0.11\pm 0.14)\times 10^{-7}$. 
The branching ratio is larger than the one reported by NA48, but
the difference is mostly due to the inclusion of
an M1 form factor, which is not used
in the NA48 analysis and which significantly increases the measured value
by reducing the acceptance.  Figure~\ref{fig:e799_ppee}$a$ shows the invariant 
mass spectrum observed by KTeV.

Both experiments observe a very large {\it CP}-violating and {\it T}-odd asymmetry.
The asymmetry is defined by
\begin{equation}
 A_\phi = {N(\sin\phi\cos\phi > 0) - N(\sin\phi\cos\phi < 0)\over
             N(\sin\phi\cos\phi > 0) + N(\sin\phi\cos\phi < 0)}. 
\end{equation}
It is important to note that the raw asymmetry may be significantly
different from the acceptance-corrected asymmetry.  This occurs not
because of any asymmetry in the detectors but because the asymmetry
varies across the phase space for the $\pi^+\pi^-e^+e^-$ final state,
and in general, acceptance is better in regions of the phase
space where the asymmetry is large.

The raw asymmetries observed by the two experiments are therefore not
directly comparable.  Nevertheless, they seem to
agree.  NA48 finds
$A_\phi({\rm raw}) = (20\pm 5)\%$ while KTeV measures
$A_\phi({\rm raw}) = (23.3\pm 2.3)\%.$
The angular distribution observed by KTeV is shown in 
Figure~\ref{fig:e799_ppee}$b$.

So far, only KTeV has reported an acceptance-corrected 
asymmetry~\cite{kpipieeasymm}. 
An important ingredient in making the acceptance correction is the
form factor in the M1 DE amplitude, which is extracted
by fitting the $M_{\pi\pi}$ and other kinematic distributions.  Using
the fitted form factor, the acceptance-corrected average asymmetry
is found to be
\begin{equation}
 A_\phi({\rm corrected}) = (13.6\pm 2.5\pm 1.2)\%, 
\end{equation}
in excellent agreement with the theoretical prediction based on the
value of the indirect {\it CP}-violation parameter $\eta_{+-}$ and
the known $\pi\pi$ phase shifts.

The NA48 experiment has reported the first observation of \ksppee
from the 1998 data sample~\cite{kspipieebr} (see Table~\ref{tab_kzppg}). 
\begin{table}[ht]
\centering
\caption{Summary of radiative K$_{\pi2}$ results }
\vskip 0.1 in
\begin{tabular}{|l||l|r|l|} \hline
Decay Mode  &  \multicolumn{1}{|c|}{Branching Ratio} & events& Experiment \\  \hline\hline
\klppg(DE) & $(2.92\pm0.07)\times10^{-5}$ & 5900 & KTeV (2000)~\cite{klppg} \\ \hline
\kppg(DE) &$(4.72\pm0.77\pm0.28)\times10^{-6}$ & 360 & E787 (2000)~\cite{kppg} \\ \hline
\klppee & $(3.63\pm0.11\pm0.14)\times10^{-7}$ & 1500 & KTeV (1998)~\cite{kpipieebr}\\ \hline
\ksppee & $(5.1\pm0.9\pm0.3)\times10^{-5}$ & 52 & NA48 (2000)~\cite{kspipieebr}\\ \hline
\klpopog & $< 5.6\times10^{-6}$ & 0 & NA31 (1994)~\cite{klp0p0g}\\ \hline
\ksppg & $(1.78\pm0.05)\times10^{-3}$ & 3700 & E731 (1993)~\cite{ksppg}\\
\ksppg(DE) & $<0.06\times10^{-3}$ & 0 & CERN (1976)~\cite{ksppg_de}\\ \hline
\end{tabular}
\label{tab_kzppg}
\end{table}
An additional
730 events were observed in a special two day $K_S$ run in 1999. The
raw asymmetry is consistent with 0, as expected for $K_S$.
Table~\ref{tab_kzppg} summarizes the current experimental status of radiative K$_{\pi2}$ decays.

\subsection{Radiative K$_{\ell2}$ Decays}
\label{sec:kln}

As with \kzppg, the radiative decays \klng and \klnll proceed via 
two separate mechanisms. The first,  IB, is the radiative version
of the familiar \kln decays. The second, structure dependent (SD),
as with the DE process in \kzppg, involves the emission of a photon
from an intermediate state and has been studied extensively within
the framework of ChPT~\cite{chpt_kmng,chpt_kmng2}. 
The IB amplitude is completely determined by the kaon decay constant
$f_K$. The SD amplitude is parametrized in terms of the three
form factors $F_V$, $F_A$ and $R$.
The vector (${\cal A}_V$) and axial-vector(${\cal A}_A$) contributions are given by
\begin{eqnarray}
{\cal A}_V & = & \frac{-eG_F \vus}{\sqrt{2}M_K} \epsilon^\mu\ell^\nu F_V 
	e_{\mu\nu\sigma\tau}q^\sigma k^\tau \\
{\cal A}_A & = & \frac{-ieG_F \vus}{\sqrt{2}M_K}
	F_A[(kq-q^2)g_{\mu\nu} -k_\mu q_\nu] + Rq^2g_{\mu\nu},
\end{eqnarray}
where $\epsilon^\mu$ is the photon polarization, $\ell^\nu$
is the lepton current and $k$ and $q$ are the kaon and photon
4-momentum.
In ${\cal O}(p^4)$ of ChPT, the form factors are independent of
$q^2$, although in ${\cal O}(p^6)$ they take on a $q^2$
dependence.

Recent measurements should allow precise 
experimental determinations of all three parameters.  The most recent
determination of $|F_V + F_A| = 0.165\pm0.007\pm0.011$ from the E787
(see Section~\ref{sec:e787}) measurement~\cite{kmng_de} of the 
DE component of \kmng is consistent
with the previous determination~\cite{keng} of 
$|F_V + F_A| = 0.148\pm0.010$ from \keng.  
The branching ratio for the structure-dependent component of \kmng, 
$B(\kmng; SD^+) = (1.33\pm0.12\pm0.18)\times10^{-5}$,
 is about 40\% higher than the ${\cal O}(p^4)$ ChPT
calculation~\cite{chpt_kmng2}, but the ${\cal O}(p^6)$ 
contributions are expected to 
increase  the calculation
by a comparable amount (based on the ${\cal O}(p^6)$ 
calculation for 
$\pi^+\rightarrow \ell^+\nu_\ell\gamma$)~\cite{chpt_kmng_p6}.
A value of $F_V - F_A = 0.102\pm0.073\pm0.044$, also derived from 
the recent E787 \kmng $(SD^+)$ measurement, is an improvement on
the previous limit of
$-0.3 < F_V - F_A < 2.5$~\cite{kmng} (see Reference~\cite{kmng_de}
for a discussion of the sign convention).  

An improved measure of $F_V - F_A$, along with measurements 
$F_V + F_A$ and the first measurement of $R$ in $K^+$ decays,
is now available from the E865 (see Section~\ref{sec:e865})
measurements of \kenee and \kmnee~\cite{e865_enee}. These data
were collected along with the \kpee data set in 1995--1996. 
The preliminary results from the combination of both modes
are (statistical errors only) 
$F_V - F_A = 0.073\pm0.033$, 
$F_V + F_A = 0.143\pm0.027$ 
and $R = 0.233\pm0.016$. 
Combining the results on radiative \kln decays,
all three form factors, 
$F_V$, $F_A$ and $R$, will be well 
determined, along with their sign relative to IB.

The E865 experiment has also observed some few dozen \kenmm events, collected
along with the \kppen data in 1997~\cite{zeller00} (see Section~\ref{sec:ppen}).
A summary of the recent radiative K$_{\ell2}$ results is presented in
Table~\ref{tab_kl2g}.
\begin{table}[ht]
\centering
\caption{Summary of radiative K$_{\ell2}$ results}
\vskip 0.1 in
\begin{tabular}{|l||l|r|l|} \hline
Decay Mode  &  \multicolumn{1}{|c|}{Branching Ratio} & events& Experiment 
\\ \hline\hline
\kmng    & $(5.50\pm0.28)\times10^{-3}$ & & KEK (1985)~\cite{kmng}\\
\kmng(DE)& $(1.33\pm0.12\pm0.18)\times10^{-5}$ & 2588 & E787 
(2000)~\cite{kmng_de}\\ \hline
\keng(DE)& $(1.52\pm0.23)\times10^{-5}$ & 51 & CERN (1979)~\cite{keng}\\ \hline
\kmnmm   & $< 4.1\times10^{-7}$ & 0 & E787 (1989)~\cite{kmnmm}\\ \hline
\kenmm   & $< 5.0\times10^{-7}$ & 0 & E787 (1998)~\cite{kenmm}\\ \hline
\kmnee   & $(6.84\pm0.40)\times10^{-8}$ & $\sim$1500 & E865 
(2000)~\cite{e865_enee}\\ \hline
\kenee   & $(2.60\pm0.15)\times10^{-8}$ & $\sim$400 & E865 
(2000)~\cite{e865_enee}\\ \hline
\end{tabular}
\label{tab_kl2g}
\end{table}

\subsection{\kzppen}

\label{sec:ppen}
The \kppen decay provides the best system for study of $\pi\pi$
scattering at low energy. The measurement of the relative
phase of the pion wave functions is an important test for ChPT,
as $\pi\pi$ scattering is uniquely sensitive to chiral symmetry
breaking in the strong interaction and can provide powerful
constraints on the parameters of ChPT.
The ChPT calculations of $\pi\pi$ scattering have been done to
${\cal O}(p^4)$~\cite{chpt_gl,chpt_ppen0} and 
${\cal O}(p^6)$~\cite{chpt_ppen}.

\begin{figure}[htbp]
\epsfig{file=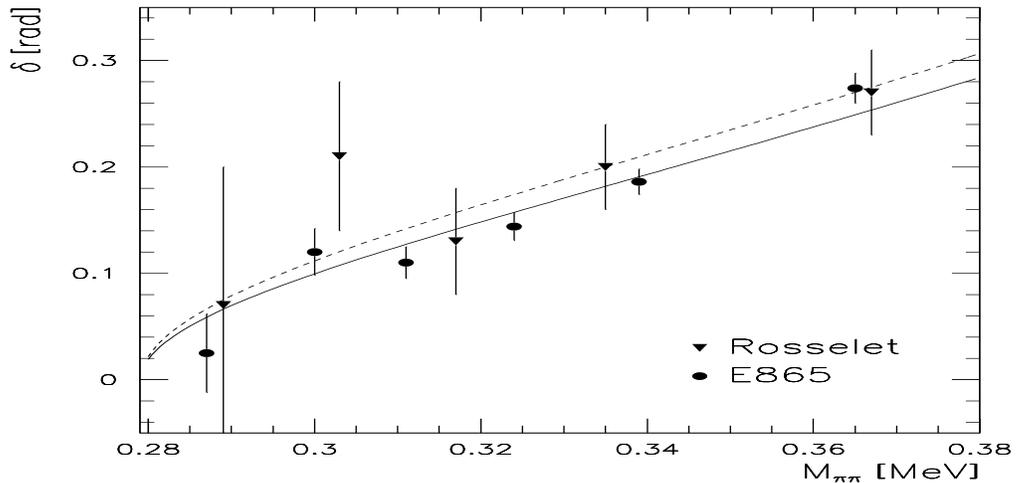,height=2.5in,width=5.25in,angle=0}
 \caption{Preliminary $\pi\pi$ phase shift ($\delta$) vs $\pi\pi$ invariant 
mass ($M_{\pi\pi}$).
Both older data from Rosselet~\cite{rosselet} and recent data from 
E865~\cite{e865_ppen1} are shown, along with fits based on the 
Roy equation formalism
of Basdevant, Froggatt \& Petersen~\cite{basdevant}.
 \label{fig_ppen} }
\end{figure}

The primary motivation for studying \kppen is to measure this $\pi\pi$ 
scattering. The previous experiment had
$\sim$30,000 events~\cite{rosselet}. The E865 experiment 
(see Section~\ref{sec:e865})
has collected $\sim$400,000 events. 
Figure~\ref{fig_ppen} shows a preliminary plot of the $\pi\pi$ phase shift 
($\delta \equiv \delta_0^0 - \delta_1^1$) as a function of $\pi\pi$
invariant mass ($M_{\pi\pi}$)
for the new E865 data~\cite{e865_ppen1} and the previous Rosselet data. 
The preliminary E865 result, from a fit to an improved functional form~\cite{schenk},
gives a 
scattering length of $a_0^0 = 0.235\pm0.013(stat)$~\cite{e865_ppen2}.
In principle, $\pi\pi$ phase shifts could be extracted from the decay
\klppen, or from the modes, \kppmn, \klppmn.  However, due to limited
statistics in these modes, they have not contributed significantly.
Table~\ref{tab_kppen} gives a summary of K$_{\ell 4}$ decays.
\begin{table}[ht]
\centering
\caption{Summary of K$_{\ell 4}$ results}
\vskip 0.1 in
\begin{tabular}{|l||l|r|l|} \hline
Decay Mode  &  \multicolumn{1}{|c|}{Branching Ratio} & events & Experiment \\ \hline\hline
\kppen   & $(4.10\pm0.01\pm0.11)\times10^{-5}$  & $>350000$ & E865 (2000)~\cite{e865_ppen2}\\ \hline
\kppmn   & $(1.4\pm0.9)\times10^{-5}$  & 7 & CERN (1967)~\cite{kppmn}\\ \hline
\klppen  & $(5.16\pm0.20\pm0.22)\times10^{-5}$  & 729  & E799 (1993)~\cite{klppen}\\ \hline
\kpopoen & $(2.1\pm0.4)\times10^{-5}$ & 10 & ITEP (1988)~\cite{kp0p0en}\\ \hline
\kpopoeng& $< 5\times10^{-6}$ & 0 & ITEP (1992)~\cite{kp0p0eng}\\ \hline
\end{tabular}
\label{tab_kppen}
\end{table}
E865 has collected \kppeng events as part
of the study of \kppen and may report on this observation in the near future.

\subsection{Other Rare or Radiative Decays}

\subsubsection{Rare $K_S$ Decays}
\label{sec:ksppp}

         The CPLEAR experiment 	studies $p\bar p$ annihilation
     at low energy leading to the final state
    $K^+\pi^-\overline K^\circ$
     and its charge conjugate $K^-\pi^+ K^\circ.$  The sign of the
     charged kaon allows the neutral kaon to be tagged as either
     a $K^\circ$ or a $\overline K^\circ.$  Detailed measurements of the
     proper time decay distributions for these states to a
     particular final state allow extraction of a variety of
     parameters describing both $K_L$ and $K_S$ decays to
     each final state.

          This technique allows the indirect determination of 
     some rare $K_S$ decays.  Most notably, CPLEAR has recently
     studied the $\pi^+\pi^-\pi^\circ$ 
     final state, a common $K_L$ decay~\cite{ksppp}.  By fitting the
     proper time decay distributions for initial $K^{\circ}$ and 
     $\overline K^{\circ},$ and fitting the Dalitz plot distributions
     of these events, one can determine both {\it CP}-conserving and
     {\it CP}-violating amplitudes for \ksppp.
     The {\it CP}-violating amplitude is found to be consistent with
     zero, but the {\it CP}-conserving amplitude is observed with a
     three- to four-sigma significance.  CPLEAR has converted this
     observation to a branching ratio~\cite{ksppp} B(\ksppp) = 
     $(2.5^{+1.3\; +0.5}_{-1.0\; -0.6})\times 10^{-7}.$
     Although it is an indirect measurement, this is the smallest
     $K_S$ branching ratio yet reported.

A new limit on \kspopopo has recently been presented by the
SND experiment at VEPP-2M in Novosibirsk: B(\kspopopo) 
$<1.4\times10^{-5}$~\cite{ksp0p0p0}.

\subsubsection{Radiative Three-Body Decays}

\label{sec:kpppg}
The experimental results for other radiative kaon decays
(e.g.\ K$_{\pi3\gamma}$ and K$_{\ell3\gamma}$)
are only sensitive to IB contributions. All of these measurements
are consistent with theoretical predictions.  A summary of these measurements
is given in Table~\ref{tab_kpppg}.
\begin{table}[ht]
\centering
\caption{Summary of radiative and rare three-body decays}
\vskip 0.1 in
\begin{tabular}{|l||l|r|l|} \hline
Decay Mode  &  \multicolumn{1}{|c|}{Branching Ratio} & events & Experiment \\ \hline\hline
\ksppp  & $(2.5^{+1.3\; +0.5}_{-1.0\; -0.6})\times 10^{-7}$ &  & CPLEAR (1997)~\cite{ksppp}\\ \hline
\kspopopo& $<1.4\times 10^{-5}$ & 0 & SND (1999)~\cite{ksp0p0p0}\\ \hline
\kpppg  & $(1.04\pm0.31)\times10^{-4}$ & 7 & ITEP (1989)~\cite{kpppg}\\ \hline
\kppopog& $(7.5^{+5.5}_{-3.0})\times10^{-6}$ & 5 & IHEP (1995)~\cite{kpp0p0g}\\ \hline
\kpmng  & $< 6.1 \times10^{-5}$ & 0 & ZGS (1973)~\cite{kpmng}\\ \hline
\klpmng & $(5.7^{+0.6}_{-0.7})\times10^{-4}$ & 252 & NA48 (1998)~\cite{klpmng}\\ \hline
\kpeng  & $(2.62\pm0.20)\times10^{-4}$ & 88 & ITEP (1991)~\cite{kpeng}\\ \hline
\kpeng(SD)& $<5.3\times10^{-5}$ & 0 & IHEP (1986)~\cite{kpeng_de}\\ \hline
\klpeng & $(3.62^{+0.26}_{-0.21})\times10^{-4}$ & 1384 & NA31 (1996)~\cite{klpeng}\\ \hline
\end{tabular}
\label{tab_kpppg}
\end{table}
The \kpmng mode should be seen for the first time in existing data
from E787. Improvements in other  modes are
likely to come from KLOE and the Institute for High Energy Physics (IHEP)
at Serpukhov.

\section{LEPTON FLAVOR VIOLATION}

All experimental evidence to date supports the exact conservation of 
an additive quantum number for each family of charged leptons.
Thus \klmm and \klee are allowed, although suppressed by the GIM mechanism
and by helicity suppression, whereas \klme appears to be absolutely
forbidden.  If neutrino masses are nonzero, some very tiny mixing
effects could permit such a decay in the standard model, but it would occur at 
unobservably small levels, many orders of magnitude beyond the present
experimental sensitivity.  Any observation of a signal for the decays
\klme, \kpme, or \klpme would thus be conclusive evidence for new physics
beyond the standard model~\cite{Wilczek}.

Although this lepton-flavor-number conservation law
appears to be respected
in the standard model, there is no fundamental reason or underlying symmetry to explain
why this should be so.  Indeed, many possible extensions to the standard model predict
new interactions involving heavy intermediate gauge bosons that could mediate
the otherwise forbidden LFV  decays.  Some of the
specific models that lead to LFV decays 
include~\cite{Eichten}
compositeness of quarks and leptons, left-right symmetric models,
technicolor, some supersymmetric models,
unified theories with horizontal gauge bosons, leptoquarks, and string theories.

It is important to look for both \klme and the modes with an extra pion,
\kpme and \klpme, because the \klme decay is sensitive to pseudoscalar and
axial vector coupling, whereas the other modes are sensitive to scalar or
vector couplings.  In both cases, the excellent sensitivity of these 
experiments
probes mass scales that are very
large.  Of course, the sensitivity of the experiments to new
interactions depends on the coupling constants involved.  If the new coupling 
for an intermediate vector boson of mass $M_X$ is $g_X,$ then the lower bound
on $M_X$ implied by an upper limit on $B(\klme)$ is given in terms of the
electroweak coupling $g$ by the approximate expression
\begin{equation}
M_X  \simeq  200 {\rm TeV}/c^2 \times \frac{g_X}{g} \times 
\left[ \frac{10^{-12}}{B(\klme)}\right]^{1/4}.
\label{eqn:kme}
\end{equation}
Thus, upper limits in the range of $10^{-12}$ yield impressive lower bounds
on $M_X,$ at least if $g_X$ is comparable to $g.$  
The following sections discuss
the present experimental limits on LFV modes and the prospects for further improvement.
While limits from rare kaon decays have provided the most stringent limits on
some models of BSM physics, there are also strong limits from neutrino-less
double beta decay and from several rare muon decays. The experimental focus
in the field has now shifted to improving some of the rare muon decay limits.

\subsection{\klme}

\begin{figure}[ht]
\epsfig{file=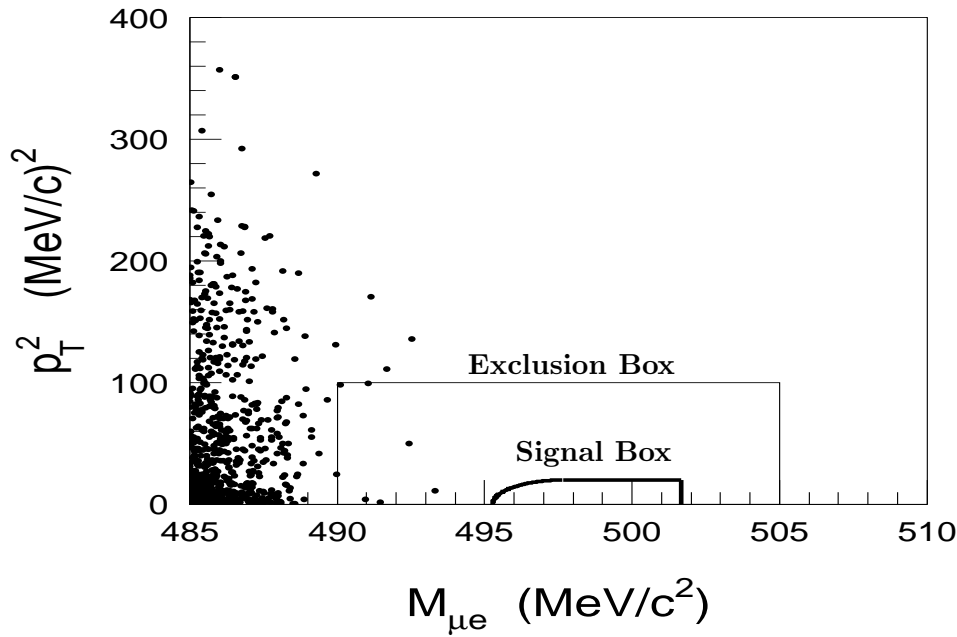,height=3.0in,width=5.25in,angle=0}

\vspace{-2.7cm}
\hbox{\bf \hspace{2.5in}  Exclusion Box} 
\vspace{0.7cm}
\hbox{\bf \hspace{2.75in}  Signal Box} 
\vspace{2.7cm}

\caption{E871: Final data sample with the reconstructed mass $M_{\mu e}$ vs
the square of the transverse momentum relative to the kaon direction,
  after all cuts there are no events in the
signal region. The exclusion box (the larger box enclosing the signal box)
was used to set cuts in an unbiased
way on data far from the signal region. The shape of the signal box
was optimized to maximize signal/background.
\label{fig:e871}}
\end{figure}

Experimental limits of \klme have steadily improved over the past
decade~\cite{e780_me,e791_me,e137_me}
from a level of about $10^{-8}$ in 
1988 to the final result
of BNL E871~\cite{e871_me}, published in 1998.  The E871
spectrometer is described in Section~\ref{sec:e871}.  It features two
analysis magnets for redundant momentum measurements.  Electrons 
and muons are each identified in two different ways to reduce
background from particle misidentification.  The analysis cuts
are then chosen to minimize the remaining backgrounds (involving
either accidental coincidences or scattered electrons) while
maintaining as much sensitivity as possible to the signal.
The final cuts correspond to a single-event sensitivity of about
$2\times 10^{-12}$ with an expected background of 0.1 event.
The cuts are set without looking at the data within the 
exclusion box shown in Figure~\ref{fig:e871}.  This ``blind'' analysis
technique ensures that the cut selection remains unbiased by
the data, and it has been adopted in many of the rare decay analyses
described in this article.

When the exclusion box is opened, no events are found in the
smaller signal region.  This allowed E871 to set a 90\%-CL
upper limit of B(\klme) $<4.7\times 10^{-12},$ the smallest upper limit
set to date on any kaon decay mode.  For an exotic boson with
electroweak coupling strength, Equation~30 then implies 
a lower bound on its mass of 150~TeV.

There are no near-term plans to pursue this decay further, as the
background from \klpen with a muon decay and a scattered electron
is difficult to reduce below a level of $10^{-13}$, which is just beyond
the E871 sensitivity.

\subsection{\kpme}

\begin{figure}[ht]
\epsfig{file=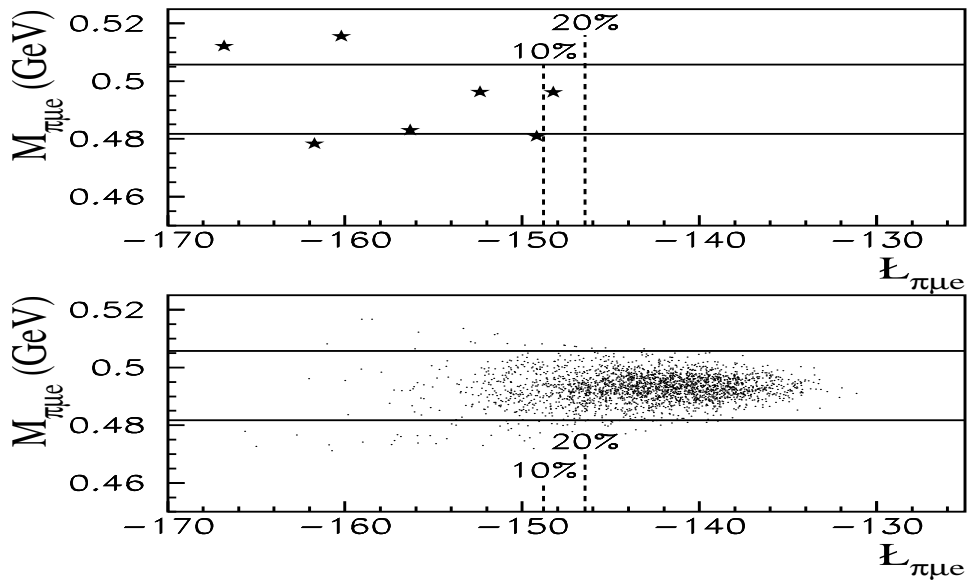,height=3.0in,width=5.in,angle=0}
\caption{E865: Final 1996 data sample after all cuts, with no events above
$\pi\mu e$ likelihood ($\L_{\pi\mu e}$) of 20\%. The $\pi\mu e$ likelihood is 
formed from
the quality of the track fits, timing, vertex, reconstruction to the target, 
and particle identification. 
Also shown are $\pi\mu e$ Monte Carlo
events passing all cuts.  \label{fig:e865}}
\end{figure}

E865 at BNL was designed to search for the LFV
decay \kpme. This decay, 
with an extra pion in the final state, is sensitive to exotic
gauge bosons with different quantum numbers from 
those that E871 could detect.
  The experiment uses $K^+$ decays in
flight, and the detector concept is similar to that
of E871, with redundant
particle identification by two Cerenkov detectors and an
electromagnetic calorimeter, and
a muon range stack (see Section~\ref{sec:e865}). Data were collected during
the 1995, 1996, and 1998 runs of the AGS. The limit on this
mode from the 1995 run~\cite{e865_95}, similar in sensitivity to the predecessor
experiment E777~\cite{e777_old}, was B(\kpme) $< 2.1\times10^{-10}$.  The limit from the
1996 run~\cite{zeller}, with no events above a likelihood to be \kpme ($\L_{\pi\mu e}$)
of 20\%, is B(\kpme) $< 3.9\times10^{-11}$ (see Figure~\ref{fig:e865}).
From the combined results from E777 and the E865 runs in 1995 and 1996, a
limit of B(\kpme) $ < 2.8\times10^{-11}$ is obtained.  The final
sensitivity, with the 1998 data included, is expected to be $\sim$3
times better.  E865 is already close to being limited by background
from accidentals; there are no plans to continue with this search.
The E865 limit implies a lower bound of several tens of TeV on exotic bosons
with electroweak coupling, depending on the exact model used.

\subsection{\klpme}

\begin{figure}[ht]
\epsfig{file=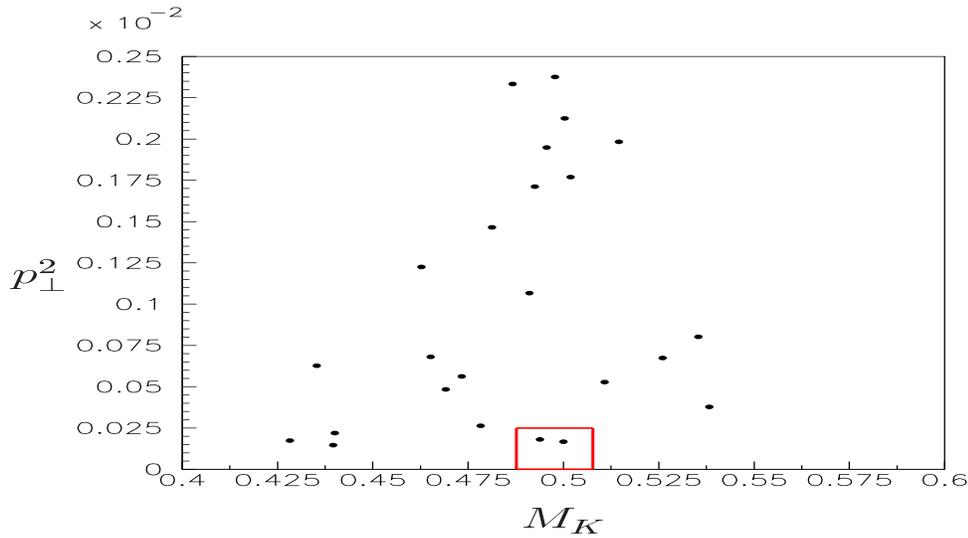,height=3.0in,width=5.25in,angle=0}
\caption{
KTeV: 1997 data sample after all cuts, with two background events 
in the signal region.
\label{fig:e799_pme}}
\end{figure}

In addition to the search for \kpme
performed by BNL E865, a search for the corresponding neutral mode
\klpme has recently been carried out by KTeV at FNAL (see Section~\ref{sec:e799}).
The main background concern was the common decay 
\klpen, in which the pion is
misidentified as a muon, with 0.6$\pm$0.6 expected events.
After all cuts, two background events were observed in the signal box, 
as can be seen
in Figure~\ref{fig:e799_pme} and the preliminary 
90\%-CL limit 
on \klpme from KTeV~\cite{e799_pme}  
is B(\klpme) $< 4.4\times10^{-10}$.
The final KTeV sensitivity, including the 1999 data, will at least double.

\subsection{Other Searches for New Physics}

In addition to the LFV searches, there have been a number
of other searches for BSM physics in recent years. These 
include  \kpx, $K^+\rightarrow\pi^-\mu^+\mu^+$, and
$K^+\rightarrow\pi^+\mu^- e^+$. Table~\ref{tab_bsm} summarizes
BSM searches reported since the previous review~\cite{litt}.
\begin{table}[ht]
\centering
\caption{  Summary of searches for physics beyond the standard model}
\vskip 0.1 in
\begin{tabular}{|l||l|l|} \hline
Decay Mode  &  \multicolumn{1}{|c|}{Branching Ratio} & Experiment \\  \hline\hline
\klme  & $< 4.7\times10^{-12}$ & E871 (1998)~\cite{e871_me}\\ \hline
\kpme  & $< 2.8\times10^{-11}$ & E865 (2000)~\cite{zeller}\\ \hline
\klpme & $< 4.4\times10^{-10}$  & KTeV (2000)~\cite{ktevpme}\\ \hline
\kpx   & $< 1.1\times10^{-10}$ & E787 (2000)~\cite{e787_pnn2}\\ \hline
$K^+\rightarrow\pi^-\mu^+\mu^+$ & $<3.0\times10^{-9}$ & E865 (2000)~\cite{kpmm_opp}\\ \hline
$K^+\rightarrow\pi^- e^+ e^+$ & $<6.4\times10^{-10}$ & E865 (2000)~\cite{kpmm_opp}\\ \hline
$K^+\rightarrow\pi^+\mu^- e^+$ & $<5.2\times10^{-10}$ & E865 (2000)~\cite{kpmm_opp}\\ \hline
$K^+\rightarrow\pi^-\mu^+ e^+$ & $<5.0\times10^{-10}$ & E865 (2000)~\cite{kpmm_opp}\\ \hline
\klbeemm & $< 1.36\times10^{-10}$  & KTeV (2000)~\cite{klmmg}\\ \hline
\end{tabular}
\label{tab_bsm}
\end{table}
The E865 limit on $K^+\rightarrow\pi^-\mu^+\mu^+$ is derived from the
\kpmm data set collected in 1997 and the other E865 lepton number violating
limits in Table~\ref{tab_bsm} derive from the \kppen data set also collected
in 1997. These modes have attracted interest in the context of heavy neutrino 
searches~\cite{shrock}.
Existing KTeV data could be used to search for some other exotic modes, such as
$K^\circ\rightarrow\pi^\pm\pi^\pm e^\mp e^\mp$ and
$K^\circ\rightarrow\pi^\pm\pi^\pm\mu^\mp\mu^\mp$, in the near future.

\section{CONCLUSIONS AND FUTURE PROSPECTS}

The unprecedented sensitivities of the rare kaon decay experiments in
setting limits on LFV have constrained many extensions of the standard model. 

The observation of \kpnn has opened the doors to measurements of the
unitarity triangle completely within the kaon system.  Significant
progress in the determination of the fundamental CKM parameters will
come from the generation of experiments that is starting now.
Comparison with the B meson system will then overconstrain the unitarity 
triangle and
test the standard-model explanation of {\it CP} violation.

The primary focus for the future of rare kaon decays is on
the measurement of the golden modes, \klpnn and \kpnn,
at sensitivities sufficient for observation of 100 SM events. Major initiatives
in this regard are underway at BNL, FNAL and KEK. At the same time
the study of a number of medium-rare and radiative modes will
be pursued, both as a by-product of the \kzpnn measurements
and current $\epsilon'/\epsilon$ measurements and in a dedicated
study at IHEP.

\subsection{Medium-Rare and Radiative Decays}

The DA$\Phi$NE $e^+e^-$ accelerator complex and KLOE detector 
at Frascati (Italy) (see Section~\ref{sec:kloe}) were
both commissioned in 1999. 
It is expected that by the time the machine  reaches 
the full luminosity of $5\times10^{32} {\rm cm}^{-2}{\rm s}^{-1}$,
KLOE will be able to observe $10^{10}$ tagged
kaons of all charges per year. In addition to the measurement
of $\epsilon'/\epsilon$, KLOE will provide a wealth of new
measurements on many rare and medium-rare decays, particularly for the
$K_S$ modes.

A new kaon decay experiment (see Section~\ref{sec:ihep}) is planned at 
the U70 accelerator in IHEP in Serpukhov. 
CERN has provided IHEP with an RF separator
to be used in a 12-GeV/$c$ separated kaon beam. This experiment
may have even greater sensitivity than KLOE to medium-rare
charged kaon decays.

The NA48 collaboration has planned a special run after the final
$\epsilon'/\epsilon$ running, with the $K_L$ beam turned off,
dedicated to improving sensitivity to rare $K_S$ decays. 
NA48 is also
considering improved measurements of some charged kaon
decays.

\subsection{\kzpnn}

The principal focus of the kaon community is the precise measurement
of the \kzpnn decay modes. These measurements will provide critical,
unambiguous determination of the standard-model {\it CP} violation
parameters. Comparison with measurements from the B meson system will then
over-constrain these parameters and test the standard-model picture of
{\it CP} violation. 

These measurements are difficult, but several clear and convincing
cases have been made for measuring up to ${\cal O}(100)$ events in
both modes. This is as far as it makes sense to go with
measurements of \kpnn, as
the charm quark uncertainty in extracting $|\vtd|$ will then
dominate. In \klpnn, experimental uncertainties will dominate
the errors.

\subsubsection{\kpnn}

A clean, convincing \kpnn event has already been seen by E787.
Building on this success, the new E949 experiment will make
modest and well-understood upgrades to the E787 detector,
which has already demonstrated sufficient background rejection
for a very precise measurement
of B(\kpnn). The experiment will make use of the entire proton flux 
from the AGS to increase its sensitivity per hour by a factor of 15 
over the sensitivity E787 achieved in 1995.
E949 is currently under construction and will
run in 2001 through 2003. The E949 sensitivity should reach  one
order of magnitude below the expectation for the signal, 
and the experiment should observe 10 standard-model events. 
The background is well-understood and is 10\% of the 
standard-model signal.
 
A proposal to improve the \kpnn sensitivity by a further factor of ten
has been initiated
at FNAL. The CKM experiment plans to collect 100 standard-model 
events, with a background to signal ratio of $\sim$10\%, in a two-year run 
starting about 2005.  This
experiment will use a new technique, with $K^+$ decay-in-flight
and momentum/velocity spectrometers. It will have significant muon veto
and photon veto capabilities and redundant tracking of both the
kaon and pion.

\subsubsection{\klpnn}

The next generation of \klpnn experiments will start with E391a at KEK,
which hopes to reach a sensitivity of $\sim$$10^{-10}$. Although the 
reach of E391a is not sufficient
to observe a signal at the standard model level
the experiment will be able to rule
out large BSM enhancements
and learn more about how to do this difficult experiment.
It is designed around a pencil $K_L$ beam, a high-resolution
crystal calorimeter, and very efficient photon veto systems. This
experiment would eventually  move to the JHF and aim for a
sensitivity of ${\cal O} (10^{-14})$.

   KAMI plans to reuse the excellent KTeV CsI calorimeter, which will
    have to be restacked to accommodate the single KAMI beam.  The
    decay volume upstream of the calorimeter will be instrumented with a
    fiber-tracker system and surrounded by a hermetic, highly efficient
    array of photon veto detectors.  An additional photon detector
    will catch photons escaping along the beam.  The initial
    KAMI run (the KAMI-far configuration) would take place with
    the target in the same location as for KTeV, some 180~m
    upstream of the CsI calorimeter.  Later, the target would be moved downstream
    (the KAMI-near configuration) in order to increase the solid angle
    and the resulting kaon intensity.  KAMI hopes to collect
    about 20 events per year of running in the KAMI-far configuration,
    increasing to 100 events per year with KAMI-near.  Backgrounds
    due to lost photons are a major concern, particularly due to
    photons escaping  down the beam hole, where there is a high neutron flux.
    If KAMI is approved, running in the KAMI-far configuration may
    begin around 2005--2006, with KAMI-near following perhaps around 2008.

KOPIO follows a different strategy.  The kaon
center of mass will be reconstructed using a bunched proton beam and a
very-low-momentum $K_L$ beam.  This technique allows for
two independent criteria to
reject background, photon veto and kinematics---allowing background
levels to be directly measured from the data---and encourages further
confidence in the signal by measuring the momentum spectrum of the decay. 
A large flux will
be obtained using the entire AGS proton current. The low-energy beam
also substantially reduces backgrounds from neutrons and other
sources. After three years of running, 65 standard-model events are expected
with a S/B $\ge$ 2:1.

\section{EXPERIMENTAL CONFIGURATIONS}

\subsection{BNL: AGS}

The Alternating Gradient Synchrotron (AGS) at Brookhaven National Laboratory 
(BNL) began operation in 1960. 
In the intervening 40 years, the intensity has increased
substantially to $\geq$65 Tp (Tp $\equiv 10^{12}$ protons per spill). The typical cycle
time of the accelerator  up to 1990 was a 1-s spill every 3~s (33\%
duty factor). 
Since that time, the duty factor has been increased as high as 55\%
(in 1998), with a 2.8-s spill every 5.1~s. The AGS has also achieved
microbunching during extraction, as required by the proposed 
E926 \klpnn experiment.

\subsubsection{E865}
\label{sec:e865}

\begin{figure}[htbp]
\epsfig{file=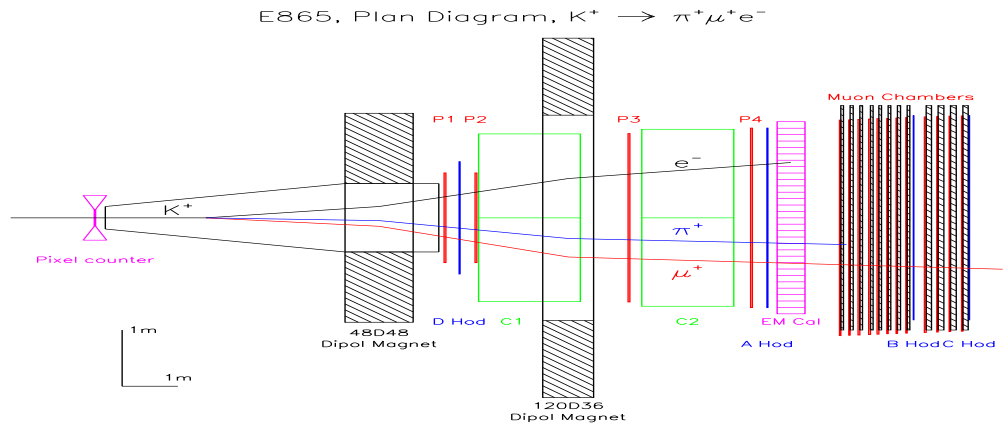,height=2.5in, width=5.25in,angle=0}
\caption{ Plan view of the E865 detector at BNL.
\label{fig:e865_det}}
\end{figure}

Experiment E777, a search for \kpme~\cite{e777_pee,e777_old}, 
ran from 1986 through 1988 and
was then modified slightly, E851, to optimize for 
$\pi^\circ\rightarrow e^+e^-$ from \kpp~\cite{pee} and
\kpee~\cite{e851_pee} and ran in 1989.
An upgraded experiment, E865~\cite{e865_nim} (see Figure~\ref{fig:e865_det}),
to  search for \kpme~\cite{e865_pee,e865_pmm,e865_95,zeller}
ran from 1995 through 1998.
The detector sits in an intense unseparated 6-GeV/$c$ $K^+$ beam, with
30 MHz of $K^+$ and 600 MHz of $\pi^+$. The first magnet separates
the charged kaon decay products, with negative particles going left; the
second magnet provides momentum analysis of these tracks with
four stations of high-rate multi-wire proportional chambers. Particle
identification consists of two sets of segmented threshold Cerenkov counters:
the left side, with a high threshold gas (H$_2$), was optimized to reject 
$\mu$'s and $\pi$'s; the right side, with a low threshold gas (CO$_2$ or
CH$_4$), was  optimized to reject e$^+$ from $\pi^\circ$ Dalitz decays. In
addition, a Pb-scintillator Shashlyk calorimeter provides electron
identification and a range stack of alternating iron plates and multi-wire
proportional chambers provides muon identification.

\subsubsection{E871}

\label{sec:e871}

\begin{figure}[htbp]
\epsfig{file=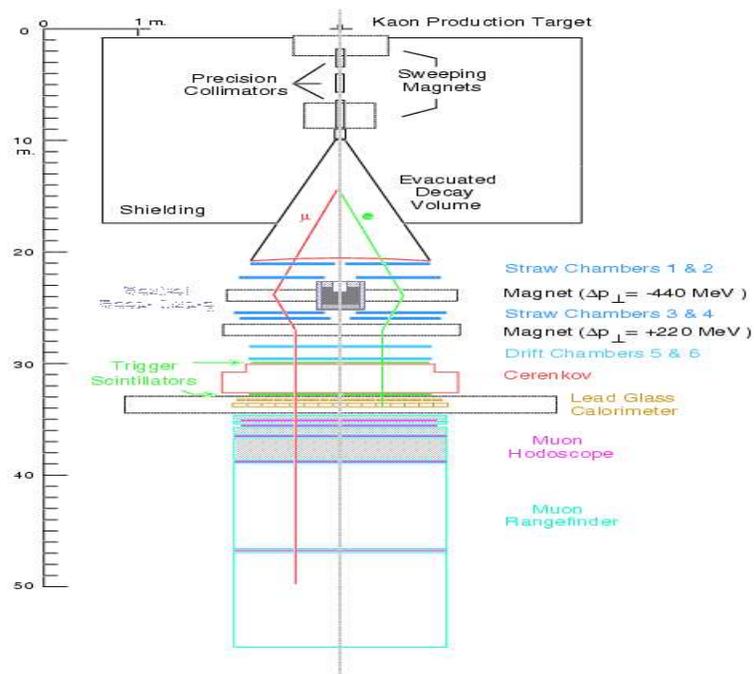,height=3.75in, width=4.25in,angle=0}
\caption{Plan view of the E871 detector at BNL. 
\label{fig:e871_det}}
\end{figure}

Experiment E791~\cite{e791_det},
a search for \klme~\cite{e791_me,e791_other},
ran from 1988 through 1990.
An upgraded experiment to continue the 
\klme~\cite{e871_mm,e871_ee,e871_me} search,
E871~\cite{e871_det}, ran during 1995--1996
(see Figure~\ref{fig:e871_det}).
The spectrometer has two arms, each with six gas-drift tracking stations and
two momentum-analyzing magnets to provide independent momentum measurements.
Each tracking station has three $x$
(bending plane) measurements to minimize the probability of tracking errors.
This is critical, as a tracking mistake in combination with a pion decay can
give a good track $\chi^2$ with a
mismeasured momentum. A novel beam plug, to stop the neutral
beam in the center of the first magnet, succeeded in reducing rates
in the downstream particle identification detectors. 
Redundant electron identification uses
an H$_2$ Cerenkov counter and a lead glass calorimeter. 
Muon identification is achieved with a range stack
of scintillator and drift tubes with an absorber of iron, marble and aluminum.

\subsubsection{E787 \& E949}
\label{sec:e787}

\begin{figure}[htbp]
\epsfig{file=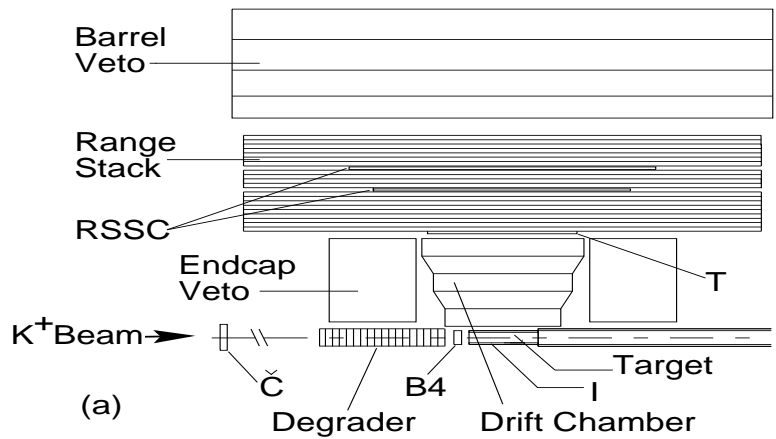,height=2.25in, width=4.in,angle=0}
\caption{Elevation view of the top half of the E787 detector at BNL.
\label{fig:e787_det}}
\end{figure}

Experiment E787~\cite{e787_det}, to search for 
\kpnn (see Figure~\ref{fig:e787_det}), ran from 1989 through 
1991~\cite{e787_pnn1,kpgg,e787_pmm,atiya}
and again, after an  upgrade~\cite{e787_det2}, 
from 1994 through 1998~\cite{e787_pnn2,kmng_de}.
The E787 detector is located in a low-energy separated $K^+$ beam.
The beam particles are tagged 
with a Cerenkov counter and tracked
with MWPC and scintillator counters until stopped
in a scintillating
fiber target inside of a 1-T magnetic field. The kaon decay
particles are tracked through the fiber target, a low-mass
central drift chamber, and into a segmented cylindrical plastic
scintillator range stack, with embedded straw tube chambers.
The $\pi^+\rightarrow\mu^+\rightarrow e^+$ decay chain is identified with
500-MHz transient digitizers recording output from the entire range stack.
The detector is surrounded by a nearly hermetic photon veto
system.

\begin{figure}[htbp]
\epsfig{file=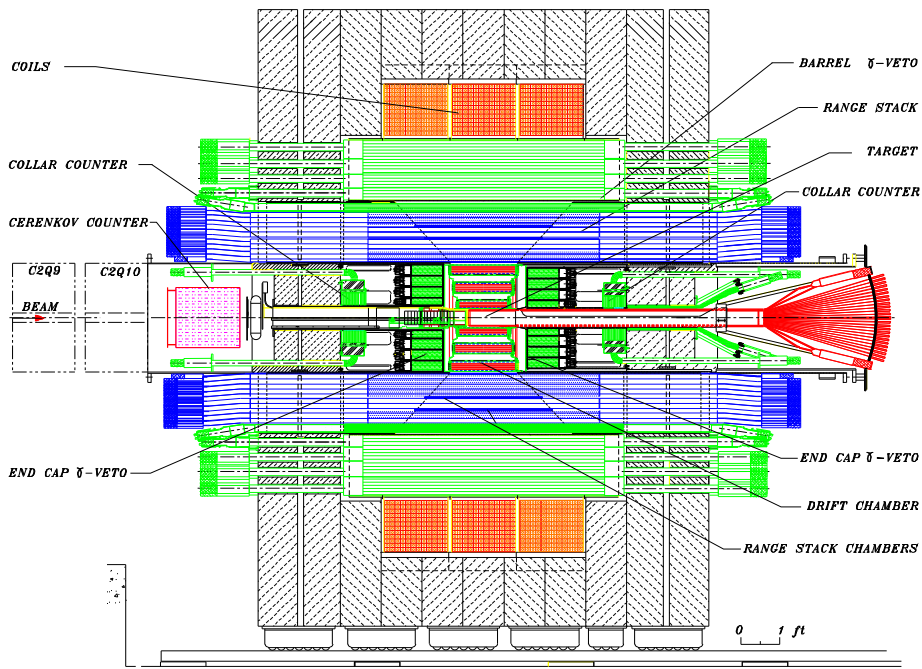, height=3.75in, width=5.25in}
\caption{Elevation view of the E949 detector at BNL.
\label{fig:e949_det}}
\end{figure}

The E949 experiment (see Figure~\ref{fig:e949_det}), 
upgrading the E787 detector, will
run from 2001 through 2003.
The E949 detector will increase
photon veto coverage with an additional Pb-scintillator
barrel veto liner and additional photon veto coverage along the beam axis.
Part of the range stack scintillator will be replaced to obtain
more light, and several non- or poorly working detectors will be
replaced. The trigger and DAQ systems will be substantially upgraded.

\subsubsection{KOPIO/E926}
\label{sec:e926}

\begin{figure}[htbp]
\epsfig{file=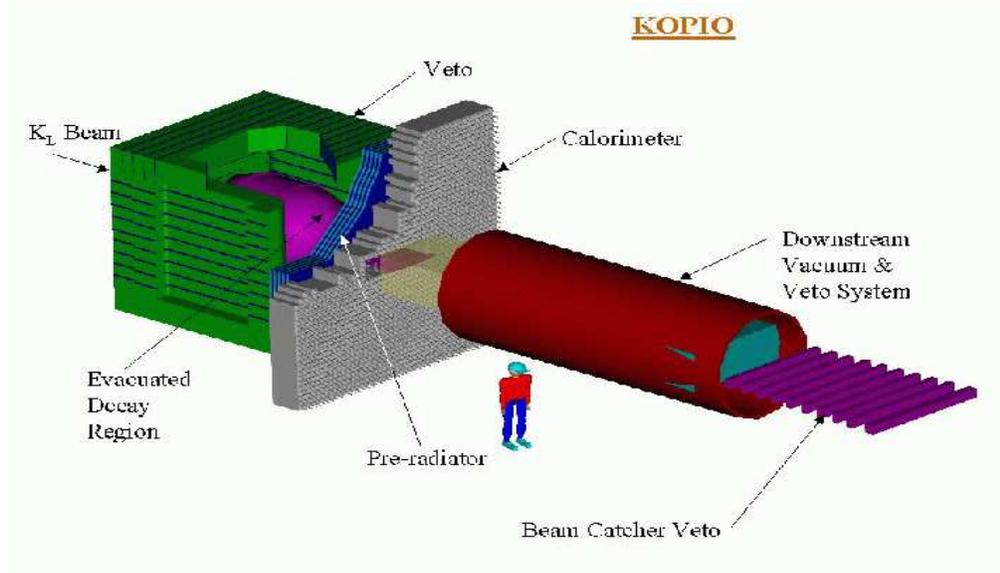,height=5.25in, width=3.0in, angle=270}
\caption{The Proposed KOPIO detector at BNL.
\label{fig:e926_det}}
\end{figure}

Experiment E926, named KOPIO, (see Figure~\ref{fig:e926_det}) 
received scientific approval at BNL in 1997 but is not yet funded.
Its proponents, together with those from the competing FNAL proposal
called KAMI (see below), are undertaking joint research and development
efforts with the aim of identifying the best technique for a future
$\klpnn$ measurement.
In the proposed KOPIO experiment, the AGS proton beam is micro-bunched,
with 200-ps bunches
every 40 ns. 
The angle of the neutral kaon beam is 45$^\circ$,
with an average $K_L$ momentum of 700 MeV/$c$. 
The kaon time of flight is
used to measure the kaon
momentum and the decay $\pi^\circ$ momentum in the kaon
center-of-mass frame.  The neutrons at this large targeting angle are mostly 
below $\pi^\circ$ production threshold. The beam is
very well collimated and flat. The detector consists
 of a Shashlyk calorimeter
and a pre-radiator of scintillator and copper drift
chambers. The vacuum decay volume is 
surrounded by  a charged
particle veto and a very thin vacuum tank. The outside of the tank
is surrounded by Pb-scintillator photon veto. 
A Pb-aerogel Cerenkov detector will be mounted in the beam hole for
additional photon veto.

\subsection{FNAL: Tevatron, Main Injector}

The Tevatron at FNAL began operation in 1984 
The typical proton intensity delivered to kaon experiments ranged
from $7\times 10^{11}$ protons per 20-s spill (60-s duty cycle) for E731,
the $\epsilon'/\epsilon$ experiment that collected data in 1987--1988, to as 
high as 10~Tp per 40-s spill (80-s duty cycle) during the 1999--2000
run of the rare decay experiment E799-II. The Main Injector was commissioned 
in 1999 and was used as an injector to the Tevatron during the 1999--2000 run.
Future fixed-target experiments
are proposed to run with a
120-GeV/$c$ beam directly from the Main Injector 
while the collider is running.  The Main Injector can supply 
30~Tp  per 3-s spill, allowing experiments to run at
much greater intensity than 
is possible at the Tevatron.

\subsubsection{KTeV}
\label{sec:e799}

Experiment E799-I ran during 1991 and early 1992, during the same 
fixed-target run as the {\it CP}-violation experiment E773, which used the same
detector.  The primary goal of E799-I was to search for \klpee and \klpmm, 
but a variety of other rare decays were 
studied~\cite{E731piee,E799oldpiee,klmmg,ksppg,kp0p0en,e799_pme,e799_eemm,e799_other}, 
including \kleeee, \klmmee and \klpme.  The single-event sensitivity achieved for
four-body modes was in the range of $10^{-9}.$

The KTeV  proposal consists of two experiments:  E832, a precision
measurement of $Re(\epsilon'/\epsilon),$ and E799-II, a second,
upgraded phase of the earlier E799-I, which had been envisioned in the
original E799 proposal. The centerpiece of the KTeV 
detector is a
large, high-precision CsI calorimeter~\cite{e799_det} capable of measuring 
photon and electron energies to better than 1\% precision.  The new
detector reuses the drift chambers from E799-I, but everything else
is new, including an extensive transition radiation detector (TRD)
system for enhanced pion-electron
separation, a greatly upgraded array of photon veto detectors, and
a trigger and data acquisition system with about 50 times the bandwidth
of the one used in E799-I.

\begin{figure}[htbp]
\epsfig{file=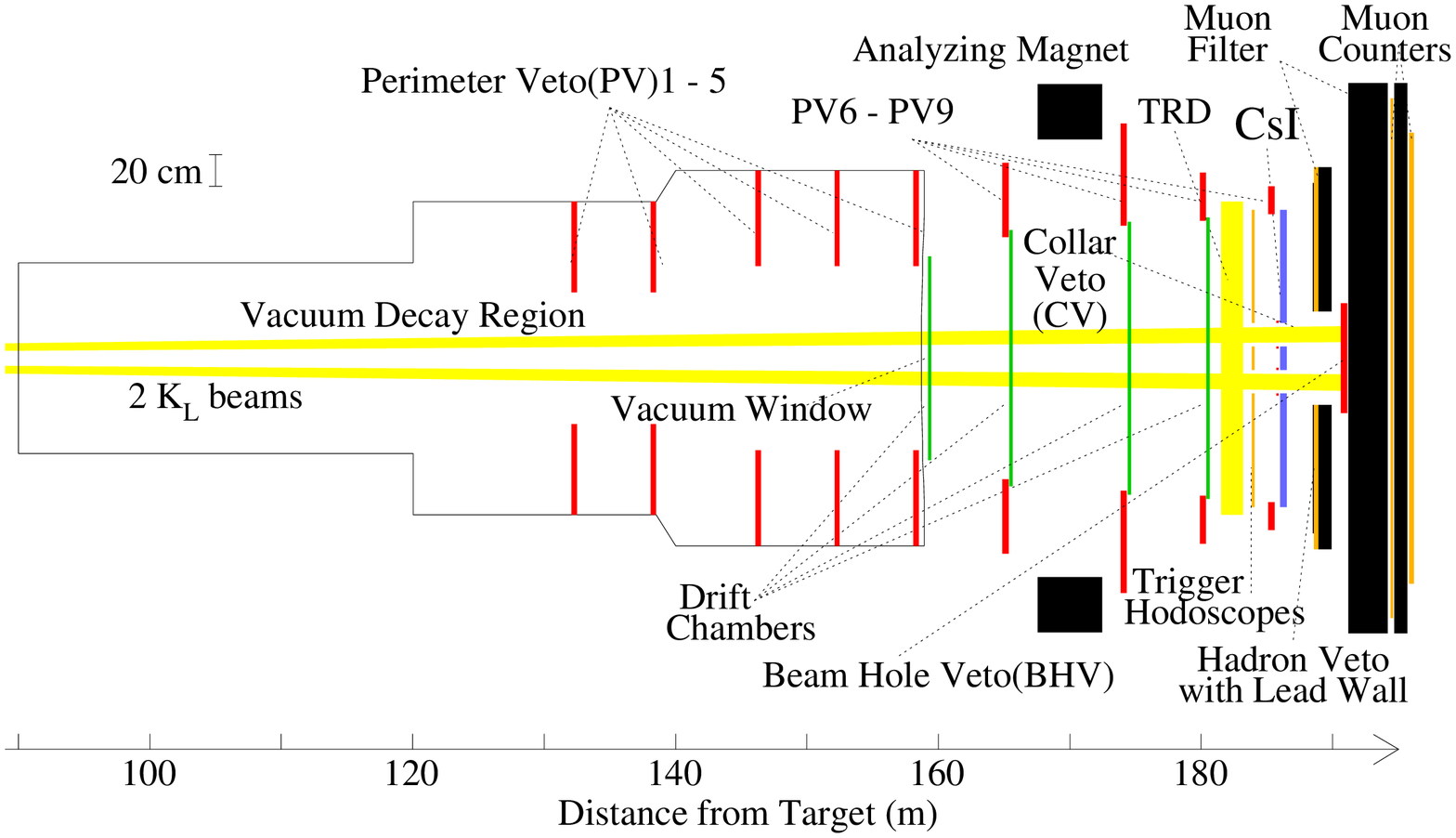,height=2.75in, width=5.25in,angle=0}
\caption{Plan view of the KTeV detector at the Tevatron at FNAL.
\label{fig:e799_det}}
\end{figure}

E799-II (see Figure~\ref{fig:e799_det}) collected data during 
fixed target runs in 1996--1997 and 1999--2000. A number of results 
from the analysis of the data from the first run have been published 
or reported at 
conferences~\cite{e799_pnn_gg,e799_pnn,klpgg,kpipieebr,kpipieeasymm}.  
Data from the 1999--2000 
run will increase the experiment's
rare kaon decay sensitivity by a 
factor of two to three,
depending on the mode.  With the complete data set,
E799-II should achieve an improvement of about
a factor of 20 over E799-I in single-event sensitivity for \klpee and \klpmm.

\subsubsection{KAMI/E804}
\label{sec:kami}

\begin{figure}[htbp]
\epsfig{file=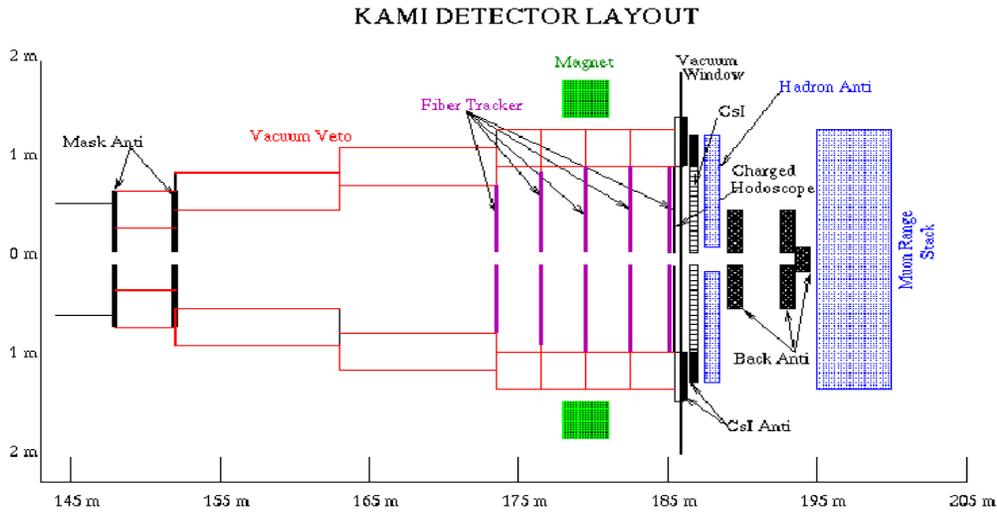,height=5.25in, width=2.75in,angle=-90}
\caption{Plan view of the proposed KAMI detector at FNAL.
\label{fig:e804_det}}
\end{figure}

KAMI (Kaons At the Main Injector) is the name of the
detector proposed in FNAL Expression of Interest 804 
(see Figure~\ref{fig:e804_det}).
As presently 
conceived, KAMI 
focuses on the difficult but rewarding mode \klpnn.  It competes with
the proposed KOPIO experiment at BNL (see above).  The KAMI proposal 
includes reusing the high-resolution
KTeV CsI calorimeter but replaces the tracking and photon veto systems.
In order to achieve the very high background rejection
needed to see a \klpnn signal at the standard-model level, KAMI uses
a completely hermetic photon veto system with good photon detection
efficiency down to energies as low as 20~MeV at large angles.  A major
challenge for KAMI is the design of a beam-hole photon detector that
would function in the presence of a very large neutron flux.  KAMI also
plans to continue the study of \klpee and \klpmm, as well as other
rare kaon decays, by building a system of fiber trackers.  KAMI is not
yet approved, but hopes to collect data beginning about 2005.

\subsubsection{CKM/E905}

\label{sec:e905}

\begin{figure}[htbp]
\epsfig{file=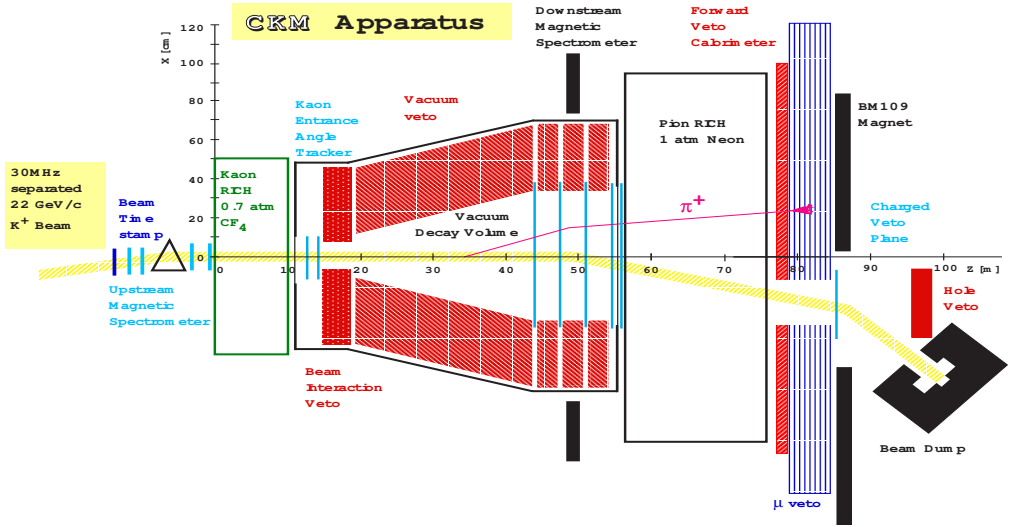,height=2.75in, width=5.25in,angle=0}
\caption{Plan view of the proposed CKM detector at FNAL.
\label{fig:e905_det}}
\end{figure}

The CKM (Charged Kaons at the Main injector) experiment (see Figure~\ref{fig:e905_det})
is proposed to run at the Main Injector at FNAL. 
In the proposed experiment,
a 22-GeV/$c$ separated $K^+$ beam, with $K/\pi$ = 2:1, and 30 MHz of
$K^+$ decays are delivered to the detector. A debunched proton 
beam of 5~Tp is extracted from the Main Injector with high
duty factor. The incident $K^+$ beam is momentum-analyzed in a Si-tracking
system before impinging on a kaon ring imaging Cerenkov hodoscope (RICH), 
where its velocity and direction are remeasured. The $K^+$ direction is
measured a final time inside the
vacuum decay volume; then the decay $\pi^+$ is momentum-analyzed and tracked
with low-mass straw tube chambers in the vacuum. A pion RICH measures the
$\pi^+$ velocity and direction. A muon veto system establishes that the
outgoing track was a $\pi^+$. The entire apparatus is surrounded by photon
veto systems.

\subsection{NA48}
\label{sec:na48}

\begin{figure}[htbp]
\epsfig{file=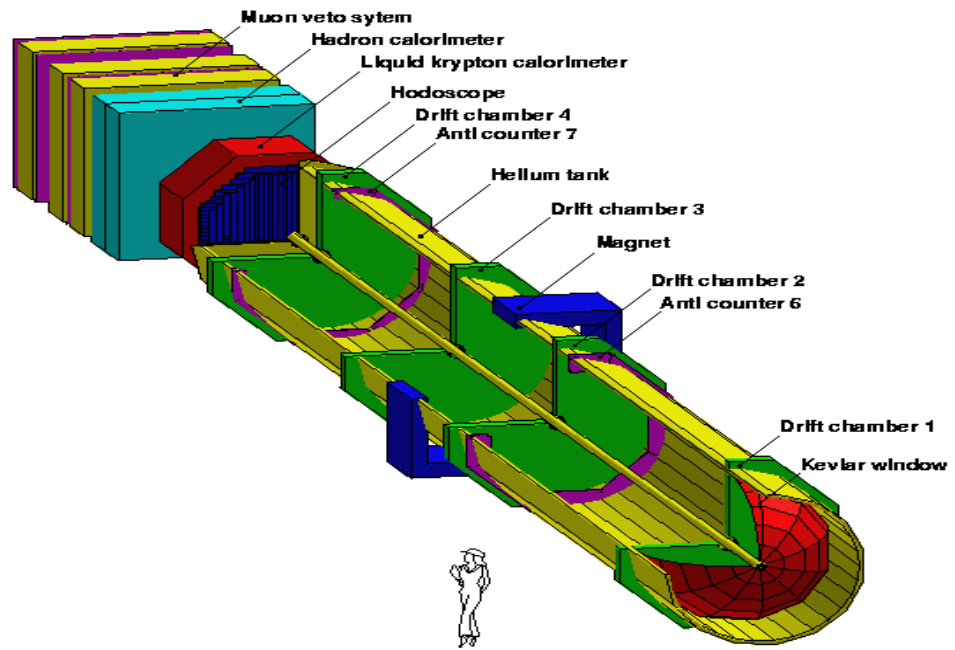,height=3.5in, width=5.25in,angle=0}
\caption{The NA48 detector at CERN.
\label{fig:na48_det}}
\end{figure}

The CERN Super Proton Synchrotron provides extracted protons
at 400--450~GeV/c for a fixed target program including the
rare kaon decay experiments NA31 and its successor, NA48.
The proton intensity was as high as 1--1.5 Tp with a 2.4-s spill
every 14.4 seconds. The NA31 experiment~\cite{na31_det},
designed to measure $\epsilon'/\epsilon$, also searched for 
rare $K_L$ and $K_S$ decays~\cite{klgg,klp0p0g,klpeng,na31_other}
from 1982 through 1991.
An upgraded experiment, NA48~\cite{na48_det}, 
again with the primary aim to
measure $\epsilon'/\epsilon$ will measure several rare 
decays~\cite{kleeg,klpmng,na48_mmg} as well (see Figure~\ref{fig:na48_det}).
The NA48 beamline is innovative in its use of a bent crystal to deflect
a small fraction of the proton beam onto a $K_S$ target just upstream of
the spectrometer.  This provides the opportunity to study rare $K_S$
decays.  The centerpiece of the NA48 detector is a 
high-precision liquid krypton calorimeter of unprecedented size.
Energy resolutions of
better than 1\% for photons and electrons have been
achieved with this device, which was completed in 1997.
NA48 had an engineering run in 1996 and data-taking
runs in 1997--1999. Unfortunately, on November 15, 1999, 
the vacuum pipe for the beam traversing the
center of the spectrometer imploded, destroying the
drift chamber spectrometer.  Current plans are for a short run in 2000
to search for \kspopopo and further running with the rebuilt
spectrometer starting in 2001.

\subsection{KLOE}
\label{sec:kloe}

The DA$\Phi$NE  $\Phi$ factory in Frascati,
is a 1.02-GeV $e^+ e^-$ collider sitting 
at the $\Phi$ peak.
The accelerator  began commissioning 
in early 1999, with a design luminosity goal of
$L = 5\times10^{32}$.

\begin{figure}[htbp]
\epsfig{file=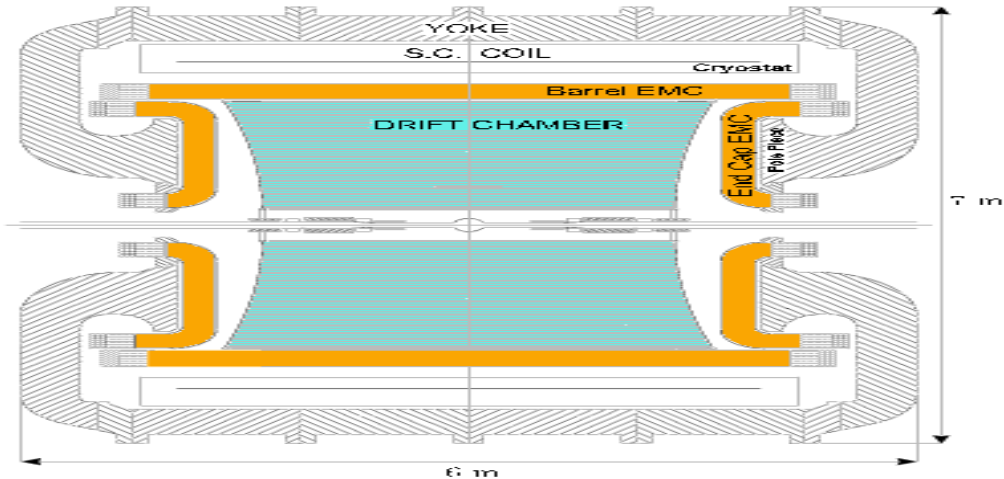,height=5.25in, width=2.5in,angle=-90}
\caption{Elevation view of the KLOE detector at Frascati.
\label{fig:kloe_det}}
\end{figure}

The KLOE 
experiment~\cite{kloe_det}
 was designed to measure $\epsilon'/\epsilon$,
although it will search for a variety of rare $K_L$, $K_S$ 
and $K^+$ decays (see Figure~\ref{fig:kloe_det}) 
with tagged kaons from copious $\Phi$ decays.
At the DA$\Phi$NE $e^+e^-$ collider KLOE is
working in the center of mass frame. It is a cylindrically symmetric 
general-purpose detector with a low-mass central drift chamber
of very large volume (4-m diameter), surrounded
by a Pb-scintillating fiber calorimeter. The detector is in a 0.6-T magnetic
field, so the calorimeter is read out with high-field fine-mesh PMTs.

\subsection{IHEP: Separated Kaon Experiment}
\label{sec:ihep}

The U-70 accelerator at IHEP in Serpukhov
has reached an intensity of 15~Tp. It runs
with a 2-s spill every 10~s, for a duty factor of 20\%.

A new experiment at IHEP is proposed 
to run in the N-21 line with a 12-GeV/$c$ separated
$K^\pm$ beam starting in 2002. This experiment will study
a variety of medium-rare (mostly radiative) kaon decays
and should be able
to substantially improve existing measurements.
This experiment will use a new RF separator from CERN~\cite{ihep} to provide
a K purity of 2:1. The experiment will reuse existing apparatus from
SPHINX, GAMS, and ISTRA-M.

\subsection{KEK}

The Proton Synchrotron (PS) at KEK
in Japan can deliver up to
6~Tp during a 0.7-s spill every 3~s.
During the past 15 years several experiments have been operating
in high intensity kaon beams, starting with E137,
E162 and now with E391a.


Experiment E137, the first $K_L$ decay experiment at the KEK PS
searched for the decay \klme~\cite{e137_me} from 1988 through 1990.
In addition to the search for \klme,
E137 set limits on \klee and
measured branching ratios for \klmm~\cite{e137_me,e137_3} and \kleeee~\cite{e137_4}.
The experiment
ran with 2~Tp of protons on target and $\sim10^7$ $K_L$ per spill.
The neutral-beam solid angle was 154 $\mu$str at production angles of 0$^\circ$ and
2$^\circ$.
The E137 detector consisted of a two-arm spectrometer with five drift chamber
stations and two dipole magnets, with a total $P_T$ of 238 MeV/$c$, in each 
arm. Particle identification consisted of
threshold gas Cerenkov counters, a Pb-scintillator 
electromagnetic calorimeter, and a muon
range stack.

\begin{figure}[htbp]
\epsfig{file=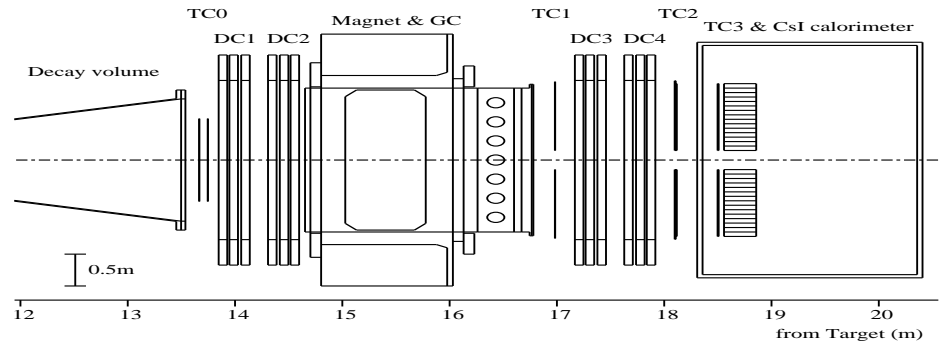,height=2.in, width=5.25in,angle=0}
\caption{Plan view of the E162 detector at KEK.
\label{fig:e162_det}}
\end{figure}

Experiment E162 was designed to search for \klpee
(see Figure~\ref{fig:e162_det}).
After an engineering run that indicated that neutron contamination
in the beam was too high, the experiment
changed focus to search for
\klppee~\cite{e162_ppee} and collected data from 1996 through 1997.
In addition to \klppee, a limit on \klpeeg~\cite{e162_3} was set.

Experiment E162
ran with a neutral kaon beam of 8 mrad $\times$ 20 mrad at a production
angle of 2$^\circ$ and 1~Tp of
protons on target in a 2-s spill every 4~s.
The spectrometer consisted of four drift chambers ($x$, $u$ and $v$) views
and a magnet with 136 MeV/$c$ $P_T$ kick. 
Particle identification included a threshold gas Cerenkov counter and an undoped
CsI calorimeter.

\label{sec:e391a}

\begin{figure}[htbp]
\epsfig{file=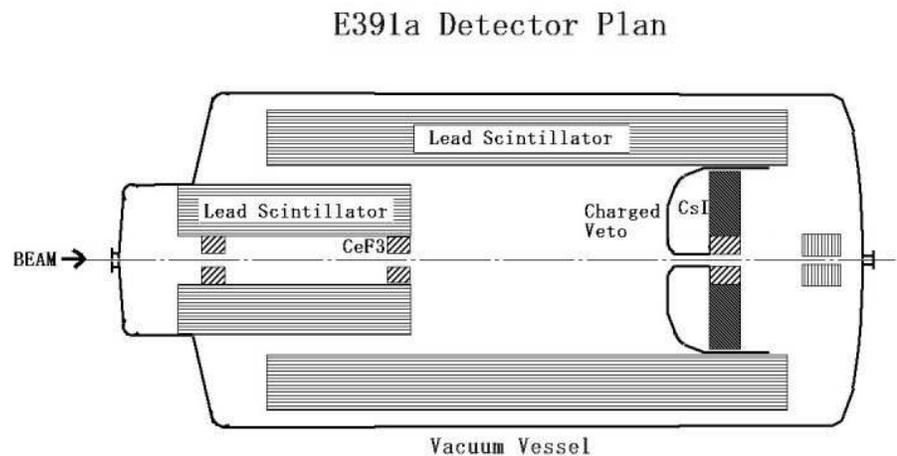,height=2.5in, width=5.25in,angle=0}
\caption{Plan view of the E391a detector at KEK.
\label{fig:e391a_det}}
\end{figure}

Experiment E391a will search for \klpnn (see Figure~\ref{fig:e391a_det}). It is
scheduled to run at the KEK PS  from 2001 through 2005.
The detector has a high-resolution crystal calorimeter and is otherwise 
surrounded by Pb-scintillator photon veto. The entire detector is
situated in vacuum. The pencil beam is incident on the detector, giving
a very small solid angle for background with a photon going down the beam
hole. There is a plan
to move this experiment to the JHF and, with the
increased flux, measure ${\cal O}(1000)$ events.

\section{Acknowledgments}

We would like to thank the many people who helped with this paper by
providing data and useful comments.
We would particularly like to acknowledge 
Laurie Littenberg, Yau Wah, Hong Ma, Stefan Pislak, Robin Appel,
Jim Lowe, Mike Zeller, Bill Molzon, Toshio Numao, Takao Inagaki, Takeshi Komatsubara,
Tadashi Nomura, Leonid Landsberg, Brad Cox, Breese Quinn, and Yoshi Kuno.
This work
was supported in part under US Department of Energy
contract \#DE-AC02-98CH10886.


\begin{thebibliography}{999}

\bibitem{litt} Littenberg L, Valencia G. {\it Annu.\ Rev.\ Nucl.\ Part.\ Sci.} 43:729 (1993)
\bibitem{hagelin} Hagelin J, Littenberg L. {\it Prog.\ Part.\ Nucl.\ Phys.} 23:1 (1989);
Ritchie J, Wojcicki S. {\it Rev.\ Mod.\ Phys.} 65:1149 (1993);
Buchholz P, Renk B. {\it Prog.\ Part.\ Nucl.\ Phys.} 39:253 (1997)
\bibitem{buchalla} Buchalla G, Buras AJ, Lautenbacher ME. {\it Rev.\ Mod.\ Phys.} 68:1125 (1996)
\bibitem{bf} Buras AJ, Fleischer R. In {\it Heavy Flavors II,} ed.\ AJ Buras,
M Lindner, pp.\ 65 Singapore: World Sci.\ (1997), also hep-ph/9704376 
\bibitem{buras} Buras AJ. In {\it Probing the Standard Model of Particle
Interactions,} ed.\ Gupta R, De Rafael E, David F, Morel A,
Vol.\ I, Ch.\ 5, New York, Elsevier Sci.\ (1999), also hep-ph/9806471
\bibitem{dambrosio} D'Ambrosio G, Isidori G. {\it Int.\ J.\ Mod.\ Phys.} A13:1 (1998)
\bibitem{winstein} Winstein B, Wolfenstein L. {\it Rev.\ Mod.\ Phys.} 65:1113 (1993); Bertolini S, Fabbrichesi M, Eeg JO {\it Rev.\ Mod.\ Phys.} 72:65 (2000)
\bibitem{fitch} Christenson JH, Cronin JW, Fitch VL, Turlay R. {\it Phys.\ Rev.\ Lett.} 13:138 (1964)
\bibitem{ktev_e} Alavi-Harati A, et al. {\it Phys.\ Rev.\ Lett.} 83:22 (1999)
\bibitem{na48_e} Fanti V, et al. {\it Phys.\ Lett.} B465:335 (1999);
Graziani G, In Ref.~\cite{moriond00}
\bibitem{brown} Brown R, et al. {\it Nature} 163:82 (1949)
\bibitem{pais} Pais A. {\it Phys.\ Rev.} 86:663 (1952);
Gell-Mann M. {\it Phys.\ Rev.} 92:833 (1953)
\bibitem{dalitz} Dalitz R. {\it Phys.\ Rev.} 94:1046 (1954)
\bibitem{lee} Lee TD, Yang CN. {\it Phys.\ Rev.} 104:254 (1956)
\bibitem{cabibbo} Cabibbo N. {\it Phys.\ Rev.\ Lett.} 10:531 (1963)
\bibitem{kobayashi} Kobayashi M, Maskawa T. {\it Prog.\ Theor.\ Phys.} 46:652 (1973)
\bibitem{glashow} Glashow SL, Iliopoulos J, Maiani L. {\it Phys.\ Rev.\ D} 2:1285 (1970)
\bibitem{GL}Gaillard MK, Lee BW. {\it Phys.\ Rev.\ D} 10:897 (1974);
Gaillard MK, Lee BW, Shrock RE {\it Phys.\ Rev.\ D} 13:2674 (1976)
\bibitem{ichep98} A. Astbury, et al., eds. {\it Proc.\ XXIX Int.\ Conf.\ High Energy Phys., Vancouver, Canada, July 1998}
Singapore: World Sci.\  (1999)
\bibitem{dpf99} Arisaka K and Bern Z, eds. {\it Proc.\ Meet.\ DPF, UCLA, Jan.\ 1999.}
http://www.dpf99.library.ucla.edu (1999)
\bibitem{moriond99} J. Tran Thanh Van, ed. {\it Proc.\ XXXIV Rencontres de Moriond, Les Arcs, France, March 1999} 
Paris: Ed. Frontieres (1999)
\bibitem{panic99} F\"{a}ldt G,  et al., eds. {\it Proc.\ XV$^{th}$ Particles and Nuclei Int.\ Conf., 
Uppsala, Sweden, June 1999} Singapore: World Sci.\ (2000)
\bibitem{epshep99} Huitu K, et al., eds. {\it Proc.\ Int.\ Europhys.\ Conf.\ High Energy Phys.,\ 
Tampere, Finland,  July 1999} Bristol, UK: IOP-Publishing
\bibitem{kaon99} Rosner JL and Winstein B, eds. 
{\it Kaon Physics,} Proc.\ Chicago Conf.\
Kaon Phys., June  1999 
Chicago: Univ.\ Chicago Press (2000)
\bibitem{hf99} Dauncey P and Sachrajda C, eds. {\it Proc.\ Heavy Flavours 8, 
Southampton, UK, July 1999} Southampton: J.\ High Energy Phys.\ (1999)
http://jhep.sissa.it/cgi-bin/PrHEP/cgi/reader/list.cgi?confid=3
\bibitem{lp99} Jaros J and Peskin M, eds. {\it Proc.\ XIX Int.\ Symp.\ Lepton and Photon Interact.\
at High Energies, Stanford, August 1999.} Singapore: World Sci.\ (2000)
\bibitem{daphne99} Bianco S, et al., eds. {\it Proc.\ 3rd DA$\Phi$NE Workshop Phys.\ and Detectors,
Frascati, Italy, Nov.\ 1999} Frascati Phys.\ Ser.\
Frascati: INFN Laboratori Nazionali di Frascati  (2000)
\bibitem{bconf99} Cheng HY and Hou WS, eds. {\it Proc.\ 3rd Int.\ Conf.\ B Phys.\ and CP Violation,
Taipei, Taiwan, Dec.\ 1999} Singapore: World Sci.\ (2000)
\bibitem{lathuile00} Bellettini G, Chiarelli G, Greco M, eds.\ {\it Proc.\ 14th Workshop in Part.\ Phys., La Thuile, Italy, Feb.\ 2000}
Frascati: INFN Laboratori Nazionali di Frascati  (2000)
\bibitem{moriond00} J. Tran Thanh Van, ed. {\it Proc.\ XXXV Rencontres de Moriond, Les Arcs, France, March 2000} 
Paris: Ed. Frontieres (2000)
\bibitem{ichep00} {\it Proc.\ XXX Int.\ Conf.\ High Energy Phys., Osaka, Japan, July 2000} Singapore: World Sci.\ (2001)
\bibitem{dpf00}  {\it Proc.\ Meet.\ DPF, Columbus OH, August 2000} Singapore: World Sci.\ (2001)
\bibitem{wolfenstein} Wolfenstein L. {\it Phys.\ Rev.\ Lett.} 51:1945 (1983)
\bibitem{buras2} Buras AJ, Lautenbacher ME, Ostermaier G. {\it Phys.\ Rev.\ D} 50:3433 (1994)
\bibitem{jarlskog} Jarlskog C. {\it Phys.\ Rev.\ Lett.} 55:1039 (1985); Jarlskog C. {\it Z.\ Phys.\ C} 29:491 (1985); Jarlskog C and Stora R. {\it Phys.\ Lett.} B208:268 (1988)
\bibitem{sinb} Bucahalla G and Buras AJ {\it Phys.\ Lett.} B333:221 (1994);
Bucahalla G and Buras AJ {\it Phys.\ Rev.\ D} 54:6782 (1996);
Nir Y and Worah MP.  {\it Phys.\ Lett.} B423:319 (1998)
\bibitem{bergmann} Bergmann S and Perez G. {\it J.\ High Energy Phys.} 0008:034 (2000)
\bibitem{bb3} Buchalla G, Buras A. {\it Nucl.\ Phys.} B548:309 (1999)\label{ref:bb3}
\bibitem{marciano2} Marciano WJ. In Ref.~\cite{kaon99} (2000)
\bibitem{chpt} Weinberg S. {\it Physica} A96:327 (1979);
Gasser J, Leutwyler H.  {\it Nucl.\ Phys.} B250:465 (1985);
Leutwyler H.  {\it Ann.\ Phys.} 235:165 (1994)
\bibitem{chpt_gl} Gasser J, Leutwyler H.  {\it Ann.\ Phys.} 158:142 (1984)
\bibitem{leutwyler} Leutwyler H, Roos M. {\it Z.\ Phys.\ C} 25:91 (1984)
\bibitem{pdg} Caso C, et al.  {\it Euro.\ Phys.\ J.} C3:1. http://pdg.lbl.gov (1998)
\bibitem{rosner} Rosner JL. In {\it Proc.\ 2${nd}$ Tropical Workshop in Part.\ Phys.\ and Cosm., San Juan,
Puerto Rico, May 2000} New York: AIP (2001)
\bibitem{ckmfit} Bargiotti M, et al. {\it Riv.\ Nuov.\ Cim.} 23:1 (2000), also hep-ph/0001293 (2000)
and references therein; Caravaglios F, Parodi F, Roudeau P, Stocchi A. In Ref.~\cite{bconf99} (2000)
\bibitem{sehgal} Sehgal LM. {\it Phys.\ Rev.} 183:1511 (1969)
\bibitem{dambrosio1}D'Ambrosio G, Isidori G, Portol{\'e}s J. {\it Phys.\ Lett.} B423:385 (1998)
\bibitem{dumm} G{\'o}mez Dumm D, Pich A. {\it Phys.\ Rev.\ Lett.} 80:4633 (1998)
\bibitem{valencia} Valencia G. {\it Nucl.\ Phys.} B517:339 (1998)
\bibitem{derafael} Knecht M, Peris S, Perrottet M, de Rafael E.  {\it Phys.\ Rev.\ Lett.} 83:5230 (1999)
\bibitem{geng} Geng CQ, Ng JN. {\it Phys.\ Rev.\ D} 41:2351 (1990)
\bibitem{inami} Inami T, Lim CS. {\it Prog.\ Theor.\ Phys.} 65:297 (1981)
\bibitem{e871_mm} Ambrose D, et al. {\it Phys.\ Rev.\ Lett.} 84:1389 (2000)
\bibitem{e871_ee} Ambrose D, et al. {\it Phys.\ Rev.\ Lett.} 81:4309 (1998)
\bibitem{DonoghueGabbiani} Donoghue JF, Gabbiani F. {\it Phys.\ Rev.\ D} 51:2187 (1995)
\bibitem{bosch} Bosch S, et al.  {\it Nucl.\ Phys.} B565:3 (2000); and other recent work, such as: Bijnens J, Prades J. {\it J.\ High Energy Phys.} 0006:035 (2000); Buras AJ, et al. hep-ph/0007313; Gardner S, Valencia G. {\it Phys.\ Rev.\ D}  62:094024 (2000)
\bibitem{kspee} Kekelidze V, et al. In Ref.~\cite{ichep00}
\bibitem{Greenlee} Greenlee HB. {\it Phys.\ Rev.\ D} 42:3724 (1990)
\bibitem{klpee} Alavi-Harati A, et al. {\it Phys.\ Rev.\ Lett.} In press  (2000), also hep-ex/0009030 (2000) 
\bibitem{E731piee} Barker A, et al. {\it Phys.\ Rev.\ D} 41:3546 (1990)
\bibitem{BNL845piee} Ohl KE, et al. {\it Phys.\ Rev.\ Lett.} 64:2755 (1990)
\bibitem{E799oldpiee} Harris DA, et al. {\it Phys.\ Rev.\ Lett.} 71:3918 (1993)
\bibitem{klpmm} Alavi-Harati A, et al. {\it Phys.\ Rev.\ Lett.} 84:5279 (2000)
\bibitem{bb1} Buchalla G, Buras A. {\it Nucl.\ Phys.} B412:106 (1994)
\bibitem{marciano} Marciano WJ, Parsa Z. {\it Phys.\ Rev.\ D} 53:R1 (1996)
\bibitem{bb2} Buchalla G, Buras A. {\it Phys.\ Rev.\ D} 57:216 (1998)
\bibitem{bsbd} Willocq S. In Ref.~\cite{bconf99} (2000), also www.cern.ch/LEPBOSC
\bibitem{buras3} Buras AJ, et al. {\it Nucl.\ Phys.}  B566:3 (2000)\label{ref:buras3} and 
references therein
\bibitem{theory_e} Pallante E, Pich A, {\it Phys.\ Rev.\ Lett.} 84:2568 (2000);
Buras AJ, In Ref.~\cite{kaon99}; 
Buras AJ, et al. {\it Nucl.\ Phys.} B408:209 (1993);
Bertolini S, Eeg JO, Fabbrichesi M, Lashin EI. {\it Nucl.\ Phys.} B514:93 (1998)
\bibitem{e787_pnn1} Adler S, et al. {\it Phys.\ Rev.\ Lett.} 79:2204 (1997)
\bibitem{e787_pnn2} Adler S, et al. {\it Phys.\ Rev.\ Lett.} 84:3768 (2000); 
Kettell SH. In Ref.~\cite{bconf99}
\bibitem{grossman} Grossman Y,  Nir Y. {\it Phys.\ Lett.} B398:163 (1997)
\bibitem{e799_pnn_gg} Adams J, et al. {\it Phys.\ Lett.} B447:240 (1999)
\bibitem{e799_pnn} Alavi-Harati A, et al. {\it Phys.\ Rev.\ D} 61:072006 (2000)
\bibitem{ppnn} Geng CQ, Hsu IJ, Lin YC. {\it Phys.\ Rev.\ D} 50:5744 (1994);
Littenberg LS, Valencia G. {\it Phys.\ Lett.} B385:379 (1996);
Gilman FJ, Chiang C-W. {\it Phys.\ Rev.\ D} 62:094026 (2000)
\bibitem{e787_ppnn} Adler S, et al. {\it Phys.\ Rev.\ D} In press (2000), also hep-ex/0009055 (2000)
\bibitem{kettell} Kettell SH. In Ref.~\cite{daphne99} (1999)
\bibitem{lowe} Lowe J. In Ref.~\cite{daphne99} (1999)
\bibitem{chpt_klpgg} Ecker G, et al. {\it Phys.\ Lett.} B189:363 (1987);
Cappiello L, D'Ambrosio G. {\it Nuov.\ Cim.} 99A:153 (1988)
\bibitem{numao} Numao T. {\it Mod.\ Phys.\ Lett.} A7:3357 (1992)
\bibitem{BMS} Bergstr{\"o}m L, Mass{\'o} E, Singer P. {\it Phys.\ Lett.} 
B131:229 (1983)
\bibitem{kleeg} Fanti V, et al.  {\it Phys.\ Lett.} B458:553 (1999)
\bibitem{klmmg} Quinn B. In Ref.~\cite{dpf00}
\bibitem{klgg} Burkhardt H, et al.  {\it Phys.\ Lett.} B199:139 (1987)
\bibitem{klmmee} Lath A. In Ref.~\cite{dpf99} (1999)
\bibitem{ksmm} Gjesdal S, et al.  {\it Phys.\ Lett.} B44:217 (1973)
\bibitem{ksee} Angelopoulos A, et al.  {\it Phys.\ Lett.} B413:232 (1997)
\bibitem{kleegg} Alavi-Harati A, et al. {\it Phys.\ Rev.\ D} In press (2000);
also hep-ex/0010059 (2000)
\bibitem{klmmgg} Alavi-Harati A, et al. {\it Phys.\ Rev.\ D} 62:112001 (2000)
\bibitem{chpt_av} Heiliger P, et al. {\it Phys.\ Rev.\ D} 47:4920 (1993)
\bibitem{klpgg} Alavi-Harati A, et al. {\it Phys.\ Rev.\ Lett.} 83:917 (1999)
\bibitem{na48_klpgg} Contalbrigo M. In Ref.~\cite{moriond00} (2000)
\bibitem{chpt_ecker} Ecker G, Pich A, de Rafael E. {\it Phys.\ Lett.} B237:481 (1990);
Cheng HY. {\it Phys.\ Rev.\ D} 42:72 (1990);
Bruno C, Prades J. {\it Z.\ Phys.\ C} 57:585 (1993)
\bibitem{chpt_ecker2} Ecker G, Kambor J, Wyler D. {\it Nucl.\ Phys.} B394:101 (1993)
\bibitem{chpt_kpgg} Ecker G, Pich A, de Rafael E. {\it Nucl.\ Phys.} B303:665 (1988)
\bibitem{chpt_unitarity} Cappiello L, D'Ambrosio G, Miragliuolo M. {\it Phys.\ Lett.} B298:423 (1993);
Cohen AG, Ecker G, Pich A. {\it Phys.\ Lett.} B304:347 (1993)
\bibitem{kpgg} Kitching P, et al. {\it Phys.\ Rev.\ Lett.} 79:4079 (1997)
\bibitem{klpeeg} Taegar S. In Ref.~\cite{dpf99} (1999)
\bibitem{chpt_pee} Ecker G, Pich A, de Rafael E. {\it Nucl.\ Phys.} B291:692 (1987)
\bibitem{chpt_damb} D'Ambrosio G, Ecker G, Isidori G, Portol{\'e}s J. {\it J.\ High Energy Phys.}  8:4 (1998)
\bibitem{cern_pee} Bloch P, et al. {\it Phys.\ Lett.} B56:201 (1975)
\bibitem{e777_pee} Alliegro C, et al. {\it Phys.\ Rev.\ Lett.} 68:278 (1992)
\bibitem{e851_pee} Deshpande AL. {\it A study of the decay of a positively
charged kaon into a positively charged pion, a positron and an electron,
and a measurement of the decay of a neutral pion into a positron and an 
electron.}
PhD thesis. Yale Univ. (1995)
\bibitem{e865_pee} Appel R, et al. {\it Phys.\ Rev.\ Lett.} 83:4482 (1999)
\bibitem{e787_pmm} Adler S, et al. {\it Phys.\ Rev.\ Lett.} 79:4756 (1997)
\bibitem{e865_pmm} Ma H, et al. {\it Phys.\ Rev.\ Lett.} 84:2580 (2000)
\bibitem{kpeeg} Kraus D. In Ref.~\cite{dpf99} (1999)
\bibitem{dambrosio2} D'Ambrosio G, Gao D-N. hep-ph/0010122;
Tandean J, Valencia G {\it Phys.\ Rev.\ D} In press (2000), 
also hep-ph/0008238;
D'Ambrosio G, Isidori G.  {\it Z.\ Phys.\ C} 65:649 (1995);
McGuigan M, Sanda AI. {\it Phys.\ Rev.\ D} 36:1413 (1987);
Ecker G, Neufeld H, Pich A. {\it Phys.\ Lett.} B278:337 (1992);
Ecker G, Neufeld H, Pich A. {\it Nucl.\ Phys.} B413:321 (1994);
Lin YCR, Valencia G. {\it Phys.\ Rev.\ D} 37:143 (1988);
Ko P, Truong TN. {\it Phys.\ Rev.\ D} 43:R4 (1991);
Picciotto C. {\it Phys.\ Rev.\ D} 45:1569 (1992);
Donoghue J, Holstein B, Lin YCR. {\it Nucl.\ Phys.} B277:651 (1986);
He X, Valencia G. {\it Phys.\ Rev.\ D} 61:075003 (2000);
Funck R, Kambor J.  {\it Nucl.\ Phys.} B396:53 (1993);
Heiliger P, Sehgal LM.  {\it Phys.\ Lett.} B307:182 (1993)
\bibitem{carroll} Carrol A, et al. {\it Phys.\ Rev.\ Lett.} 44:529 (1980)
\bibitem{ksppg_de} Taureg H, et al. {\it Phys.\ Lett.} B65:92 (1976)
\bibitem{ksppg} Ramberg EJ, et al. {\it Phys.\ Rev.\ Lett.} 70:2525 (1993)
\bibitem{klppg} Alavi-Harati A, et al. {\it Phys.\ Rev.\ Lett.} In press (2000),
also  hep-ex/0008045 (2000)
\bibitem{abrams} Abrams , et al. {\it Phys.\ Rev.\ Lett.} 29:1118 (1972)
\bibitem{smith} Smith KM, et al. {\it Nucl.\ Phys.} B109:173 (1976)
\bibitem{bolotov} Bolotov VN, et al. {\it Sov.\ J.\ Nucl.\ Phys.} 45:1023 (1987)
\bibitem{kppg} Adler S, et al. {\it Phys.\ Rev.\ Lett.} In press (2000),
also hep-ex/0007021 (2000)
\bibitem{ppee_asym1} Sehgal LM, Wanniger M. {\it Phys.\ Rev.\ D} 46:1035; 
{\it Phys.\ Rev.\ D} 46:5209 (E) (1992)
\bibitem{ppee_asym2} Heiliger P, Sehgal LM.  {\it Phys.\ Rev.\ D} 48:4146 (1993);
Elwood JK, Wise MB, Savage MJ.  {\it Phys.\ Rev.\ D} 52:5095 (1995);
{\it Phys.\ Rev.\ D} 53:2855 (E) (1996);
Elwood JK, Wise MB, Savage MJ, Walden JW.  {\it Phys.\ Rev.\ D} 53:4078 (1996);
Sehgal LM, van Leusen J.  {\it Phys.\ Rev.\ Lett.} 83:4933 (1999)
\bibitem{lubranohf8} Lubrano P. In Ref.~\cite{hf99} (2000)
\bibitem{kpipieebr} Adams J, et al. {\it Phys.\ Rev.\ Lett.} 80:4123 (1998)
\bibitem{barkerhf8} Barker AR. In Ref.~\cite{hf99} (2000); 
Senyo K. In Ref.~\cite{epshep99} (2000)
\bibitem{kpipieeasymm} Alavi-Harati A, et al. {\it Phys.\ Rev.\ Lett.} 84:408 (2000)
\bibitem{kspipieebr} Contalbrigo M. In Ref~\cite{moriond00}
\bibitem{klp0p0g} Barr GD, et al. {\it Phys.\ Lett.} B328:528 (1994)
\bibitem{chpt_kmng} Donoghue J, Holstein B. {\it Phys.\ Rev.\ D} 50:3700 (1989);
Ametller Ll. {\it Phys.\ Lett.} B303:140 (1993)
\bibitem{chpt_kmng2} Bijnens J, Ecker G, Gasser J. {\it Nucl.\ Phys.} B396:81 (1993)
\bibitem{kmng_de}Adler S, et al. {\it Phys.\ Rev.\ Lett.} 85:2256 (2000)
\bibitem{keng} Heintze J, et al. {\it Nucl.\ Phys.} B149:365 (1979)
\bibitem{chpt_kmng_p6} Bijnens J, Talavera P. {\it Nucl.\ Phys.} B489:387 (1997)
\bibitem{kmng} Akiba Y, et al.  {\it Phys.\ Rev.\ D} 32:2911 (1985)
\bibitem{e865_enee} Poblaguev A, private communication;
Zeller M. In Ref.~\cite{kaon99} (2000)
\bibitem{zeller00} Ma H, private communication
\bibitem{kmnmm} Atiya MS, et al. {\it Phys.\ Rev.\ Lett.} 63:2177 (1989)
\bibitem{kenmm} Adler S, et al. {\it Phys.\ Rev.\ D} 58:012003-1 (1998)
\bibitem{chpt_ppen0} Weinberg S. {\it Phys.\ Rev.\ Lett.} 17:616  (1966);
Gasser J, Leutwyler H. {\it Phys.\ Lett.} B125:325 (1983)
\bibitem{chpt_ppen} Bijnens J, et al. {\it Phys.\ Lett.} B374:210 (1996);
Knecht M, Moussallam B, Stern J,  Fuchs NH. {\it Nucl.\ Phys.} B457:513 (1995);
Bijnens J, et al.  {\it Nucl.\ Phys.} B508:263 (1997);
Amoros J, Bijnens J.  {\it J.\ Phys.} G25:1607 (1999);
Amoros J, Bijnens J, Talavera P. hep-ph/9912398 (1999)
\bibitem{rosselet} Rosselet L, et al. {\it Phys.\ Rev.\ D} 15:574 (1977)
\bibitem{e865_ppen1} Pislak S. {\it Proc.\ Workshop on Hadronic Atoms, Bern, Oct.\ 1999} 
\bibitem{schenk} Schenk A. {\it Nucl.\ Phys.\ } B363:97 (1991); 
Ananthanarayan B, Colangelo G, Gasser J, Leutwyler H. hep-ph/0005297 (2000)
\bibitem{e865_ppen2} Pislak S, private communication
\bibitem{basdevant} Basdevant JL, Froggatt CD, Petersen JL. {\it Nucl.\ Phys.} B72:413 (1974)
\bibitem{kppmn} Bisi V, Cester R, Chiesa AM, Vigone M. {\it Phys.\ Lett.} B25:572 (1967)
\bibitem{klppen} Makoff G, et al. {\it Phys.\ Rev.\ Lett.} 70:1591 (1993)
\bibitem{kp0p0en} Barmin VV, et al. {\it Sov.\ J.\ Nucl.\ Phys.} 48:1032 (1988)
\bibitem{kp0p0eng} Barmin VV, et al. {\it Sov.\ J.\ Nucl.\ Phys.} 55:547 (1992)
\bibitem{ksppp} Adler R, et al. {\it Phys.\ Lett.} B407:193 (1997)
\bibitem{ksp0p0p0} Achasov, MN et al. {\it Phys.\ Lett.} B459:674 (1999)
\bibitem{kpppg} Barmin VV, et al.  {\it Sov.\ J.\ Nucl.\ Phys.} 50:421 (1989)
\bibitem{kpp0p0g} Bolotov VN, et al.  {\it J.\ Exp.\ Theor.\ Phys.\ Lett.} 42:481 (1995)
\bibitem{kpmng} Ljung D, Cline D.  {\it Phys.\ Rev.\ D} 8:1307 (1973)
\bibitem{klpmng} Bender M, et al. {\it Phys.\ Lett.} B418:411 (1998)
\bibitem{kpeng} Barmin VV, et al. {\it Sov.\ J.\ Nucl.\ Phys.} 53:606 (1991)
\bibitem{kpeng_de} Bolotov VN, et al.  {\it Sov.\ J.\ Nucl.\ Phys.} 44:68 (1986)
\bibitem{klpeng} Leber F, et al. {\it Phys.\ Lett.} B369:69 (1996)
\bibitem{Wilczek} Wilczek F, Zee A. {\it Phys.\ Rev.\ Lett.} 42:421 (1979);
Cahn R, Harari H. {\it Nucl.\ Phys.} B176:135 (1980)
\bibitem{Eichten} Eichten E, Lane K. {\it Phys.\ Lett.} B90:125 (1980);
Dimopoulous S, Ellis J. {\it Nucl. Phys.} B182:505 (1981);
Shanker O. {\it Nucl.\ Phys.} B206:253 (1982);
Haber HE, Kane GL. {\it Phys.\ Rep.} 117:75 (1985);
Bigi II. {\it Phys.\ Lett.} B166:238 (1986);
Pati J, Stremnitzer H. {\it Phys.\ Lett.} B172:441 (1986);
Campbell BA, et al. {\it Int.\ J.\ Mod.\ Phys.} A2:831 (1987);
Langacker P, Sankar SU, Schilcher K. {\it Phys.\ Rev.\ D} 38:2841 (1988);
Mukhopadhyaya B, Raychaudhuri A. {\it Phys.\ Rev.\ D} 42:3515 (1990);
Gagyi-Palffy Z, Pilaftsis A, Schilcher K. {\it Nucl.\ Phys.} B513:517 (1998);
Lee BW, Shrock RE,  {\it Phys.\ Rev.\  D} 16:1444 (1977)
\bibitem{e780_me} Greenlee HB, et al. {\it Phys.\ Rev.\ Lett.} 60:893 (1988);
Schaffner SF, et al. {\it Phys.\ Rev.\ D}  39:990 (1989)
\bibitem{e791_me} Cousins RD, et al. {\it Phys.\ Rev.\ D} 38:2914 (1988);
Mathiazhagan C, et al. {\it Phys.\ Rev.\ Lett.} 63:2181 (1989);
Arisaka K, et al. {\it Phys.\ Rev.\ Lett.} 70:1049 (1993)
\bibitem{e137_me} Inagaki T, et al. {\it Phys.\ Rev.\ D} 40:1712 (1989);
Akagi T, et al. {\it Phys.\ Rev.\ Lett.} 67:2614 (1991);
Akagi T, et al. {\it Phys.\ Rev.\ D} 51:2061 (1995)
\bibitem{e871_me} Ambrose D, et al. {\it Phys.\ Rev.\ Lett.} 81:5734 (1998)
\bibitem{e865_95} Bergman DR.  {\it A search for the decay \kpme}.
PhD thesis. Yale Univ. (1998);
Pislak S. {\it Experiment E865 at BNL: a search for the decay \kpme}.
PhD thesis. Univ. Z\"{u}rich (1998)
\bibitem{e777_old} Lee AM, et al. {\it Phys.\ Rev.\ Lett.} 64:165 (1990);
Campagnari C, et al. {\it Phys.\ Rev.\ Lett.} 61:2062 (1988);
Baker NJ, et al. {\it Phys.\ Rev.\ Lett.} 59:2832 (1987)
\bibitem{zeller} Appel R, et al. {\it Phys.\ Rev.\ Lett.} 85:2450 (2000)
\bibitem{e799_pme} Arisaka K, et al. {\it Phys.\ Lett.} B432:230 (1998)
\bibitem{ktevpme} Bellavance A. In Ref.~\cite{dpf00} (2000)
\bibitem{kpmm_opp} Appel R, et al. {\it Phys.\ Rev.\ Lett.} 85:2877 (2000)
\bibitem{shrock} Littenberg LS, Shrock RE {\it Phys.\ Lett.} B In Press, also hep-ph/0005285;
Zuber K {\it Phys.\ Lett.} B479:33 (2000);
Littenberg LS, Shrock RE {\it Phys.\ Lett.} B68:443 (1992)
\bibitem{e799_eemm} Gu P, et al. {\it Phys.\ Rev.\ Lett.} 76:4312 (1996)
\bibitem{pee}  Deshpande A, et al. {\it Phys.\ Rev.\ Lett.} 71:27 (1993)
\bibitem{e865_nim} Appel R, et al. submitted to {\it Nucl.\ Instrum.\ Methods Phys.\ Res.,\ Sect.\ A}  (2000)
\bibitem{e791_det} Lee DM, et al. {\it Nucl.\ Instrum.\ Methods Phys.\ Res.,\ Sect.\ A} 256:329 (1987);
Kenney CJ, et al. {\it IEEE Trans.\ Nucl.\ Sci.} 36:74 (1989);
Frank J, et al. {\it IEEE Trans.\ Nucl.\ Sci.} 36:79 (1989);
Cousins RD, et al. {\it IEEE Trans.\ Nucl.\ Sci.} 36:646 (1989);
Biery KA, et al. {\it IEEE Trans.\ Nucl.\ Sci.} 36:650 (1989);
Cousins RD, et al. {\it Nucl.\ Instrum.\ Methods Phys.\ Res.,\ Sect.\ A} 277:517 (1989)
\bibitem{e791_other} Mathiazhagan C, et al. {\it Phys.\ Rev.\ Lett.} 63:2185 (1989);
Heinson AP, et al. {\it Phys.\ Rev.\ D} 44:R1 (1991);
Arisaka K, et al. {\it Phys.\ Rev.\ Lett.} 71:3910 (1993);
Heinson AP, et al. {\it Phys.\ Rev.\ D} 51:985 (1995)
\bibitem{e871_det} Belz J, et al. {\it Nucl.\ Instrum.\ Methods Phys.\ Res.,\ Sect.\ A} 428:239 (1999)
\bibitem{e787_det} Ahmad S, et al. {\it IEEE Trans.\ Nucl.\ Sci.} 33:178 (1986);
Cresswell JV, et al. {\it IEEE Trans.\ Nucl.\ Sci.} 35:460 (1988);
Atiya MS, et al. {\it IEEE Trans.\ Nucl.\ Sci.} 36:813 (1989);
Atiya MS, et al. {\it Nucl.\ Instrum.\ Methods Phys.\ Res.,\ Sect.\ A} 279:180 (1989);
Atiya MS, et al. {\it Nucl.\ Instrum.\ Methods Phys.\ Res.,\ Sect.\ A} 321:129 (1992)
\bibitem{atiya} Atiya MS, et al. {\it Phys.\ Rev.\ Lett.} 64:21 (1990);
Atiya MS, et al. {\it Nucl.\ Phys.\ B Proc.\ Suppl.} 13:568 (1990);
Atiya MS, et al. {\it Phys.\ Rev.\ Lett.} 65:1188 (1990);
Atiya MS, et al. {\it Phys.\ Rev.\ Lett.} 66:2189 (1991);
Atiya MS, et al. {\it Phys.\ Rev.\ Lett.} 69:733 (1992);
Atiya MS, et al. {\it Nucl.\ Phys.} A527:727c (1991);
Atiya MS, et al. {\it Phys.\ Rev.\ Lett.} 70:2521 (1993);
Atiya MS, et al. {\it Phys.\ Rev.\ D} 48:R1 (1993);
Adler S, et al. {\it Phys.\ Rev.\ Lett.} 76:1421 (1996)
\bibitem{e787_det2} Burke M, et al. {\it IEEE Trans.\ Nucl.\ Sci.} 41:131 (1946); 
Kobayashi M, et al. {\it Nucl.\ Instrum.\ Methods Phys.\ Res.,\ Sect.\ A} 337:355 (1994);
Chiang IH, et al. {\it IEEE Trans.\ Nucl.\ Sci.} 42:394 (1995);
Bryman DA, et al. {\it Nucl.\ Instrum.\ Methods Phys.\ Res.,\ Sect.\ A} 396:394 (1997);
Komatsubara TK, et al. {\it Nucl.\ Instrum.\ Methods Phys.\ Res.,\ Sect.\ A} 404:315 (1998);
Blackmore EW, et al. {\it Nucl.\ Instrum.\ Methods Phys.\ Res.,\ Sect.\ A} 404:295 (1998);
Doornbos J, et al. {\it Nucl.\ Instrum.\ Methods Phys.\ Res.,\ Sect.\ A} 444:546 (2000)
\bibitem{e799_other} Papadimitriou V, et al. {\it Phys.\ Rev.\ D} 44:573 (1991);
Graham G, et al. {\it Phys.\ Lett.} B295:169 (1992);
Krolak P, et al. {\it Phys.\ Lett.} B320:407 (1994);
Gu P, et al. {\it Phys.\ Rev.\ Lett.} 72:3000 (1994);
Weaver M, et al. {\it Phys.\ Rev.\ Lett.} 72:3758 (1994);
Roberts D, et al. {\it Phys.\ Rev.\ D} 50:1874 (1994);
Nakaya T, et al. {\it Phys.\ Rev.\ Lett.} 73:2169 (1994)
\bibitem{e799_det} Whitmore J. {\it Nucl.\ Instrum.\ Methods Phys.\ Res.,\ Sect.\ A} 409:687 (1998);
Bown C, et al. {\it Nucl.\ Instrum.\ Methods Phys.\ Res.,\ Sect.\ A} 369:248 (1996)
\bibitem{na31_det} Burkhardt H, et al. {\it Nucl.\ Instrum.\ Methods Phys.\ Res.,\ Sect.\ A} 268:116 (1988);
Barr GD, et al. {\it IEEE Trans.\ Nucl.\ Sci.} 36:66 (1989);
Barr G, et al. {\it Nucl.\ Instrum.\ Methods Phys.\ Res.,\ Sect.\ A} 294:465 (1990)
\bibitem{na31_other} Barr GD, et al. {\it Phys.\ Lett.} B235:356 (1990);
Barr GD, et al. {\it Phys.\ Lett.} B240:283 (1990);
Barr GD, et al. {\it Phys.\ Lett.} B242:523 (1990);
Barr GD, et al. {\it Phys.\ Lett.} B259:389 (1991);
Barr GD, et al. {\it Phys.\ Lett.} B284:440 (1992);
Barr GD, et al. {\it Phys.\ Lett.} B304:381 (1993);
Kreutz A, et al. {\it Z.\ Phys.\ C} 65:67 (1995);
Barr GD, et al. {\it Z.\ Phys.\ C} 65:361 (1995);
Barr GD, et al. {\it Phys.\ Lett. }  B351:579 (1995)
Barr GD, et al. {\it Phys.\ Lett.} B358:339 (1995)
\bibitem{na48_det} Buchholz P, et al. {\it Nucl.\ Instrum.\ Methods Phys.\ Res.,\ Sect.\ A} 316:1 (1992);
Barr GD, et al. {\it Nucl.\ Instrum.\ Methods Phys.\ Res.,\ Sect.\ A} 323:393 (1992);
Fanti V, et al. {\it Nucl.\ Instrum.\ Methods Phys.\ Res.,\ Sect.\ A} 344:507 (1994);
Ceccucci A, et al. {\it Nucl.\ Instrum.\ Methods Phys.\ Res.,\ Sect.\ A} 360:224 (1995);
Barr GD, et al. {\it Nucl.\ Instrum.\ Methods Phys.\ Res.,\ Sect.\ A} 370:413 (1996);
Gorini B, et al. {\it IEEE Trans.\ Nucl.\ Sci.} 45:1771 (1998);
Anvar S, et al. {\it Nucl.\ Instrum.\ Methods Phys.\ Res.,\ Sect.\ A} 419:686 (1998);
Bergauer H, et al. {\it Nucl.\ Instrum.\ Methods Phys.\ Res.,\ Sect.\ A} 419:623 (1998);
Hallgren B, et al. {\it Nucl.\ Instrum.\ Methods Phys.\ Res.,\ Sect.\ A} 419:680 (1998);
Fischer G, et al. {\it Nucl.\ Instrum.\ Methods Phys.\ Res.,\ Sect.\ A} 419:695 (1998);
Schinzel D, et al. {\it Nucl.\ Instrum.\ Methods Phys.\ Res.,\ Sect.\ A} 419:217 (1998);
Palestini S, et al. {\it Nucl.\ Instrum.\ Methods Phys.\ Res.,\ Sect.\ A} 421:75 (1998)
\bibitem{na48_mmg} Fanti V, et al. {\it Z.\ Phys.\ C} 76:653 (1998)
\bibitem{kloe_det} Finocchiaro G, et al. {\it Nucl.\ Instrum.\ Methods Phys.\ Res.,\ Sect.\ A} 360:48 (1997);
Franzini J, et al. {\it Nucl.\ Instrum.\ Methods Phys.\ Res.,\ Sect.\ A} 360:201 (1997);
Calcaterra A, et al. {\it Nucl.\ Instrum.\ Methods Phys.\ Res.,\ Sect.\ A} 367:104 (1997);
Grangnolo F, et al. {\it Nucl.\ Instrum.\ Methods Phys.\ Res.,\ Sect.\ A} 367:108 (1997);
Antonelli M, et al. {\it Nucl.\ Instrum.\ Methods Phys.\ Res.,\ Sect.\ A} 368:352 (1997);
Bossi F, et al. {\it Nucl.\ Instrum.\ Methods Phys.\ Res.,\ Sect.\ A} 379:536 (1997);
Antonelli M, et al. {\it Nucl.\ Phys.\ B Proc.\ Suppl.} 54:14 (1997);
Dell'Agnello S, et al. {\it Nucl.\ Phys.\ B Proc.\ Suppl.} 54:57 (1997);
Elia V, et al. {\it Nucl.\ Phys.\ B Proc.\ Suppl.} 54:66 (1997);
Spagnolo S, et al. {\it Nucl.\ Phys.\ B Proc.\ Suppl.} 54:70 (1997);
Lacava F, et al. {\it Nucl.\ Phys.\ B Proc.\ Suppl.} 54:327 (1997)
\bibitem{ihep} Landsberg L. In Ref.~\cite{kaon99} (2000)
\bibitem{e137_3} Akagi T, et al. {\it Phys.\ Rev.\ Lett.} 67:2618 (1991)
\bibitem{e137_4} Akagi T, et al. {\it Phys.\ Rev.\ D} 47:R2644 (1993)
\bibitem{e162_ppee} Nomura T, et al. {\it Phys.\ Lett.} B408:445 (1997);
Takeuchi Y, et al. {\it Phys.\ Lett.} B443:409 (1998)
\bibitem{e162_3} Murakami K, et al. {\it Phys.\ Lett.} B463:333 (1999)
\end{thebibliography}
\end{document}